\documentclass{article}
\usepackage{amssymb}
\usepackage{amsmath}
\usepackage{cprotect}
\usepackage{cite}
\usepackage{hyperref}
\usepackage{amsfonts}
\usepackage{comment}
\usepackage{bbm}

\usepackage{soul}
\usepackage{tikz}
\usepackage{tikz-cd}
\usepackage{extarrows}
\usepackage{tensor}
\usepackage{subcaption}
\usetikzlibrary{arrows,decorations.markings,decorations.pathreplacing}
\usepackage{bbm}
\usepackage{cancel}

\input xy
\xyoption{all}
\usepackage{enumitem}

\usepackage{graphicx} 

\usepackage{arydshln}

\newcommand{\be}{\begin{equation}}
	\newcommand{\ee}{\end{equation}}

\newcommand{\al}{\alpha}
\renewcommand{\d}{\delta}
\newcommand{\e}{\epsilon}

\newcommand{\g}{\gamma}

\newcommand{\la}{\lambda}
\newcommand{\m}{\mu}

\newcommand{\Om}{\Omega}
\newcommand{\om}{\omega}

\newcommand{\cat}{\mathcal{C}}

\newcommand{\id}{\mathbbm{1}}

\newcommand{\C}{\mathbb{C}}
\newcommand{\Z}{\mathbb{Z}}

\newcommand{\Hom}{\operatorname{Hom}}
\newcommand{\Rep}{\operatorname{Rep}}

\newcommand{\hlf}{\frac{1}{2}}

\newcommand{\non}{\nonumber}

\newcommand{\GL}{\operatorname{GL}}

\newcommand{\lp}{\left(}
\newcommand{\rp}{\right)}
\newcommand{\ls}{\left[}
\newcommand{\rs}{\right]}

\newcommand{\btau}{\bar{\tau}}

\newcommand{\ov}[1]{{\overline{#1}}}

\def\sqr#1#2{{\vcenter{\vbox{\hrule height.#2pt
            \hbox{\vrule width.#2pt height#1pt \kern#1pt
                  \vrule width.#2pt}\hrule height.#2pt}}}}

\def\sqra#1#2#3{{\vcenter{\vbox{\hrule height.#2pt
            \hbox{\vrule width.#2pt height#1pt \kern5pt 
#3
                  \vrule width.#2pt}\hrule height.#2pt}}}}

\tikzset{
	partial ellipse/.style args={#1:#2:#3}{
		insert path={+ (#1:#3) arc (#1:#2:#3)}
	}
}

\numberwithin{equation}{section}

\numberwithin{table}{section}

\begin{document}

\begin{center}

{\large\bf Notes on gauging noninvertible symmetries, part 2: higher multiplicity cases}

A.~Perez-Lona$^1$, D.~Robbins$^2$, E.~Sharpe$^1$, T.~Vandermeulen$^3$, X.~Yu$^1$

        \vspace*{0.1in}
        
        \begin{tabular}{cc}
                {\begin{tabular}{l}
                $^1$ Department of Physics MC 0435\\
                                850 West Campus Drive\\
                                Virginia Tech\\
                                Blacksburg, VA  24061 \end{tabular}}
                                 &
                {\begin{tabular}{l}
                $^2$ Department of Physics\\
                                University at Albany\\
                                Albany, NY 12222 \end{tabular}}
        \end{tabular}
        
                {\begin{tabular}{l}
                                $^3$ George P.~and Cynthia W.~Mitchell Institute\\
                                for Fundamental Physics and Astronomy\\
                                Texas A\&M University\\
                                College Station, TX 77843 \end{tabular}}
                        
        \vspace*{0.2in}

        {\tt aperezl@vt.edu},
        {\tt dgrobbins@albany.edu},
        {\tt ersharpe@vt.edu},
        {\tt tvand@tamu.edu},
        {\tt xingyangy@vt.edu}
        
\end{center}

In this paper we discuss gauging noninvertible zero-form symmetries in two dimensions, extending our previous work.  Specifically, in this work we discuss more general gauged noninvertible symmetries in which the noninvertible symmetry is not multiplicity free, and discuss the case of Rep$(A_4)$ in detail. We realize Rep$(A_4)$ gaugings for the $c=1$ CFT at the exceptional point in the moduli space and find new self-duality under gauging a certain non-group algebra object, leading to a larger noninvertible symmetry Rep$(SL(2,\Z_3))$. 
We also discuss more general examples of decomposition in two-dimensional gauge theories with trivially-acting
gauged noninvertible symmetries.

\begin{flushleft}
August 2024
\end{flushleft}

\newpage

\tableofcontents

\newpage

\section{Introduction}

Recently there has been a great deal of interest in global noninvertible symmetries, see for example 
\cite{McGreevy:2022oyu,Cordova:2022ruw,Schafer-Nameki:2023jdn,Brennan:2023mmt,Bhardwaj:2023kri,Luo:2023ive,Shao:2023gho} for some overviews.
Gauging of such noninvertible symmetries in two-dimensional theories was described in
\cite{Perez-Lona:2023djo,CLS23,Diatlyk:2023fwf}.  These papers only considered gauging ``multiplicity-free'' cases,
meaning, cases in which the space of junction operators at an intersection of any three simple lines is at most one-dimensional.
In this paper, we generalize to consider gauging noninvertible symmetries in two dimensions which are not multiplicity-free -- for which the Hilbert space of junctions is greater than one dimension.  

We also discuss more general decompositions.  Decomposition, first discussed in \cite{Hellerman:2006zs}, is the statement that a local quantum field theory in $d$ dimensions with a global $(d-1)$-form symmetry is equivalent to (`decomposes' into) a disjoint union of local quantum field theories, and arises in, for example, two-dimensional gauge theories in which a subgroup of the gauge group acts trivially.  (See for example \cite{Sharpe:2022ene} for a recent overview.) Our previous work \cite{Perez-Lona:2023djo} discussed decomposition in two dimensions in gauged noninvertible symmetries in the special case that the entire gauged noninvertible symmetry acted trivially.  In this paper, we study more general examples arising in gauged noninvertible symmetries, in which only a subsymmetry acts nontrivially.

We begin in section~\ref{sect:genl} with a general overview of gauging noninvertible symmetries in two dimensions,
both reviewing the discussion of our previous paper \cite{Perez-Lona:2023djo} and also extending to non-multiplicity-free cases.  We also review notions of discrete torsion in gauged noninvertible symmetries, as discussed
recently in \cite{Perez-Lona:2024yih}.  For example, we review that there are two distinct generalizations, one of which naturally generalizes the old picture of discrete torsion as a group action on $B$ fields in ordinary orbifolds \cite{Sharpe:2000ki}.

In section~\ref{sect:ex:rep-a4} we work through the details of gauging in a non-multiplicity-free example,
namely Rep$(A_4)$.  This is, to our knowledge, the simplest nontrivial non-multiplicity-free example.  That said, the analysis is somewhat lengthy. We present a complete expression of partition functions for various gaugings and discuss possible discrete torsions.

In section~\ref{sec: c=1} we apply our Rep$(A)_4$ analysis to $c=1$ CFTs. We find resulting theories under various Rep$(A_4)$ gaugings for the $SU(2)_1/A_4$ theory, which is at a exceptional point of the $c=1$ moduli space. Looking at the dual categories of gaugings, we find other $c=1$ CFTs enjoying Rep$(A_4)$ symmetry, even on the circle branch. We find that in general, if a theory is self-dual under gauging a noninvertible symmetry, there can be multiple new topological defects show up in order to obtain a associative fusion algebra. We utilize our methods into the self-dual gauging of Rep$(A_4)$ in $SU(2)_1/A_4$, and find out the theory enjoys a larger noninvertible symmetry Rep($SL(2,\Z_3)$).

In section~\ref{sec:repd4} we compare gaugings of Rep$(D_4)$ to gaugings of $D_4$, extending an analysis in our previous paper 
\cite{Perez-Lona:2023djo} to include contributions due to discrete torsion.

In section~\ref{sect:decomp} we discuss examples of decomposition \cite{Hellerman:2006zs,Sharpe:2022ene}.  Two-dimensional gauge theories in which a subgroup of the gauge group acts trivially are examples, as a gauged trivially-acting $p$-form symmetry yields a global $(p+1)$-form symmetry.  Our previous paper
\cite{Perez-Lona:2023djo} discussed decomposition in two-dimensional theories with gauged noninvertible symmetries in the special case that the entire noninvertible symmetry acted trivially.  Here, we consider more general cases in which a subset of the gauged noninvertible symmetry acts trivially.  We recover the results of \cite{Perez-Lona:2023djo} as special cases.

In appendix~\ref{app:a4-details} we collect various technical results relevant to the Rep$(A_4)$ gauging, which are essential to the computation but which are somewhat too technical for the readability of the main argument of the paper, and so are banished to this appendix.  In appendix~\ref{app: BrPic} we review a few facts about the Brauer-Picard group, which makes an appearance.

\section{Gauging: general principles} \label{sect:genl}

\subsection{Review of basics of gauging}

In this section we review the basics of gauging noninvertible symmetries in 2d QFT. We will concentrate on symmetries described by fusion categories. In particular, we will mainly focus on fusion categories of the form $\mathcal{C}=\text{Rep}(\mathcal{H})$ for $\cal H$ a finite-dimensional semisimple Hopf algebra over the complex numbers. Our fusion category $\mathcal{C}$, whose objects describe topological line operators in a 2d QFT $\mathcal{T}$, comes equipped with a finite set of (isomorphism classes of) simple objects $A,B,C,\cdots$, in the present case the irreducible complex representations of $\cal H$. In this way, every object in the category (i.e.~every topological line operator in our theory) is isomorphic to a finite sum of simple objects (the category is semisimple).  There is also a conjugation operation sending an object $A$ to its conjugate $\ov{A}$ which represents the orientation reversal of the corresponding line operator.  As representations in $\Rep(\mathcal{H})$, $\ov{A}$ is the dual representation to $A$.  We also have a definition of fusion on objects, which for simple objects means that we have an expansion
\begin{equation}\label{eq:fusionring}
    A\otimes B\cong\sum_CN_{A,B}^CC,
\end{equation}
where $N_{A,B}^C\in\Z_{\ge 0}$ are fusion coefficients.  In particular, there is always a simple object $\mathbbm{1}$, called the monoidal unit, with the property that $N_{\mathbbm{1},A}^B = N_{A,\mathbbm{1}}^B=\delta_{A,B}$, as well as $N_{A,\ov{A}}^\mathbbm{1}=1$ for all simple $A$. These coefficients $N_{A,B}^C$ are the dimensions of the vector spaces $\Hom(A\otimes B,C)$, which we can think of as the space of topological operators which can be placed at a junction with $A$ and $B$ lines coming in and a $C$ line going out.  Generalizing \cite{Perez-Lona:2023djo}, we consider categories with multiplicity, meaning these hom-spaces can have dimension greater than one.  For each such space $\Hom(A\otimes B,C)$ of junction operators, we can choose a basis $(\la_{A,B}^C)_i$, $i=1,\cdots,N_{A,B}^C$, a fusion basis. For fusions involving the identity object there are canonical choices for $\la_{A,\mathbbm{1}}^A$ and $\la_{\mathbbm{1},A}^A$ (this will be explicit in specific examples), but the rest of the basis is simply a choice. 

\begin{figure}[ht]
    \centering
    \includegraphics[width=10cm]{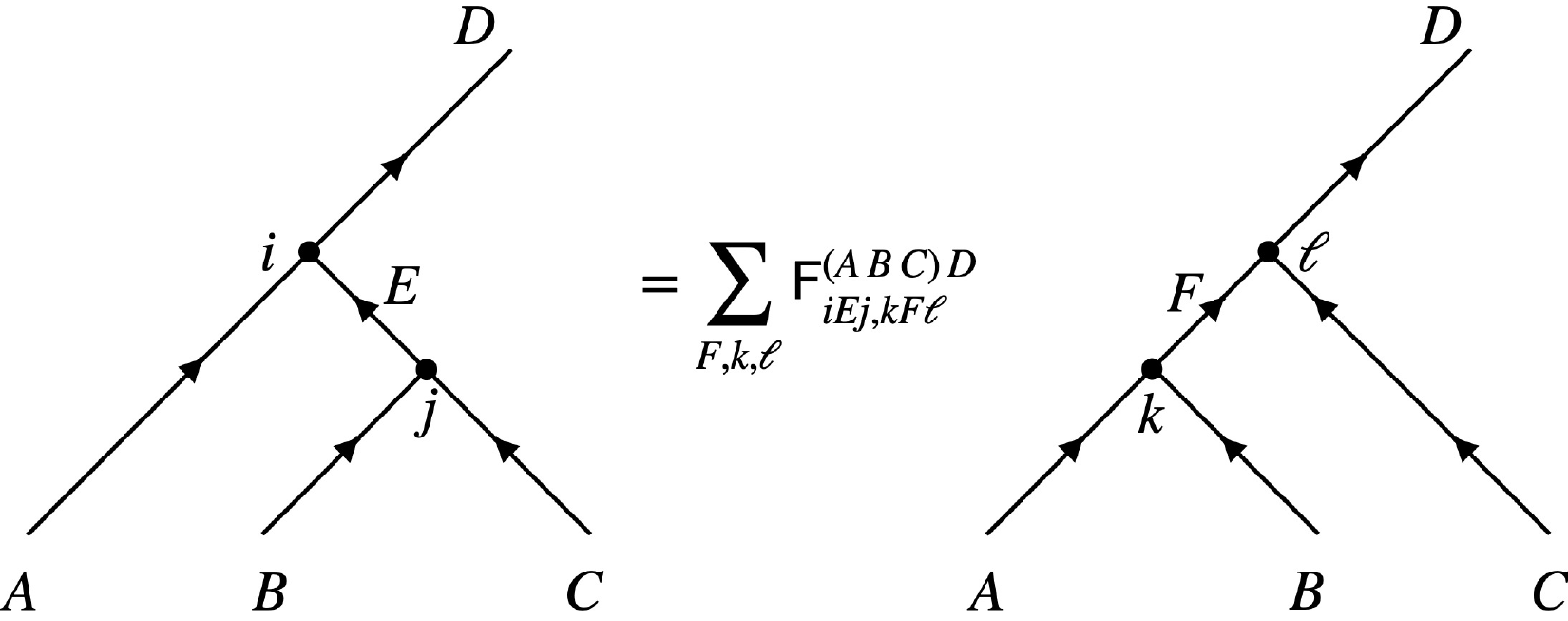}
    \caption{The associator for topological lines are encoded in F-symbols.  The labels on the vertices represent insertions of our fusion basis vectors, i.e.~$(\la_{A,E}^D)_i$ and $(\la_{B,C}^E)_j$ on the left and $(\la_{A,B}^F)_k$ and $(\la_{F,C}^D)_\ell$ on the right.}
    \label{fig:Fsymbol}
\end{figure}

Our category also comes equipped with an associator structure, a set of maps $\al_{A,B,C}\in\Hom((A\otimes B)\otimes C,A\otimes(B\otimes C))$.  With the fusion bases chosen above, we can expand these associator maps as
\begin{equation}\label{eq:fsymbols}
    \lp\la_{A,E}^D\rp_i\circ\lp 1_A\otimes\lp\la_{B,C}^E\rp_j\rp\circ\al_{A,B,C}=\sum_{F,k,\ell}\mathsf{F}^{(A\,B\,C)\,D}_{iEj,kF\ell}\lp\la_{F,C}^D\rp_\ell\circ\lp\lp\la_{A,B}^F\rp_k\otimes 1_C\rp.
\end{equation}
The coefficients $\mathsf{F}^{(A\,B\,C)\,D}_{iEj,kF\ell}$ are known as F-symbols, and they encode the information of the associator. See Figure \ref{fig:Fsymbol} for an illustration.

In~\cite{Perez-Lona:2023djo}, we often used the notation of the crossing kernels $\widetilde{K}^{A\,D}_{B,C}(E,F)$~\cite{Chang:2018iay} to describe the components of the associator in the multiplicity-free case.  In the present paper we choose to switch to the slightly more common convention of F-symbols, which are related to the crossing kernels by
\begin{equation}
    \widetilde{K}^{A\,D}_{B,C}(E,F)=\lp\mathsf{F}^{(\ov{A}\,\ov{B}\,\ov{C})\,D}\rp^{-1}_{\ov{E},\ov{F}}.
\end{equation}

The associator obeys a consistency condition known as the pentagon identity, which when translated to a statement about the F-symbols becomes
\begin{equation}
    \sum_t\mathsf{F}^{(A\,B\,F)\,E}_{kGj,qJt}\,\mathsf{F}^{(J\,C\,D)\,E}_{tFi,rIp}=\sum_{H;\ell,m,n}\mathsf{F}^{(B\,C\,D)\,G}_{jFi,\ell Hm}\,\mathsf{F}^{(A\,H\,D)\,E}_{kGm,nIp}\,\mathsf{F}^{(A\,B\,C)\,I}_{nH\ell,qJr}.
\end{equation}
The fusion ring (\ref{eq:fusionring}) and the F-symbols (\ref{eq:fsymbols}) together determine the fusion category $\cal C$.

As with fusion, a co-fusion basis can be chosen for each hom-space $\Hom(A,B\otimes C)$, which likewise are allowed to have dimension greater than one (indeed $\dim(\Hom(A,B\otimes C))=N_{\ov{C},\ov{B}}^{\ov{A}}$).  In principle the co-fusion basis can be chosen independently of the fusion basis, but we prefer to take the following set of conventions.  We define evaluation maps for each simple object $A$ by $\e_A:=\la_{\ov{A},A}^1\in\Hom(\ov{A}\otimes A,\mathbbm{1})$, $\ov{\e}_A:=\la_{A,\ov{A}}^1\in\Hom(A\otimes\ov{A},\mathbbm{1})$.  Then we define co-evaluation maps $\g_A\in\Hom(\mathbbm{1},A\otimes\ov{A})$, $\ov{\g}_A\in\Hom(\mathbbm{1},\ov{A}\otimes A)$ uniquely by the requirement that
\begin{equation}
    \lp \ov{\e}_A\otimes 1\rp\circ\al_{A,\ov{A},A}^{-1}\circ\lp 1\otimes \ov{\g}_A\rp=1,\qquad\lp 1\otimes\e_A\rp\circ\al_{A,\ov{A},A}\circ\lp\g_A\otimes 1\rp=1.
\end{equation}
Finally, for the co-fusion basis we again have canonical choices for $\d_A^{\mathbbm{1},A}$ and $\d_A^{A,\mathbbm{1}}$ and then we define the remainder of the basis by
\begin{equation}
\label{eq:CoFusionConvention}
    \lp\d_A^{B,C}\rp_i=\lp\lp\la_{A,\ov{C}}^B\rp_i\otimes 1_C\rp\circ\al^{-1}_{A,\ov{C},C}\circ\lp 1_A\otimes\ov{\g}_C\rp\circ\d_A^{A,\mathbbm{1}}.
\end{equation}

In the present context, the notion of gauging depends on a choice of a symmetric special Frobenius algebra $\cal A$ in the fusion category $\cal C$. Such a choice is understood as specifying a subsymmetry of $\cal C$, which often but not necessarily corresponds to a fusion subcategory $\cal C'\subset \cal C$. An algebra $\cal A$ of such kind is specified by an object $\cal A\in \text{ob}(\cal C)$, along with morphisms $\mu:\cal A\otimes \cal A\to \cal A$ (multiplication), $u:\mathbbm{1}\to \cal A$ (unit), $\Delta: \cal A\to \cal A\otimes \cal A$ (comultiplication), and $u^o:\cal A\to \mathbbm{1}$ (counit), satisfying a series of identities described in e.g.~\cite[Appendix A.2]{Perez-Lona:2023djo}. Such identities ensure that the gauged theory $\cal T / \cal A$ is well-defined. One should note that for a fixed object $\cal A \in \text{ob}(\cal C)$ there can exist different morphisms $(\mu,u,\Delta,u^o)$ that make it into a symmetric special Frobenius algebra. The physically-meaningful information, however, only depends on the Morita equivalence class of such algebra structures.

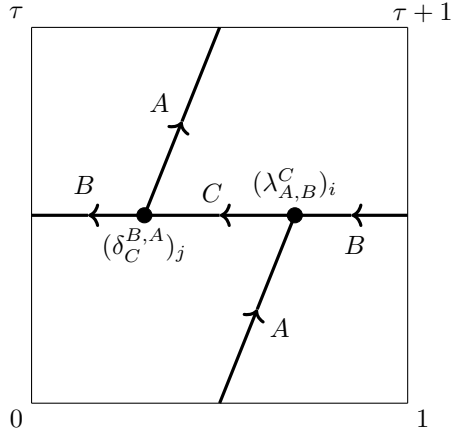
\begin{figure}
		\centering
		\begin{tikzpicture}
			\draw[thin] (0,0)--(5,0);
			\draw[thin] (5,0)--(5,5);
			\draw[thin] (0,0)--(0,5);
			\draw[thin] (0,5)--(5,5);
			\draw[very thick,->] (2.5,0)--(3,1.25);
			\draw[very thick] (3,1.25)--(3.5,2.5);
			\draw[very thick,->] (1.5,2.5)--(2,3.75);
			\draw[very thick] (2,3.75)--(2.5,5);
			\draw[very thick,->] (5,2.5)--(4.25,2.5);
			\draw[very thick] (4.25,2.5)--(3.5,2.5);
			\draw[very thick,->] (3.5,2.5)--(2.5,2.5);
			\draw[very thick] (2.5,2.5)--(1.5,2.5);
			\draw[very thick,->] (1.5,2.5)--(0.75,2.5);
			\draw[very thick] (0.75,2.5)--(0,2.5);
			\node at (3.3,1) {$A$};
			\node at (1.7,4) {$A$};
			\node at (0.7,2.9) {$B$};
			\node at (4.3,2.1) {$B$};
			\node at (2.4,2.8) {$C$};
            \filldraw (3.5,2.5) circle (0.1);
            \node at (3.5,2.9) {$(\la_{A,B}^C)_i$};
            \filldraw (1.5,2.5) circle (0.1);
            \node at (1.5,2.1) {$(\d_C^{B,A})_j$};
			\node at (-0.2,5.2) {$\tau$};
            \node at (5.2,-0.2) {$1$};
            \node at (5.2,5.2) {$\tau+1$};
            \node at (-0.2,-0.2) {$0$};
		\end{tikzpicture}
  \caption{\label{fig:PartialTraceDef}
  Definition of the partial trace $(Z_{A,B}^C)_{ij}$ as a $T^2$ correlation function with topological lines inserted as shown and our chosen fusion and co-fusion basis vectors at the junctions.
  }
  \end{figure}

In particular, the partition function of the ${\cal A}$-gauged theory can be constructed using the algebra structure of $\cal A$, using the same procedure as described in our previous paper \cite{Perez-Lona:2023djo}. For this, a triangulation of the two-dimensional spacetime is chosen, which in a prescribed way gives rise to a combination of (co)multiplications of $\cal A$. The formalism involved ensures that the final result is independent of the chosen triangulation.  For example, the partition function for a genus one surface ($T^2$) is
\begin{equation}\label{eq:partitionfunction}
    Z(\tau,\bar{\tau}) = \sum_{A,B,C,i,j} (\mu_{A,B}^C)^i\,(\Delta_C^{B,A})^j \,(Z_{A,B}^C)_{ij}(\tau,\bar{\tau}).
\end{equation}
In the expression above,
the $A, B, C$ are the simple objects (simple line operators) of the parent theory $\cal C$, and the
coefficients $(\mu^C_{A,B})^i$ and $(\Delta^{B,A}_C)^j$ are obtained by expanding the morphisms $\mu:\cal A\otimes\cal A\to \cal A$ and $\Delta:\cal A\to \cal A\otimes \cal A$ in terms of a fixed (co-)fusion basis of hom-spaces $\text{Hom}(A\otimes B,C)$ and $\text{Hom}(C,B\otimes A)$, respectively.
Thus, for instance, the coefficient $(\mu_{A,B}^C)^i$ corresponds to expanding the multiplication morphism $\mu:\cal A\otimes\cal A\to\cal A$ in terms of the basis junction operators and exctracting the coefficient of $(\la_{A,B}^C)_i$. The
partial trace $(Z_{A,B}^C)_{ij}(\tau,\bar{\tau})$ is defined as the correlation function of the theory on a $T^2$ with modular parameter $\tau$, and with insertions of topological lines $A$ wrapping the vertical cycle from bottom to top, joining on the right with $B$ wrapping the horizontal cycle from right to left.  The two lines meet and form a $C$ line which then splits again into a $B$ line and an $A$ line.  At the junction where $A$ and $B$ join into $C$ we put an operator $(\la_{A,B}^C)_i$, and at the junction where $C$ splits into $B$ and $A$ we put an operator $(\d_C^{B,A})_j$. This is sketched in Figure~\ref{fig:PartialTraceDef}.  This is understood as the generalization of the partition function expression \cite[Equation (2.91)]{Perez-Lona:2023djo} to fusion categories that are not necessarily multiplicity-free.

We can rewrite the co-fusion junction $(\d_C^{B,A})_j$ as a combination of a fusion junction $(\la_{C,\ov{A}}^B)_j$ and a co-evaluation $\ov{\g}_A$ using (\ref{eq:CoFusionConvention}).  Then, we can use familiar manipulations along with applications of the associator to derive the modular transformation properties of our partial traces,
\begin{align}
    \lp Z_{A,B}^C\rp_{ij}(\tau+1,\bar{\tau}+1)=\ & \sum_{D,k,\ell}\ls \mathsf{F}^{(A\,B\,\ov{A})\,B}\rs^{-1}_{iCj,kD\ell}\,\lp Z_{A,D}^B\rp_{k\ell}(\tau,\bar{\tau}),\\
    \lp Z_{A,B}^C\rp_{ij}(-1/\tau,-1/\bar{\tau})=\ & \sum_{D,k,\ell,m}\ls\mathsf{F}^{(A\,B\,\ov{A})\,B}\rs^{-1}_{iCj,kD\ell}\,\ls\mathsf{F}^{(A\,D\,\ov{B})\,1}\rs^{-1}_{kB,\ov{A}m}\,\lp Z_{B,\ov{A}}^D\rp_{\ell m}(\tau,\bar{\tau}).
\end{align}
The steps in this derivation are nearly identical to those in the multiplicity-free case which is presented in~\cite{Perez-Lona:2023djo} and so are omitted here.

\subsection{Discrete torsion}

Discrete torsion was first described in \cite{Vafa:1986wx} as a choice of modular-invariant phases that could be added to orbifolds by finite groups, to generate new theories -- essentially, a choice of
physically-inequivalent ways to gauge a finite group. These different gaugings were argued to be classified by the second cohomology group $H^2(G,U(1))$ of the orbifold group $G$. In \cite{Sharpe:2000ki}, discrete torsion was described as a consequence of a choice of $G$-action on the $B$ field and corresponding gerbe, the higher-categorical generalization of principal $U(1)$ bundles,
or more precisely, discrete torsion is the difference between group actions on the $B$ field. 
In this section, for completeness, we briefly review its noninvertible version.

In the context of noninvertible symmetries described by a fusion category $\cat$, more than one kind of structure has been referred to as the generalization of discrete torsion, most notably either as part of the information specified by a choice of symmetric special Frobenius algebra $(A,\mu,\Delta)$ in $\cat$ \cite[Equation 4.4]{Bhardwaj:2017xup}, \cite[Section 4.2]{Putrov24}, or as a choice of a fiber functor for $\cat$ \cite[slide 26]{Thornslides}. While these two characterizations are clearly applicable to a wider variety of symmetries, they do not obviously generalize some of the properties of group-like discrete torsion, including in particular its classification by some cohomology group, and its interpretation as a difference of actions on the $B$ field.

In \cite{Perez-Lona:2024yih}, these issues are resolved. The generalization above, characterized as the Morita equivalence classes of symmetric special Frobenius algebra structures on a given object $A$ in the fusion category $\cal C$ (which for $A=R$ the regular object in $\cal C$ this is the same as the set of equivalence classes of fiber functors on $\cal C$), is referred to as the \textit{set} of discrete torsion \textit{choices} on $A$. Generally-speaking, these only form a set, hence the name. In the current literature, what is predominantly referred to as ``discrete torsion'' is this notion. A distinct yet complementary generalization of discrete torsion to noninvertible symmetries has a cohomological classification by lazy cohomology group $H^2_{\ell}(\cat)$, a cohomological notion due to \cite{PSV10} that is intrinsic to the fusion category $\cat$. This is referred to as the (cohomology) \textit{group} of discrete torsion \textit{twists} which, as its name suggests, always forms a group. This cohomology group is the group of equivalence classes of natural isomorphisms of the tensor product functor of $\cat$ satisfying a 2-cocycle condition, where the equivalence is given by some appropriate notion of 2-coboundary natural isomorphism. In particular, this recovers the group cohomology classification of discrete torsion for group-like symmetries by letting $\cat=\text{Vec}_G$, but also provides a classification of discrete torsion for categories of comodules $\cat=\text{Comod}(\mathcal{H})$ of a Hopf algebra $\mathcal{H}$ in terms of the more standard notion of the lazy cohomology group $H^2_{\ell}(\mathcal{H})$ of $\mathcal{H}$ \cite{BC06}, which has been computed for several such algebras (see e.g. \cite{GK10}).

These two notions are complementary in the sense that discrete torsion choices on any object $A$ in $\cal C$ form not only a set but more precisely a set equipped with a well-defined group action by the discrete torsion twists. Indeed, different discrete torsion choices can turn out to be related by these categorical twists, an example of which is the discrete torsion choices on the regular object in $\text{Rep}(A_4)$ (c.f. Section~\ref{sssec:dta4}).

Furthermore, it was shown in \cite{Perez-Lona:2024yih} that this second generalization of discrete torsion, classified by lazy cohomology, also classifies (differences between) actions of a noninvertible symmetry on a $B$ field,
generalizing the picture of \cite{Sharpe:2000ki} from ordinary orbifolds to noninvertible ones. 

These cocycles interact with a variety of mathematical structures relevant to global symmetries and their gaugings, such as symmetric special Frobenius algebras (discrete torsion choices) mentioned above, as well as fiber functors, and noninvertible actions on gerbes with connection. The key observation is that discrete torsion acts on any choice of such structure by what can be understood as a twist. In the context of actions on gerbes, this is compatible with the characterization of discrete torsion as differences of actions previously described. 

Strictly speaking, it was argued in \cite{Perez-Lona:2024yih} that the discrete torsion described by lazy cohomology is not a choice of a symmetric special Frobenius algebra, or of a fiber functor, but the possible twists that the category-theoretic cocycles can produce on such choices. (This is closely analaogous to ordinary orbifolds, where as discussed in \cite{Sharpe:2000ki}, discrete torsion in ordinary orbifolds is technically the difference between group actions on a $B$ field.)
As described in \cite[Section 3]{Perez-Lona:2024yih}, this subtle point is visible in many cases. For example, the category $\text{Rep}(D_4)$ admits several inequivalent fiber functors, which in turn, and as described further in Section~\ref{sec:repd4}, implies that its regular object admits several Morita-inequivalent symmetric special Frobenius algebra structures. However, the corresponding discrete torsion group $H^2_{\ell}(\text{Rep}(D_4))$ vanishes, meaning there are no nontrivial discrete torsion twists. On the other hand, categories that do not admit a fiber functor such as $\text{Vec}_G^{\alpha}$ for $[\alpha]\in H^3(G,\mathbb{C}^{\times})$ a nontrivial associator class in general have a nontrivial lazy cohomology group. Such cocycles can nevertheless still twist algebra structures in these categories and even quasi-fiber functors, which in particular are used to describe anomalous actions on other monoidal categories and gerbes.

\section{Higher multiplicity example: Rep$(A_4)$ gaugings}  \label{sect:ex:rep-a4}

\subsection{Rep$(A_4)$}

We'll take the following presentation of $A_4$,
\be
\left\langle a,b\mid a^3=b^2=(ab)^3=1\right\rangle.
\ee
In terms of familiar cycle notation, we can think of $a=(1\,2\,3)$ and $b=(1\,2)(3\,4)$.  The group splits into four conjugacy classes,
\begin{align}
[1]=\ & \{1\},\\
[a]=\ & \{a,ab,ba,bab\}=\{(1\,2\,3),(1\,3\,4),(2\,4\,3),(1\,4\,2)\},\\
[a^2]=\ & \{a^2,a^2b,aba,ba^2\}=\{(1\,3\,2),(2\,3\,4),(1\,2\,4),(1\,4\,3)\},\\
[b]=\ & \{b,a^2ba,aba^2\}=\{(1\,2)(3\,4),(1\,3)(2\,4),(1\,4)(2\,3)\}.
\end{align}
This means that $A_4$ should also have four distinct irreducible representations which we will label $1,X,Y,Z$.  The character table for these irreps is
\begin{center}
    \begin{tabular}{c|c|c|c|c|}
    & $[1]$ & $[a]$ & $[a^2]$ & $[b]$ \\
    \hline
    $\chi_1$ & $1$ & $1$ & $1$ & $1$ \\
    \hline
    $\chi_X$ & $1$ & $\zeta$ & $\zeta^2$ & $1$ \\
    \hline
    $\chi_Y$ & $1$ & $\zeta^2$ & $\zeta$ & $1$ \\
    \hline
    $\chi_Z$ & $3$ & $0$ & $0$ & $-1$ \\
    \hline
    \end{tabular}
\end{center}
where $\zeta=e^{2\pi i/3}$.  From the character table we can obtain the fusion algebra,
\be
X^2=Y,\quad Y^2=X,\quad XY=YX=1,\quad XZ=ZX=YZ=ZY=Z,\quad Z^2=1+X+Y+2Z.
\ee
Crucially, there is a coefficient of $2$ appearing in the $Z^2$ fusion.  This means that we are not in the multiplicity-free case anymore.

We'll need more explicit realizations of the different irreps.  We can specify\footnote{To obtain the non-trivial form of the $\rho_Z$ matrices, we can realize the action of $A_4$ geometrically as the rotational symmetries of a regular tetrahedron whose vertices (labeled $1$, $2$, $3$, $4$ respectively) sit at the points $(-\hlf,-\frac{1}{2\sqrt{3}},-\frac{1}{2\sqrt{6}})$, $(\hlf,-\frac{1}{2\sqrt{3}},-\frac{1}{2\sqrt{6}})$, $(0,\frac{1}{\sqrt{3}},-\frac{1}{2\sqrt{6}})$, and $(0,0,\frac{\sqrt{3}}{2\sqrt{2}})$.} each one by its action on the generators $a$ and $b$,
\begin{align}
    \rho_1(a)=\ & 1,\qquad\rho_1(b)=1,\\
    \rho_X(a)\ & =\zeta,\qquad\rho_X(b)=1,\\
    \rho_Y(a)\ & =\zeta^2,\quad\rho_Y(b)=1,\\
    \rho_Z(a)\ & =\lp\begin{matrix} -\hlf & -\frac{\sqrt{3}}{2} & 0 \\ \frac{\sqrt{3}}{2} & -\hlf & 0 \\ 0 & 0 & 1 \end{matrix}\rp,\qquad\rho_Z(b)=\lp\begin{matrix} -1 & 0 & 0 \\ 0 & \frac{1}{3} & \frac{2\sqrt{2}}{3} \\ 0 & \frac{2\sqrt{2}}{3} & -\frac{1}{3} \end{matrix}\rp.
\end{align}
We'll denote the basis vectors for which the irreps take this form as $e$, $e_X$, $e_Y$ for the one-dimensional irreps, and $\{e_1,e_2,e_3\}$ for $Z$.

Next we need to pick bases for the fusion intertwiners.  It is a somewhat lengthy task to determine the general form of these intertwiners, so we have relegated the calculations to appendix~\ref{appsub:A4FusionIntertwiners}.  Apart from making a canonical choice for fusions involving the trivial irrep,
\be
\la_{1,R}^R(ev)=\la_{R,1}^R(ve)=v,
\ee
we make the most general choice possible.  For the cases where the complex vector space $\Hom(A\otimes B,C)$ of intertwiners is one-dimensional, the specification of our single basis vector is determined up to a $\C^\times$ constant.  This is then precisely the same procedure as followed in~\cite{Perez-Lona:2023djo}, and results in $\la_{X,X}^Y$, $\la_{X,Y}^1$, $\la_{Y,X}^1$, $\la_{Y,Y}^X$, $\la_{X,Z}^Z$, $\la_{Z,X}^Z$, $\la_{Y,Z}^Z$, $\la_{Z,Y}^Z$, $\la_{Z,Z}^1$, $\la_{Z,Z}^X$, and $\la_{Z,Z}^Y$, being fixed in terms of parameters $\beta_1,\cdots,\beta_{11}$ respectively.

On the other hand $\Hom(Z\otimes Z,Z)$ is a two-dimensional vector space.  In this case we need to specify a pair of basis intertwiners, $(\la_{Z,Z}^Z)_1$ and $(\la_{Z,Z}^Z)_2$, and they are determined up to a $\GL(2,\C)$ matrix $\lp\begin{matrix} \beta_{12} & \beta_{13} \\ \beta_{14} & \beta_{15} \end{matrix}\rp$.  These intertwiners are also worked out in appendix~\ref{appsub:A4FusionIntertwiners} and given in equations (\ref{eq:A4lambdaZZZ1e1e1}) through (\ref{eq:A4lambdaZZZ1e3e3}).

With this basis of fusion intertwiners established, we can construct additional intertwiners for evaluation maps $\e_A\in\Hom(\ov{A}\otimes A,1)$, $\ov{\e}_A\in\Hom(A\otimes\ov{A},1)$, co-evaluation maps $\g_A\in\Hom(1,A\otimes\ov{A})$, $\ov{\g}_A\in\Hom(1,\ov{A}\otimes A)$, and a basis ($\d_A^{B,C})_i$ for co-fusion intertwiners that span the vector spaces $\Hom(A,B\otimes C)$.  Following the conventions in~\cite{Perez-Lona:2023djo}, these maps do not require any additional parameters and are computed in appendix~\ref{appsub:A4CoFusionMaps}.

Next we turn to computing the associator.  These are determined by our choice of basis $(\la_{A,B}^C)_i$ for the fusion intertwiners, and are defined by the requirement
\be
\label{eq:FundamentalAssociatorRelation}
\lp\la_{A,E}^D\rp_i\circ\lp 1_A\otimes\lp\la_{B,C}^E\rp_j\rp\circ\al_{A,B,C}=\sum_{F,k,\ell}\mathsf{F}^{(A\,B\,C)\,D}_{iEj,kF\ell}\lp\la_{F,C}^D\rp_\ell\circ\lp\lp\la_{A,B}^F\rp_k\otimes 1_C\rp,
\ee
as maps in $\Hom((A\otimes B)\otimes C,D)$.  The $\mathsf{F}^{(A\,B\,C)\,D}_{iEj,kF\ell}$ are complex coefficients.  In $\Rep(G)$ these maps are intertwiners, and the associator map $\al_{A,B,C}\in\Hom((A\otimes B)\otimes C,A\otimes(B\otimes C))$ is canonical, in the sense that simply
\be
\al_{A,B,C}((v_Av_B)v_C)=v_A(v_Bv_C).
\ee

This is enough information to compute the components of the $\mathsf{F}$ symbols, and this is done in appendix~\ref{appsub:A4Associators}.

\subsection{Subgroups of $A_4$ and gaugings of $\Rep(A_4)$}

Up to conjugation, $A_4$ has five distinct subgroups (conjugation relates all four $\Z_2$ subgroups to each other, and all four $\Z_3$ subgroups to each other), $1$, $\langle b\rangle\cong\Z_2$, $\langle a\rangle\cong\Z_3$, $\{1,b,a^2ba,aba^2\}\cong\Z_2\times\Z_2$, and $A_4$ itself.  The cosets of each of these give gaugeable subsymmetries of $\Rep(A_4)$.

\begin{itemize}
    \item $H=A_4$, $A_4/H=1$.  This is the trivial subsymmetry of $\Rep(A_4)$.  Gauging it of course does nothing.
    \item $H=\Z_2\times\Z_2$.  The cosets are $A_4/H=\{H,aH,a^2H\}$.  To identify the corresponding representation we note that on the listed basis of cosets the action of $a$ and $b$ is (e.g.~$b\cdot aH=baH=a(a^2ba)H=aH$),
    \be
    \rho_{A_4/H}(a)=\lp\begin{matrix} 0 & 0 & 1 \\ 1 & 0 & 0 \\ 0 & 1 & 0 \end{matrix}\rp,\qquad\rho_{A_4/H}(b)=\lp\begin{matrix} 1 & 0 & 0 \\ 0 & 1 & 0 \\ 0 & 0 & 1 \end{matrix}\rp.
    \ee
    Taking inner products with the irrep characters, we deduce that this is the $1+X+Y$ representation.  This is a group-like $\Z_3$ sub-symmetry.
    \item $H=\langle a\rangle=\Z_3$.  The cosets are $A_4/H=\{H,bH,abH,a^2bH\}$.  Here we have
    \be
    \rho_{A_4/H}(a)=\lp\begin{matrix} 1 & 0 & 0 & 0 \\ 0 & 0 & 0 & 1 \\ 0 & 1 & 0 & 0 \\ 0 & 0 & 1 & 0 \end{matrix}\rp,\qquad\rho_{A_4/H}(b)=\lp\begin{matrix} 0 & 1 & 0 & 0 \\ 1 & 0 & 0 & 0 \\ 0 & 0 & 0 & 1 \\ 0 & 0 & 1 & 0 \end{matrix}\rp.
    \ee
    This implies that the representation is $1+Z$.
    \item $H=\langle b\rangle=\Z_2$.  The cosets are $A_4/H=\{H,aH,a^2H,baH,abaH,a^2baH\}$.  Then
    \be
    \rho_{A_4/H}(a)=\lp\begin{matrix} 0 & 0 & 1 & 0 & 0 & 0 \\ 1 & 0 & 0 & 0 & 0 & 0 \\ 0 & 1 & 0 & 0 & 0 & 0 \\ 0 & 0 & 0 & 0 & 0 & 1 \\ 0 & 0 & 0 & 1 & 0 & 0 \\ 0 & 0 & 0 & 0 & 1 & 0 \end{matrix}\rp,\qquad\rho_{A_4/H}(b)=\lp\begin{matrix} 1 & 0 & 0 & 0 & 0 & 0 \\ 0 & 0 & 0 & 1 & 0 & 0 \\ 0 & 0 & 0 & 0 & 1 & 0 \\ 0 & 1 & 0 & 0 & 0 & 0 \\ 0 & 0 & 1 & 0 & 0 & 0 \\ 0 & 0 & 0 & 0 & 0 & 1 \end{matrix}\rp.
    \ee
    Here we find the representation $1+X+Y+Z$.
    \item $H=1$.  The cosets $A_4/H\cong A_4$ transform as the regular representation $1+X+Y+3Z$.
\end{itemize}

\subsubsection{$1+X+Y$}

The coset representation has vectors $v_H$, $v_{aH}$, and $v_{a^2H}$.  When we decompose into irreps, we can write
\begin{align}
    e=\ & v_H+v_{aH}+v_{a^2H},\\
    e_X=\ & v_H+\zeta^2v_{aH}+\zeta v_{a^2H},\\
    e_Y=\ & v_H+\zeta v_{aH}+\zeta^2v_{a^2H},
\end{align}
where $\zeta=e^{2\pi i/3}$.  From $\m(v_{gH},v_{kH})=\d_{gH,kH}v_{gH}$ (so in particular $\m(uv)=\m(vu)$), we have $\m(ev)=\m(ve)=v$ as well as
\begin{align}
    \m(e_Xe_X)=\ & e_Y,\\
    \m(e_Xe_Y)=\ & e,\\
    \m(e_Ye_Y)=\ & e_X,
\end{align}
or in components, $\m_{1,A}^A=\m_{A,1}^A=1$ and 
\begin{align}
    \m_{X,X}^Y  =\ & \beta_1^{-1},\\
    \m_{X,Y}^1  =\ & \beta_2^{-1},\\
    \m_{Y,X}^1  =\ & \beta_3^{-1},\\
    \m_{Y,Y}^X =\ & \beta_4^{-1}.
\end{align}

Similarly for co-multiplication, $\Delta(v_{gH})=v_{gH}v_{gH}$, we get
\begin{align}
    \Delta(e)=\ & \frac{1}{3}\lp ee+e_Xe_Y+e_Ye_X\rp,\\
    \Delta(e_X)=\ & \frac{1}{3}\lp ee_X+e_Xe+e_Ye_Y\rp,\\
    \Delta(e_Y)=\ & \frac{1}{3}\lp ee_Y+e_Xe_X+e_Ye\rp,
\end{align}
hence $\Delta_A^{1,A}=\Delta_A^{A,1}=\frac{1}{3}$ and
\begin{align}
    \Delta_1^{X,Y}=\ & \frac{\beta_3}{3},\\
    \Delta_1^{Y,X}=\ & \frac{\beta_2}{3},\\
    \Delta_X^{Y,Y}=\ & \frac{\beta_3}{3\beta_1},\\
    \Delta_Y^{X,X}=\ & \frac{\beta_2}{3\beta_4}.
\end{align}
Putting this together in $Z=\sum_{A,B,C}\m_{A,B}^C\Delta_C^{B,A}Z_{A,B}^C$ gives
\begin{equation}
    Z_{1+X+Y}=\frac{1}{3}\ls Z_{1,1}^1+Z_{1,X}^X+Z_{1,Y}^Y+Z_{X,1}^X+\frac{\beta_2}{\beta_1\beta_4}Z_{X,X}^Y+Z_{X,Y}^1+Z_{Y,1}^Y+Z_{Y,X}^1+\frac{\beta_3}{\beta_1\beta_4}Z_{Y,Y}^X\rs.
\end{equation}

In the simplifying gauge choice of appendix~\ref{appsub:A4GaugeChoice}, this simply looks like a group-like $\Z_3$ orbifold, as we expect,
\begin{equation}
    Z_{1+X+Y}=\frac{1}{3}\ls Z_{1,1}^1+Z_{1,X}^X+Z_{1,Y}^Y+Z_{X,1}^X+Z_{X,X}^Y+Z_{X,Y}^1+Z_{Y,1}^Y+Z_{Y,X}^1+Z_{Y,Y}^X\rs.
\end{equation}

\subsubsection{$1+Z$}

The subgroup in this case is $H=\{1,a,a^2\}\cong\Z_3$.  We have four vectors in the coset representation, $v_H$, $v_{bH}$, $v_{abH}$, and $v_{a^2bH}$.  To collect them into irreps we have
\begin{align}
    e=\ & v_H+v_{bH}+v_{abH}+v_{a^2bH},\\
    e_1=\ & \sqrt{2}\lp v_{abH}-v_{a^2bH}\rp,\\
    e_2=\ & \sqrt{\frac{2}{3}}\lp 2v_{bH}-v_{abH}-v_{a^2bH}\rp,\\
    e_3=\ & \frac{1}{\sqrt{3}}\lp 3v_H-v_{bH}-v_{abH}-v_{a^2bH}\rp.
\end{align}
From here we can compute the Frobenius algebra multiplication and co-multiplication, namely $\m(ev)=\m(ve)=v$ and
\begin{align}
    \m(e_1e_1)=\ & e-\sqrt{\frac{2}{3}}e_2-\frac{1}{\sqrt{3}}e_3,\\
    \m(e_1e_2)=\m(e_2e_1)=\ & -\sqrt{\frac{2}{3}}e_1,\\
    \m(e_1e_3)=\m(e_3e_1)=\ & -\frac{1}{\sqrt{3}}e_1,\\
    \m(e_2e_2)=\ & e+\sqrt{\frac{2}{3}}e_2-\frac{1}{\sqrt{3}}e_3,\\
    \m(e_2e_3)=\m(e_3e_2)=\ & -\frac{1}{\sqrt{3}}e_2,\\
    \m(e_3e_3)=\ & e+\frac{2}{\sqrt{3}}e_3,
\end{align}
and
\begin{align}
    \Delta(e)=\ & \frac{1}{4}\lp ee+e_1e_1+e_2e_2+e_3e_3\rp,\\
    \Delta(e_1)=\ & \frac{1}{4}\lp ee_1+e_1e\rp-\frac{1}{2\sqrt{6}}\lp e_1e_2+e_2e_1\rp-\frac{1}{4\sqrt{3}}\lp e_1e_3+e_3e_1\rp,\\
    \Delta(e_2)=\ & \frac{1}{4}\lp ee_2+e_2e\rp-\frac{1}{2\sqrt{6}}e_1e_1+\frac{1}{2\sqrt{6}}e_2e_2-\frac{1}{4\sqrt{3}}\lp e_2e_3+e_3e_2\rp,\\
    \Delta(e_3)=\ & \frac{1}{4}\lp ee_3+e_3e\rp-\frac{1}{4\sqrt{3}}e_1e_1-\frac{1}{4\sqrt{3}}e_2e_2+\frac{1}{2\sqrt{3}}e_3e_3.
\end{align}
In components, $\m_{1,A}^A=\m_{A,1}^A=1$, $\Delta_A^{1,A}=\Delta_A^{A,1}=\frac{1}{4}$, and
\begin{align}
    \m_{Z,Z}^1=\ & \beta_9^{-1},\\
    \lp\m_{Z,Z}^Z\rp^1=\ & \frac{1}{\Om}\lp\frac{2i}{\sqrt{3}}\beta_{15}+\frac{1}{\sqrt{3}}\beta_{14}\rp,\\
    \lp\m_{Z,Z}^Z\rp^2=\ & \frac{1}{\Om}\lp -\frac{2i}{\sqrt{3}}\beta_{13}-\frac{1}{\sqrt{3}}\beta_{12}\rp,
\end{align}
where $\Om=\beta_{12}\beta_{15}-\beta_{13}\beta_{14}$, and
\begin{align}
    \Delta_1^{Z,Z}=\ & \frac{1}{4}\beta_9,\\
    \lp\Delta_Z^{Z,Z}\rp^1=\ & \frac{\beta_9}{\Om}\lp\frac{1}{4\sqrt{3}}\beta_{14}+\frac{i}{2\sqrt{3}}\beta_{15}\rp,\\
    \lp\Delta_Z^{Z,Z}\rp^2=\ & \frac{\beta_9}{\Om}\lp-\frac{1}{4\sqrt{3}}\beta_{12}-\frac{i}{2\sqrt{3}}\beta_{13}\rp.
\end{align}
Thus we obtain
\begin{align}
    Z_{1+Z}=\ & \frac{1}{4}\ls Z_{1,1}^1+Z_{1,Z}^Z+Z_{Z,1}^Z+Z_{Z,Z}^1+\frac{\beta_9}{3\Om^2}\lp\lp\beta_{14}+2i\beta_{15}\rp^2\lp Z_{Z,Z}^Z\rp_{11}\right.\right.\non\\
    & \qquad\quad\left.\left. -\lp\beta_{12}+2i\beta_{13}\rp\lp\beta_{14}+2i\beta_{15}\rp\lp\lp Z_{Z,Z}^Z\rp_{12}+\lp Z_{Z,Z}^Z\rp_{21}\rp+\lp\beta_{12}+2i\beta_{13}\rp^2\lp Z_{Z,Z}^Z\rp_{22}\rp\rs.
\end{align}

Making the simple gauge choice of appendix~\ref{appsub:A4GaugeChoice}, this becomes
\begin{equation}
    Z_{1+Z}=\frac{1}{4}\ls Z_{1,1}^1+Z_{1,Z}^Z+Z_{Z,1}^Z+Z_{Z,Z}^1+\frac{2}{\sqrt{3}}\lp Z_{Z,Z}^Z\rp_{11}\rs.
\end{equation}

\subsubsection{$1+X+Y+Z$}

For $H=\{1,b\}\cong\Z_2$, the coset representation is six-dimensional, with basis $v_H$, $v_{aH}$, $v_{a^2H}$, $v_{baH}$, $v_{abaH}$, and $v_{a^2baH}$.  To organize into irreps, we change basis to
\begin{align}
    e=\ & v_H+v_{aH}+v_{a^2H}+v_{baH}+v_{abaH}+v_{a^2baH},\\
    e_X=\ & v_H+\zeta^2v_{aH}+\zeta v_{a^2H}+\zeta^2v_{baH}+\zeta v_{abaH}+v_{a^2baH},\\
    e_Y=\ & v_H+\zeta v_{aH}+\zeta^2v_{a^2H}+\zeta v_{baH}+\zeta^2v_{abaH}+v_{a^2baH},\\
    e_1=\ & \sqrt{\frac{3}{2}}\lp -v_{aH}+v_{a^2H}+v_{baH}-v_{abaH}\rp,\\
    e_2=\ & \frac{1}{\sqrt{2}}\lp 2v_H-v_{aH}-v_{a^2H}+v_{baH}+v_{abaH}-2v_{a^2baH}\rp,\\
    e_3=\ & v_H+v_{aH}+v_{a^2H}-v_{baH}-v_{abaH}-v_{a^2baH}.
\end{align}
This leads to the Frobenius algebra structures $\m(ev)=\m(ve)=v$ and (we won't compute all components, just enough to fix the Frobenius multiplication and co-multiplication coefficients)
\begin{align}
    \m(e_Xe_X)=\ & e_Y,\\
    \m(e_Xe_Y)=\m(e_Ye_X)=\ & e,\\
    \m(e_Xe_1)=\m(e_1e_X)=\ & -\hlf e_1-\frac{i}{2}e_2+\frac{i}{\sqrt{2}}e_3,\\
    \m(e_Ye_Y)=\ & e_X,\\ \m(e_Ye_1)=\m(e_1e_Y)=\ & -\hlf e_1+\frac{i}{2}e_2-\frac{i}{\sqrt{2}}e_3,\\
    \m(e_1e_1)=\ & e-\hlf e_X-\hlf e_Y,\\
    \m(e_1e_2)=\m(e_2e_1)=\ & -\frac{i}{2}e_X+\frac{i}{2}e_Y,
\end{align}
as well as
\begin{align}
    \Delta(e)=\ & \frac{1}{6}\lp ee+e_Xe_Y+e_Ye_X+e_1e_1+e_2e_2+e_3e_3\rp,\\
    \Delta(e_X)=\ & \frac{1}{6}\lp ee_X+e_Xe+e_Ye_Y\vphantom{\frac{i}{\sqrt{2}}}\right.\non\\
    & \qquad\left. -\hlf e_1e_1-\frac{i}{2}e_1e_2+\frac{i}{\sqrt{2}}e_1e_3-\frac{i}{2}e_2e_1+\hlf e_2e_2+\frac{1}{\sqrt{2}}e_2e_3+\frac{i}{\sqrt{2}}e_3e_1+\frac{1}{\sqrt{2}}e_3e_2\rp,\\
    \Delta(e_Y)=\ & \frac{1}{6}\lp ee_Y+e_Xe_X+e_Ye\vphantom{\frac{i}{\sqrt{2}}}\right.\non\\
    & \qquad\left. -\hlf e_1e_1+\frac{i}{2}e_1e_2-\frac{i}{\sqrt{2}}e_1e_3+\frac{i}{2}e_2e_1+\hlf e_2e_2+\frac{1}{\sqrt{2}}e_2e_3-\frac{i}{\sqrt{2}}e_3e_1+\frac{1}{\sqrt{2}}e_3e_2\rp,\\
    \Delta(e_1)=\ & \frac{1}{6}\lp ee_1-\hlf e_Xe_1+\frac{i}{2}e_Xe_2-\frac{i}{\sqrt{2}}e_Xe_3-\hlf e_Ye_1-\frac{i}{2}e_Ye_2+\frac{i}{\sqrt{2}}e_Ye_3\right.\non\\
    & \qquad\left. +e_1e-\hlf e_1e_X-\hlf e_1e_Y+\frac{i}{2}e_2e_X-\frac{i}{2}e_2e_Y-\frac{i}{\sqrt{2}}e_3e_X+\frac{i}{\sqrt{2}}e_3e_Y\rp.
\end{align}
From these we extract the coefficients, $\m_{1,A}^A=\m_{A,1}^A=1$, $\Delta_A^{1,A}=\Delta_A^{A,1}=\frac{1}{6}$,
\begin{align}
    \m_{X,X}^Y=\ & \beta_1^{-1},\\
    \m_{X,Y}^1=\ & \beta_2^{-1},\\
    \m_{X,Z}^Z=\ & -\beta_5^{-1},\\
    \m_{Y,X}^1=\ & \beta_3^{-1},\\
    \m_{Y,Y}^X=\ & \beta_4^{-1},\\
    \m_{Y,Z}^Z=\ & -i\beta_7^{-1},\\
    \m_{Z,X}^Z=\ & -\beta_6^{-1},\\
    \m_{Z,Y}^Z=\ & -i\beta_8^{-1},\\
    \m_{Z,Z}^1=\ & \beta_9^{-1},\\
    \m_{Z,Z}^X=\ & -i\beta_{10}^{-1},\\
    \m_{Z,Z}^Y=\ & -\beta_{11}^{-1},
\end{align}
\begin{align}
    \Delta_1^{X,Y}=\ & \frac{\beta_3}{6},\\
    \Delta_1^{Y,X}=\ & \frac{\beta_2}{6},\\
    \Delta_1^{Z,Z}=\ & \frac{\beta_9}{6},\\
    \Delta_X^{Y,Y}=\ & \frac{\beta_3}{6\beta_1},\\
    \Delta_X^{Z,Z}=\ & -\frac{\beta_9}{6\beta_5},\\
    \Delta_Y^{X,X}=\ & \frac{\beta_2}{6\beta_4},\\
    \Delta_Y^{Z,Z}=\ & -\frac{i\beta_9}{6\beta_7},\\
    \Delta_Z^{X,Z}=\ & -\frac{i\beta_9}{6\beta_{10}},\\
    \Delta_Z^{Y,Z}=\ & -\frac{\beta_9}{6\beta_{11}},\\
    \Delta_Z^{Z,X}=\ & -\frac{i\beta_2}{6\beta_8},\\
    \Delta_Z^{Z,Y}=\ & -\frac{\beta_3}{6\beta_6}.
\end{align}
Note that $(\m_{Z,Z}^Z)^i=(\Delta_Z^{Z,Z})^j=0$.

Assembling the ingredients, we obtain the partition function
\begin{align}
    Z_{1+X+Y+Z}=\ & \frac{1}{6}\ls Z_{1,1}^1+Z_{1,X}^X+Z_{1,Y}^Y+Z_{1,Z}^Z+Z_{X,1}^X+\frac{\beta_2}{\beta_1\beta_4}Z_{X,X}^Y+Z_{X,Y}^1+\frac{i\beta_2}{\beta_5\beta_8}Z_{X,Z}^Z\right.\non\\
    & \qquad\left. +Z_{Y,1}^Y+Z_{Y,X}^1+\frac{\beta_3}{\beta_1\beta_4}Z_{Y,Y}^X+\frac{i\beta_3}{\beta_6\beta_7}Z_{Y,Z}^Z+Z_{Z,1}^Z+\frac{i\beta_9}{\beta_6\beta_{10}}Z_{Z,X}^Z+\frac{i\beta_9}{\beta_8\beta_{11}}Z_{Z,Y}^Z\right.\non\\
    & \qquad\left. +Z_{Z,Z}^1+\frac{i\beta_9}{\beta_5\beta_{10}}Z_{Z,Z}^X+\frac{i\beta_9}{\beta_7\beta_{11}}Z_{Z,Z}^Y\rs.
\end{align}
It is a highly nontrivial check on our methods that this (as with the other orbifold partition functions listed in this section) is a sum of the modular invariant combinations found in Appendix~\ref{appsub:A4ModTrans}.

In our simplified gauge of appendix~\ref{appsub:A4GaugeChoice} we get
\begin{align}
    Z_{1+X+Y+Z}=\ & \frac{1}{6}\ls Z_{1,1}^1+Z_{1,X}^X+Z_{1,Y}^Y+Z_{1,Z}^Z+Z_{X,1}^X+Z_{X,X}^Y+Z_{X,Y}^1+Z_{X,Z}^Z+Z_{Y,1}^Y+Z_{Y,X}^1+Z_{Y,Y}^X+Z_{Y,Z}^Z\right.\non\\
    & \qquad\left. +Z_{Z,1}^Z+Z_{Z,X}^Z+Z_{Z,Y}^Z+Z_{Z,Z}^1+Z_{Z,Z}^X+Z_{Z,Z}^Y\rs.
\end{align}

\subsubsection{$1+X+Y+3Z$}

Finally we move to the case with the full regular represenation of $A_4$.  In this case we actually have three different copies of the $Z$ irrep appearing in the algebra object.  We'll label the vectors for this as $e_{ij}$, where this means the $e_j$ vector in the $i$th copy $Z_i$.  We can assemble the vectors of the regular representation into component irreps (not uniquely, indeed there is a $(C^\times)^3\times\GL(3,\C)$ worth of ambiguity in these choices, but that ambiguity drops out of the final expressions for the partition function),
\begin{align}
    e=\ & \sum_{g\in A_4}v_g,\\
    e_X=\ & v_1+\zeta^2v_a+\zeta v_{a^2}+v_b+\zeta^2v_{ab}+\zeta v_{a^2b}+\zeta^2v_{ba}+\zeta v_{aba}+v_{a^2ba}+\zeta v_{ba^2}+v_{aba^2}+\zeta^2v_{a^2ba^2},\\
    e_Y=\ & v_1+\zeta v_a+\zeta^2v_{a^2}+v_b+\zeta v_{ab}+\zeta^2v_{a^2b}+\zeta v_{ba}+\zeta^2v_{aba}+v_{a^2ba}+\zeta^2v_{ba^2}+v_{aba^2}+\zeta v_{a^2ba^2},\\
    e_{11}=\ & \sqrt{\frac{3}{2}}\lp -v_a+v_{a^2}-v_{ab}+v_{a^2b}+v_{ba}-v_{aba}-v_{ba^2}+v_{a^2ba^2}\rp,\\
    e_{12}=\ & \frac{1}{\sqrt{2}}\lp 2v_1-v_a-v_{a^2}+2v_b-v_{ab}-v_{a^2b}+v_{ba}+v_{aba}-2v_{a^2ba}+v_{ba^2}-2v_{aba^2}+v_{a^2ba^2}\rp,\\
    e_{13}=\ & v_1+v_a+v_{a^2}+v_b+v_{ab}+v_{a^2b}-v_{ba}-v_{aba}-v_{a^2ba}-v_{ba^2}-v_{aba^2}-v_{a^2ba^2},\\
    e_{21}=\ & \sqrt{\frac{3}{2}}\lp v_1-v_{a^2}-v_b+v_{a^2b}-v_{aba}+v_{a^2ba}+v_{ba^2}-v_{aba^2}\rp,\\
    e_{22}=\ & \frac{1}{\sqrt{2}}\lp -v_1+2v_a-v_{a^2}+v_b-2v_{ab}+v_{a^2b}+2v_{ba}-v_{aba}-v_{a^2ba}+v_{ba^2}+v_{aba^2}-2v_{a^2ba^2}\rp,\\
    e_{23}=\ & v_1+v_a+v_{a^2}-v_b-v_{ab}-v_{a^2b}+v_{ba}+v_{aba}+v_{a^2ba}-v_{ba^2}-v_{aba^2}-v_{a^2ba^2},\\
    e_{31}=\ & \sqrt{\frac{3}{2}}\lp -v_1+v_a+v_b-v_{ab}-v_{ba}+v_{a^2ba}-v_{aba^2}+v_{a^2ba^2}\rp,\\
    e_{32}=\ & \frac{1}{\sqrt{2}}\lp -v_1-v_a+2v_{a^2}+v_b+v_{ab}-2v_{a^2b}+v_{ba}-2v_{aba}+v_{a^2ba}+2v_{ba^2}-v_{aba^2}-v_{a^2ba^2}\rp,\\
    e_{33}=\ & v_1+v_a+v_{a^2}-v_b-v_{ab}-v_{a^2b}-v_{ba}-v_{aba}-v_{a^2ba}+v_{ba^2}+v_{aba^2}+v_{a^2ba^2}.
\end{align}

The multiplication has as usual $\m(ev)=\m(ve)=v$ and (only including enough examples to determine all components)
\begin{align}
    \m(e_Xe_X)=\ & e_Y,\\
    \m(e_Xe_Y)=\m(e_Ye_X)=\ & e,\\
    \m(e_Xe_{11})=\m(e_{11}e_X)=\ & -\hlf e_{11}-\frac{i}{2}e_{12}+\frac{i}{\sqrt{2}}e_{13},\\
    \m(e_Xe_{21})=\m(e_{21}e_X)=\ & -\hlf\zeta^2e_{21}-\frac{i}{2}\zeta^2e_{22}+\frac{i}{\sqrt{2}}\zeta^2e_{23},\\
    \m(e_Xe_{31})=\m(e_{31}e_x)=\ & -\hlf\zeta e_{31}-\frac{i}{2}\zeta e_{32}+\frac{i}{\sqrt{2}}e_{33},
\end{align}
\begin{align}
    \m(e_Ye_Y)=\ & e_X,\\
    \m(e_Ye_{11})=\m(e_{11}e_Y)=\ & -\hlf e_{11}+\frac{i}{2}e_{12}-\frac{i}{\sqrt{2}}e_{13},\\
    \m(e_Ye_{21})=\m(e_{21}e_Y)=\ & -\hlf\zeta e_{21}+\frac{i}{2}\zeta e_{22}-\frac{i}{\sqrt{2}}\zeta e_{23},\\
    \m(e_Ye_{31})=\m(e_{31}e_Y)=\ & -\hlf\zeta^2e_{31}+\frac{i}{2}\zeta^2e_{32}-\frac{i}{\sqrt{2}}\zeta^2e_{33},
\end{align}
\begin{align}
    \m(e_{11}e_{11})=\ & e-\hlf e_X-\hlf e_Y,\\
    \m(e_{11}e_{12})=\m(e_{12}e_{11})=\ & \frac{i}{2}e_X-\frac{i}{2}e_Y,\\
    \m(e_{11}e_{21})=\m(e_{21}e_{11})=\ & -\frac{1}{\sqrt{2}}e_{32}-\hlf e_{33},\\
    \m(e_{11}e_{22})=\m(e_{22}e_{11})=\ & -\frac{1}{\sqrt{2}}e_{31}-\frac{\sqrt{3}}{2}e_{33},\\
    \m(e_{11}e_{31})=\m(e_{31}e_{11})=\ & -\frac{1}{\sqrt{2}}e_{22}-\hlf e_{23},\\
    \m(e_{11}e_{32})=\m(e_{32}e_{11})=\ & -\frac{1}{\sqrt{2}}e_{21}+\frac{\sqrt{3}}{2}e_{23},
\end{align}
\begin{align}
    \m(e_{21}e_{21})=\ & e-\hlf\zeta e_X-\hlf\zeta^2e_Y,\\
    \m(e_{21}e_{22})=\m(e_{22}e_{21})=\ & \frac{i}{2}\zeta e_X-\frac{i}{2}\zeta^2e_Y,\\
    \m(e_{21}e_{31})=\m(e_{31}e_{21})=\ & -\frac{1}{\sqrt{2}}e_{12}-\hlf e_{13},\\
    \m(e_{21}e_{32})=\m(e_{32}e_{21})=\ & -\frac{1}{\sqrt{2}}e_{11}-\frac{\sqrt{3}}{2}e_{13},\\
    \m(e_{31}e_{31})=\ & e-\hlf\zeta^2e_X-\hlf\zeta e_Y,\\
    \m(e_{31}e_{32})=\m(e_{32}e_{31})=\ & \frac{i}{2}\zeta^2e_X-\frac{i}{2}\zeta e_Y.
\end{align}

The non-zero coefficients are $\m_{1,A}^A=\m_{A,1}^A=1$, and
\begin{equation}
    \m_{X,X}^Y=\beta_1^{-1},\quad\m_{X,Y}^1=\beta_2^{-1},\quad\m_{Y,X}^1=\beta_3^{-1},\quad\m_{Y,Y}^X=\beta_4^{-1},
\end{equation}
\begin{equation}
    \m_{X,Z_1}^{Z_1}=-\beta_5^{-1},\quad\m_{X,Z_2}^{Z_2}=-\zeta^2\beta_5^{-1},\ \m_{X,Z_3}^{Z_3}=-\zeta\beta_5^{-1},\ \m_{Y,Z_1}^{Z_1}=-i\beta_7^{-1},\ \m_{Y,Z_2}^{Z_2}=-i\zeta\beta_7^{-1},\ \m_{Y,Z_3}^{Z_3}=-i\zeta^2\beta_7^{-1},
    \nonumber
\end{equation}
\begin{equation}
    \m_{Z_1,X}^{Z_1}=-\beta_6^{-1},\ \m_{Z_2,X}^{Z_2}=-\zeta^2\beta_6^{-1},\ \m_{Z_3,X}^{Z_3}=-\zeta\beta_6^{-1},\ \m_{Z_1,Y}^{Z_1}=-i\beta_8^{-1},\ \m_{Z_2,Y}^{Z_2}=-i\zeta\beta_8^{-1},\ \m_{Z_3,Y}^{Z_3}=-i\zeta^2\beta_8^{-1},
    \nonumber
\end{equation}
\begin{equation}
    \m_{Z_1,Z_1}^1=\m_{Z_2,Z_2}^1=\m_{Z_3,Z_3}^1=\beta_9^{-1},\quad\m_{Z_1,Z_1}^X=-i\beta_{10}^{-1},\quad\m_{Z_2,Z_2}^X=-i\zeta\beta_{10}^{-1},\quad\m_{Z_3,Z_3}^X=-i\zeta^2\beta_{10}^{-1},
\end{equation}
\begin{equation}
    \quad\m_{Z_1,Z_1}^Y=-\beta_{11}^{-1},\quad\m_{Z_2,Z_2}^Y=-\zeta^2\beta_{11}^{-1},\quad\m_{Z_3,Z_3}^Y=-\zeta\beta_{11}^{-1},
\end{equation}
\begin{align}
    \lp\m_{Z_1,Z_2}^{Z_3}\rp^1=\lp\m_{Z_2,Z_3}^{Z_1}\rp^1=\lp\m_{Z_3,Z_1}^{Z_2}\rp^1=\ & \frac{1}{\Om}\lp -\zeta^2\beta_{14}+i\beta_{15}\rp,\\
    \lp\m_{Z_1,Z_2}^{Z_3}\rp^2=\lp\m_{Z_2,Z_3}^{Z_1}\rp^2=\lp\m_{Z_3,Z_1}^{Z_2}\rp^2=\ & \frac{1}{\Om}\lp\zeta^2\beta_{12}-i\beta_{13}\rp,\\
    \lp\m_{Z_1,Z_3}^{Z_2}\rp^1=\lp\m_{Z_2,Z_1}^{Z_3}\rp^1=\lp\m_{Z_3,Z_2}^{Z_1}\rp^1=\ & \frac{1}{\Om}\lp -\zeta\beta_{14}+i\beta_{15}\rp,\\
    \lp\m_{Z_1,Z_3}^{Z_2}\rp^2=\lp\m_{Z_2,Z_1}^{Z_3}\rp^2=\lp\m_{Z_3,Z_2}^{Z_1}\rp^2=\ & \frac{1}{\Om}\lp\zeta\beta_{12}-i\beta_{13}\rp.
\end{align}

Similarly, for co-multiplication we find
\begin{align}
    \Delta(e)=\ & \frac{1}{12}\ls ee+e_Xe_Y+e_Ye_X+e_{11}e_{11}+e_{12}e_{12}+e_{13}e_{13}+e_{21}e_{21}+e_{22}e_{22}+e_{23}e_{23}\right.\non\\
    & \qquad\left. +e_{31}e_{31}+e_{32}e_{32}+e_{33}e_{33}\rs,\\
    \Delta(e_X)=\ & \frac{1}{12}\ls ee_X+e_Xe+e_Ye_Y-\hlf e_{11}e_{11}-\frac{i}{2}e_{11}e_{12}+\frac{i}{\sqrt{2}}e_{11}e_{13}-\frac{i}{2}e_{12}e_{11}+\hlf e_{12}e_{12}+\frac{1}{\sqrt{2}}e_{12}e_{13}\right.\non\\
    & \qquad\left. +\frac{i}{\sqrt{2}}e_{13}e_{11}+\frac{1}{\sqrt{2}}e_{13}e_{12}-\hlf\zeta^2e_{21}e_{21}-\frac{i}{2}\zeta^2e_{21}e_{22}+\frac{i}{\sqrt{2}}\zeta^2e_{21}e_{23}-\frac{i}{2}\zeta^2e_{22}e_{21}+\hlf\zeta^2e_{22}e_{22}\right.\non\\
    & \qquad\left. +\frac{1}{\sqrt{2}}\zeta^2e_{22}e_{23}+\frac{i}{\sqrt{2}}\zeta^2e_{23}e_{21}+\frac{1}{\sqrt{2}}\zeta^2e_{23}e_{22}-\hlf\zeta e_{31}e_{31}-\frac{i}{2}\zeta e_{31}e_{32}+\frac{i}{\sqrt{2}}\zeta e_{31}e_{33}\right.\non\\
    & \qquad\left. -\frac{i}{2}\zeta e_{32}e_{31}+\hlf\zeta e_{32}e_{32}+\frac{1}{\sqrt{2}}\zeta e_{32}e_{33}+\frac{i}{\sqrt{2}}\zeta e_{33}e_{31}+\frac{1}{\sqrt{2}}\zeta e_{33}e_{32}\rs,\\
    \Delta(e_Y)=\ & \frac{1}{12}\ls ee_Y+e_Xe_X+e_Ye-\hlf e_{11}e_{11}+\frac{i}{2}e_{11}e_{12}-\frac{i}{\sqrt{2}}e_{11}e_{13}+\frac{i}{2}e_{12}e_{11}+\hlf e_{12}e_{12}+\frac{1}{\sqrt{2}}e_{12}e_{13}\right.\non\\
    & \qquad\left. -\frac{i}{\sqrt{2}}e_{13}e_{11}+\frac{1}{\sqrt{2}}e_{13}e_{12}-\hlf\zeta e_{21}e_{21}+\frac{i}{2}\zeta e_{21}e_{22}-\frac{i}{\sqrt{2}}\zeta e_{21}e_{23}+\frac{i}{2}\zeta e_{22}e_{21}+\hlf\zeta e_{22}e_{22}\right.\non\\
    & \qquad\left. +\frac{1}{\sqrt{2}}\zeta e_{22}e_{23}-\frac{i}{\sqrt{2}}\zeta e_{23}e_{21}+\frac{1}{\sqrt{2}}\zeta e_{23}e_{22}-\hlf\zeta^2e_{31}e_{31}+\frac{i}{2}\zeta^2e_{31}e_{32}-\frac{i}{\sqrt{2}}\zeta^2e_{31}e_{33}\right.\non\\
    & \qquad\left. +\frac{i}{2}\zeta^2e_{32}e_{31}+\hlf\zeta^2e_{32}e_{32}+\frac{1}{\sqrt{2}}\zeta^2e_{32}e_{33}-\frac{i}{\sqrt{2}}\zeta^2e_{33}e_{31}+\frac{1}{\sqrt{2}}\zeta^2e_{33}e_{32}\rs,
\end{align}
\begin{align}
    \Delta(e_{11})=\ & \frac{1}{12}\ls ee_{11}-\hlf e_Xe_{11}+\frac{i}{2}e_Xe_{12}-\frac{i}{\sqrt{2}}e_Xe_{13}-\hlf e_Ye_{11}-\frac{i}{2}e_Ye_{12}+\frac{i}{\sqrt{2}}e_Ye_{13}+e_{11}e-\hlf e_{11}e_X\right.\non\\
    & \qquad\left. +\frac{i}{2}e_{12}e_X-\frac{i}{\sqrt{2}}e_{13}e_X-\hlf e_{11}e_Y-\frac{i}{2}e_{12}e_Y+\frac{i}{\sqrt{2}}e_{13}e_Y-\frac{1}{\sqrt{2}}e_{21}e_{32}-\hlf e_{21}e_{33}-\frac{1}{\sqrt{2}}e_{22}e_{31}\right.\non\\
    & \qquad\left. -\frac{\sqrt{3}}{2}e_{22}e_{33}-\hlf e_{23}e_{31}+\frac{\sqrt{3}}{2}e_{23}e_{32}-\frac{1}{\sqrt{2}}e_{31}e_{22}-\hlf e_{31}e_{23}-\frac{1}{\sqrt{2}}e_{32}e_{21}+\frac{\sqrt{3}}{2}e_{32}e_{23}\right.\non\\
    & \qquad\left. -\hlf e_{33}e_{21}-\frac{\sqrt{3}}{2}e_{33}e_{22}\rs,\\
    \Delta(e_{21})=\ & \frac{1}{12}\ls ee_{21}-\hlf\zeta e_Xe_{21}+\frac{i}{2}\zeta e_Xe_{22}-\frac{i}{\sqrt{2}}\zeta e_Xe_{23}-\hlf\zeta^2e_Ye_{21}-\frac{i}{2}\zeta^2e_Ye_{22}+\frac{i}{\sqrt{2}}\zeta^2e_Ye_{23}\right.\non\\
    & \qquad\left. -\frac{1}{\sqrt{2}}e_{11}e_{32}-\hlf e_{11}e_{33}-\frac{1}{\sqrt{2}}e_{12}e_{31}+\frac{\sqrt{3}}{2}e_{12}e_{33}-\hlf e_{13}e_{31}-\frac{\sqrt{3}}{2}e_{13}e_{32}+e_{21}e\right.\non\\
    & \qquad\left. -\hlf\zeta e_{21}e_X+\frac{i}{2}\zeta e_{22}e_X-\frac{i}{\sqrt{2}}\zeta e_{23}e_X-\hlf\zeta^2e_{21}e_Y-\frac{i}{2}\zeta^2e_{12}e_Y+\frac{i}{\sqrt{2}}\zeta^2e_{23}e_Y\right.\non\\
    & \qquad\left. -\frac{1}{\sqrt{2}}e_{31}e_{12}-\hlf e_{31}e_{13}-\frac{1}{\sqrt{2}}e_{32}e_{11}-\frac{\sqrt{3}}{2}e_{32}e_{13}-\hlf e_{33}e_{11}+\frac{\sqrt{3}}{2}e_{33}e_{12}\rs,\\
    \Delta(e_{31})=\ & \frac{1}{12}\ls ee_{31}-\hlf\zeta^2e_Xe_{31}+\frac{i}{2}\zeta^2e_Xe_{32}-\frac{i}{\sqrt{2}}\zeta^2e_Xe_{33}-\hlf\zeta e_Ye_{31}-\frac{i}{2}\zeta e_Ye_{32}+\frac{i}{\sqrt{2}}\zeta e_Ye_{33}\right.\non\\
    & \qquad\left. -\frac{1}{\sqrt{2}}e_{11}e_{22}-\hlf e_{11}e_{23}-\frac{1}{\sqrt{2}}e_{12}e_{21}-\frac{\sqrt{3}}{2}e_{12}e_{23}-\hlf e_{13}e_{21}+\frac{\sqrt{3}}{2}e_{13}e_{22}\right.\non\\
    & \qquad\left. -\frac{1}{\sqrt{2}}e_{21}e_{12}-\hlf e_{21}e_{13}-\frac{1}{\sqrt{2}}e_{22}e_{11}+\frac{\sqrt{3}}{2}e_{22}e_{13}-\hlf e_{23}e_{11}-\frac{\sqrt{3}}{2}e_{23}e_{12}+e_{31}e\right.\non\\
    & \qquad\left. -\hlf\zeta^2e_{31}e_X+\frac{i}{2}\zeta^2e_{32}e_X-\frac{i}{\sqrt{2}}\zeta^2e_{33}e_X-\hlf\zeta e_{31}e_Y-\frac{i}{2}\zeta e_{32}e_Y+\frac{i}{\sqrt{2}}\zeta e_{33}e_Y\rs.
\end{align}
The corresponding nonzero coefficients are $\Delta_A^{1,A}=\Delta_A^{A,1}=\frac{1}{12}$ and
\begin{equation}
    \Delta_1^{X,Y}=\frac{\beta_3}{12},\quad\Delta_1^{Y,X}=\frac{\beta_2}{12},\quad\Delta_1^{Z_1,Z_1}=\Delta_1^{Z_2,Z_2}=\Delta_1^{Z_3,Z_3}=\frac{\beta_9}{12},
\end{equation}
\begin{equation}
    \Delta_X^{Y,Y}=\frac{\beta_3}{12\beta_1},\quad\Delta_X^{Z_1,Z_1}=-\frac{\beta_9}{12\beta_5},\quad\Delta_X^{Z_2,Z_2}=-\frac{\zeta^2\beta_9}{12\beta_5},\quad\Delta_X^{Z_3,Z_3}=-\frac{\zeta\beta_9}{12\beta_5},
\end{equation}
\begin{equation}
    \Delta_Y^{X,X}=\frac{\beta_2}{12\beta_4},\quad\Delta_Y^{Z_1,Z_1}=-\frac{i\beta_9}{12\beta_7},\quad\Delta_Y^{Z_2,Z_2}=-\frac{i\zeta\beta_9}{12\beta_7},\quad\Delta_Y^{Z_3,Z_3}=-\frac{i\zeta^2\beta_9}{12\beta_7},
\end{equation}
\begin{equation}
    \Delta_{Z_1}^{X,Z_1}=-\frac{i\beta_9}{12\beta_{10}},\quad\Delta_{Z_1}^{Y,Z_1}=-\frac{\beta_9}{12\beta_{11}},\quad\Delta_{Z_1}^{Z_1,X}=-\frac{i\beta_2}{12\beta_8},\quad\Delta_{Z_1}^{Z_1,Y}=-\frac{\beta_3}{12\beta_6},
\end{equation}
\begin{equation}
    \Delta_{Z_2}^{X,Z_2}=-\frac{i\zeta\beta_9}{12\beta_{10}},\quad\Delta_{Z_2}^{Y,Z_2}=-\frac{\zeta^2\beta_9}{12\beta_{11}},\quad\Delta_{Z_2}^{Z_2,X}=-\frac{i\zeta\beta_2}{12\beta_8},\quad\Delta_{Z_2}^{Z_2,Y}=-\frac{\zeta^2\beta_3}{12\beta_6},
\end{equation}
\begin{equation}
    \Delta_{Z_3}^{X,Z_3}=-\frac{i\zeta^2\beta_9}{12\beta_{10}},\quad\Delta_{Z_3}^{Y,Z_3}=-\frac{\zeta\beta_9}{12\beta_{11}},\quad\Delta_{Z_3}^{Z_3,X}=-\frac{i\zeta^2\beta_2}{12\beta_8},\quad\Delta_{Z_3}^{Z_3,Y}=-\frac{\zeta\beta_3}{12\beta_6},
\end{equation}
\begin{align}
    \lp\Delta_{Z_1}^{Z_2,Z_3}\rp^1=\lp\Delta_{Z_2}^{Z_3,Z_1}\rp^1=\lp\Delta_{Z_3}^{Z_1,Z_2}\rp^1=\ & \frac{\beta_9}{12\Om}\lp -\zeta\beta_{14}+i\beta_{15}\rp,\\
    \lp\Delta_{Z_1}^{Z_2,Z_3}\rp^2=\lp\Delta_{Z_2}^{Z_3,Z_1}\rp^2=\lp\Delta_{Z_3}^{Z_1,Z_2}\rp^2=\ & \frac{\beta_9}{12\Om}\lp\zeta\beta_{12}-i\beta_{13}\rp,\\
    \lp\Delta_{Z_1}^{Z_3,Z_2}\rp^1=\lp\Delta_{Z_2}^{Z_1,Z_3}\rp^1=\lp\Delta_{Z_3}^{Z_2,Z_1}\rp^1=\ & \frac{\beta_9}{12\Om}\lp -\zeta^2\beta_{14}+i\beta_{15}\rp,\\
    \lp\Delta_{Z_1}^{Z_3,Z_2}\rp^2=\lp\Delta_{Z_2}^{Z_1,Z_3}\rp^2=\lp\Delta_{Z_3}^{Z_2,Z_1}\rp^2=\ & \frac{\beta_9}{12\Om}\lp\zeta^2\beta_{12}-i\beta_{13}\rp.
\end{align}

Finally, assembling all the pieces gives us
\begin{align}
    Z_{1+X+Y+3Z}=\ & \frac{1}{12}\ls Z_{1,1}^1+Z_{1,X}^X+Z_{1,Y}^Y+3Z_{1,Z}^Z+Z_{X,1}^X+\frac{\beta_2}{\beta_1\beta_4}Z_{X,X}^Y+Z_{X,Y}^1+\frac{3i\beta_2}{\beta_5\beta_8}Z_{X,Z}^Z+Z_{Y,1}^Y+Z_{Y,X}^1\right.\non\\
    & \qquad\left. +\frac{\beta_3}{\beta_1\beta_4}Z_{Y,Y}^X+\frac{3i\beta_3}{\beta_6\beta_7}Z_{Y,Z}^Z+3Z_{Z,1}^Z+\frac{3i\beta_9}{\beta_6\beta_{10}}Z_{Z,X}^Z+\frac{3i\beta_9}{\beta_8\beta_{11}}Z_{Z,Y}^Z+3Z_{Z,Z}^1+\frac{3i\beta_9}{\beta_5\beta_{10}}Z_{Z,Z}^X\right.\non\\
    & \qquad\left. +\frac{3i\beta_9}{\beta_7\beta_{11}}Z_{Z,Z}^Y-\frac{3\beta_9}{\Om^2}\lp\lp\beta_{14}^2-2i\beta_{14}\beta_{15}+2\beta_{15}^2\rp\lp Z_{Z,Z}^Z\rp_{11}\right.\right.\non\\
    & \qquad\quad\left.\left. -\lp\beta_{12}\beta_{14}-i\beta_{12}\beta_{15}-i\beta_{13}\beta_{14}+2\beta_{13}\beta_{15}\rp\lp\lp Z_{Z,Z}^Z\rp_{12}+\lp Z_{Z,Z}^Z\rp_{21}\rp\right.\right.\non\\
    & \qquad\quad\left.\left. +\lp\beta_{12}^2-2i\beta_{12}\beta_{13}+2\beta_{13}^2\rp\lp Z_{Z,Z}^Z\rp_{22}\rp\rs.
\end{align}
Again, a very nontrivial check is that this combination of partial traces is completely modular invariant.

In the nice gauge of appendix~\ref{appsub:A4GaugeChoice},
\begin{align}
    Z_{1+X+Y+3Z}=\ & \frac{1}{12}\ls Z_{1,1}^1+Z_{1,X}^X+Z_{1,Y}^Y+3Z_{1,Z}^Z+Z_{X,1}^X+Z_{X,X}^Y+Z_{X,Y}^1+3Z_{X,Z}^Z+Z_{Y,1}^Y+Z_{Y,X}^1+Z_{Y,Y}^X\right.\non\\
    & \qquad\left. +3Z_{Y,Z}^Z+3Z_{Z,1}^Z+3Z_{Z,X}^Z+3Z_{Z,Y}^Z+3Z_{Z,Z}^1+3Z_{Z,Z}^X+3Z_{Z,Z}^Y\right.\non\\
    & \qquad\left. +3\sqrt{3}\lp\lp Z_{Z,Z}^Z\rp_{11}-\lp Z_{Z,Z}^Z\rp_{22}\rp\rs.
\end{align}

\subsubsection{Discrete torsion in Rep$(A_4)$ gaugings}\label{sssec:dta4}

So far, we have found five algebra objects for gauging Rep$(A_4)$, but this does not exhaust all possibilities. Recall that our strategy is that every gaugeable subgroup $H\subseteq G$ of $G=A_4$ has an associated algebra object, thus a possible gauging of Rep$(A_4)$. In the case of subgroup gauging $H=\Z_2\times \Z_2$ and $A_4$, there exist discrete torsion choices due to the fact $H^2(\Z_2\times \Z_2, U(1))=H^2(A_4, U(1))=\Z_2$. Therefore, it is easy to observe the other two gaugings:
\begin{itemize}
    \item $H=(\Z_2\times \Z_2)_\text{d.t.}$, namely gauging $\Z_2\times \Z_2\subset A_4$ with discrete torsion turned on. The associated algebra object is $(1+X+Y+3Z)_\text{d.t.}$, namely gauging the full Rep$(A_4)$ category but with the discrete torsion turned on. This matches the fact that Rep$(A_4)$ has two fiber functors\footnote{We thank R.~Radhakrishnan for valuable discussions on this point.} \cite{Putrov24} (physically speaking, two SPT phases)\footnote{In general, if a group $G$ has a $\Z_2\times \Z_2$ subgroup, there is an additional fiber functor for Rep$(G)$ \cite{ostrik1}.}\footnote{Alternatively, one sees that the equivalence classes of fiber functors are in one-to-one correspondence with the twists of $\mathbb{C}[G]$ modulo trivial twists \cite{bontea}. In the case of $G=A_4$, this class is specified by the lazy cohomology with order 2 \cite[Proposition 7.7]{GK10}.}.

    From knowledge of the algebra object, and by knowing the relations between the partition functions for orbifolds by $A_4$ subgroups with or without discrete torsion, it is possible to reverse engineer the form of the partition function in this case,
\begin{align}
    Z_{1+X+Y+3Z,\mathrm{d.t.}}=\ & \frac{1}{12}\ls Z_{1,1}^1+Z_{1,X}^X+Z_{1,Y}^Y+3Z_{1,Z}^Z+Z_{X,1}^X+\frac{\beta_2}{\beta_1\beta_4}Z_{X,X}^Y+Z_{X,Y}^1+\frac{3i\beta_2}{\beta_5\beta_8}Z_{X,Z}^Z+Z_{Y,1}^Y\right.\non\\
    & \qquad\left. +Z_{Y,X}^1+\frac{\beta_3}{\beta_1\beta_4}Z_{Y,Y}^X+\frac{3i\beta_3}{\beta_6\beta_7}Z_{Y,Z}^Z+3Z_{Z,1}^Z+\frac{3i\beta_9}{\beta_6\beta_{10}}Z_{Z,X}^Z+\frac{3i\beta_9}{\beta_8\beta_{11}}Z_{Z,Y}^Z+3Z_{Z,Z}^1\right.\non\\
    & \qquad\left. +\frac{3i\beta_9}{\beta_5\beta_{10}}Z_{Z,Z}^X+\frac{3i\beta_9}{\beta_7\beta_{11}}Z_{Z,Z}^Y+\frac{3\beta_9}{\Om^2}\lp\lp\beta_{14}^2-2i\beta_{14}\beta_{15}+2\beta_{15}^2\rp\lp Z_{Z,Z}^Z\rp_{11}\right.\right.\non\\
    & \qquad\quad\left.\left. -\lp\beta_{12}\beta_{14}-i\beta_{12}\beta_{15}-i\beta_{13}\beta_{14}+2\beta_{13}\beta_{15}\rp\lp\lp Z_{Z,Z}^Z\rp_{12}+\lp Z_{Z,Z}^Z\rp_{21}\rp\right.\right.\non\\
    & \qquad\quad\left.\left. +\lp\beta_{12}^2-2i\beta_{12}\beta_{13}+2\beta_{13}^2\rp\lp Z_{Z,Z}^Z\rp_{22}\rp\rs\\
    =\ & \frac{1}{12}\ls Z_{1,1}^1+Z_{1,X}^X+Z_{1,Y}^Y+3Z_{1,Z}^Z+Z_{X,1}^X+Z_{X,X}^Y+Z_{X,Y}^1+3Z_{X,Z}^Z+Z_{Y,1}^Y+Z_{Y,X}^1\right.\non\\
    & \qquad\left. +Z_{Y,Y}^X+3Z_{Y,Z}^Z+3Z_{Z,1}^Z+3Z_{Z,X}^Z+3Z_{Z,Y}^Z+3Z_{Z,Z}^1+3Z_{Z,Z}^X+3Z_{Z,Z}^Y\right.\non\\
    & \qquad\left. -3\sqrt{3}\lp\lp Z_{Z,Z}^Z\rp_{11}-\lp Z_{Z,Z}^Z\rp_{22}\rp\rs.\non
\end{align}
    This differs from the $1+X+Y+3Z$ result without discrete torsion only in the sign of the $(Z_{Z,Z}^Z)_{ij}$ pieces.  Note however, that we have not given the full Frobenius algebra structure in this case.
    \item $H=(A_4)_\text{d.t.}$, namely gauging $A_4$ with discrete torsion turned on. The associated algebra object is $(1+Z)_\text{d.t.}$, namely gauging $1+Z$ with discrete torsion turned on. Again, combining knowledge of the algebra object with knowledge of the form of the $A_4$ orbifold partition functions allows us to write
    \begin{align}
        Z_{1+Z,\mathrm{d.t.}}=\ & \frac{1}{4}\ls Z_{1,1}^1+Z_{1,Z}^Z+Z_{Z,1}^Z+Z_{Z,Z}^1\vphantom{\frac{\beta_9}{\Om^2}}\right.\non\\
        & \qquad\left. +\frac{\beta_9}{\Om^2}\lp\beta_{14}^2\lp Z_{Z,Z}^Z\rp_{11}-\beta_{12}\beta_{14}\lp\lp Z_{Z,Z}^Z\rp_{12}+\lp Z_{Z,Z}^Z\rp_{21}\rp+\beta_{12}^2\lp Z_{Z,Z}^Z\rp_{22}\rp\rs\\
        =\ & \frac{1}{4}\ls Z_{1,1}^1+Z_{1,Z}^Z+Z_{Z,1}^Z+Z_{Z,Z}^1+\frac{2}{\sqrt{3}}\lp Z_{Z,Z}^Z\rp_{22}\rs.\non
    \end{align}
\end{itemize}

In the language of \cite{Perez-Lona:2024yih}, the algebra structures $(1+X+Y+3Z)_{\rm d.t.}\equiv(1+X+Y+3Z,\mu_{\rm d.t.},\Delta_{\rm d.t.})$, and $(1+Z_{\rm d.t.})\equiv(1+Z,\mu_{\rm d.t.},\Delta_{\rm d.t.})$ correspond to picking a different element of the set of discrete torsion choices available for the objects $1+X+Y+3Z$ and $1+Z$, respectively. In particular, for the object $1+X+Y+3Z$, this set of choices is equivalently the set of equivalence classes of fiber functors on $\text{Rep}(A_4)$. However, here the story is richer in the sense that these choices are moreover related by discrete torsion \textit{twists}, which are classified by the lazy cohomology group $H^2_{\ell}(\text{Rep}(A_4))=\Z_2$ \cite[Proposition 7.7]{GK10}. In other words, not only do we have a set of two distinct choices of algebra structures, but these choices are furthermore in the same orbit of the twisting action by the cohomology group of discrete torsion twists. We remind the reader that this twisting action is defined for all algebra objects in the fusion category, not only for the regular object.

\subsection{Summary of Rep$(A_4)$ gaugings}

Now we are ready to summarize the gaugings of Rep$(A_4)$. We present all possible gaugings of Rep$(A_4)$,  and their corresponding subgroup gaugings of $A_4$, as well as the dual categorical symmetries, respectively, in Table \ref{table:a4_duality}.
\begin{table}[h]
    \centering
    \begin{tabular}{|c | c | c|}
    \hline
    $H\subset A_4$ & Rep$(A_4)$ Algebra Object & Dual Categorical Symmetry \\\hline
    $\id$ & $1+X+Y+3Z$ & $A_4$ \\
    \hline
    $\Z_2$ & $1+X+Y+Z$ & $\mathcal{C}^\text{T}$ \\
     \hline
    $\Z_3$ & $1+Z$ & Rep$(A_4)$ \\
     \hline
    $\Z_2\times \Z_2$ & $1+X+Y$ & $A_4$ \\
     \hline
    $(\Z_2\times \Z_2)_\text{d.t.}$ & $(1+X+Y+3Z)_\text{d.t.}$ & $A_4$\\
     \hline
    $A_4$ & $1$ & Rep$(A_4)$ \\
     \hline
    $(A_4)_\text{d.t.}$ & $(1+Z)_\text{d.t.}$ & Rep$(A_4)$ \\
    \hline
    \end{tabular}
    \caption{Algebra objects for gaugings of Rep$(A_4)$, their associated subgroups of $A_4$ as well as the resulting dual categorical symmetries.}
    \label{table:a4_duality}
\end{table}

The rightmost column presents the dual categorical symmetries under gauging the $H$ subgroup of a $A_4$ symmetric theory or gauging an associated algebra object of the Rep$(A_4)$ symmetric theory. The dual categorical symmetry from gauging an algebra object of Rep$(A_4)$ can be obtained from the corresponding subgroup $H$ gauging of $A_4$. Starting with a $A_4$ symmetry, gauging the full $A_4$ and one of the non-normal subgroups $\Z_3$, all lead to noninvertible Rep$(A_4)$ symmetry. When it comes to the $\Z_2\times \Z_2$ gauging, notice that the $A_4$ is a semi-direct product:
\begin{equation}
    A_4=\Z_2^2\rtimes \Z_3.
\end{equation}
In this case, gauging the $\Z_2\times \Z_2$, one has a dual $A_4$ symmetry. This is similar to gauging a $\Z_3$ subgroup of a $S_3$ leads to a dual $S_3$ symmetry discussed in \cite{Bhardwaj:2017xup}\footnote{One should distinguish this case from the one in \cite{Tachikawa:2017gyf}, where gauging $H\subset G$ leads to a $\hat{H}\times (G/H)$ symmetry with a mixed anomaly ($\hat{H}\cong H$ is the quantum symmetry from gauging $H$).  This mixed anomaly shows up when 
$G$ \emph{cannot} be written as a semi-direct product by $H$ and $G/H$. For $H=\Z_2^2\subset G=A_4$, one has the anomaly counted by $H^3(\Z_2^2\times \Z_3, U(1))=\Z_2^3\times \Z_3=H^3(\Z_2^2,U(1))\times H^3(\Z_3, U(1))$. This implies all possible anomalies for a $\Z_2^2\times \Z_3$ symmetry are purely for $\Z_2^2$ or $\Z_3$, but not mixed. Therefore, the resulting dual symmetry cannot be a mixed anomalous abelian one, but just an $A_4$. This distinction can be shown explicitly via SymTFTs for non-abelian finite symmetries \cite{heckmanwip}.}. 

Gauging $1+X+Y+Z$ of Rep$(A_4)$, which is associated to gauging $\Z_2\subset A_4$, is special, since it is the only case when the dual symmetry is neither $A_4$ nor Rep$(A_4)$. Given that $\Z_2$ is a non-normal subgroup of $A_4$, one expects it leading to a non-group categorical symmetry, which we denote as $\mathcal{C}^T$. It was shown in \cite{Thorngren:2021yso} that $\mathcal{C}^T$ is a noninvertible triality symmetry with group-like $\Z_2\times \Z_2$ subcategory. The fusion rules are 
\begin{equation}\label{eq: triality}
\begin{split}
    &\mathcal{D}_3\times \bar{\mathcal{D}}_3=\sum g, ~g\in \Z_2\times \Z_2,\\
    &\mathcal{D}_3\times \mathcal{D}_3=2\bar{\mathcal{D}}_3,\\
    &\mathcal{D}_3\times g=g\times \mathcal{D}_3=\mathcal{D}_3,\\
    &\bar{\mathcal{D}}_3\times g=g\times  \bar{\mathcal{D}}_3=\bar{\mathcal{D}}_3
\end{split}
\end{equation}
where $\mathcal{D}_3$ and $\bar{\mathcal{D}}_3$ are triality defect and its orientation reversal. Physically speaking, this categorical symmetry implies a triality symmetry for a theory under gauging $\Z_2\times \Z_2$.

\begin{figure}[h]
    \centering
    \includegraphics[width=8.5cm]{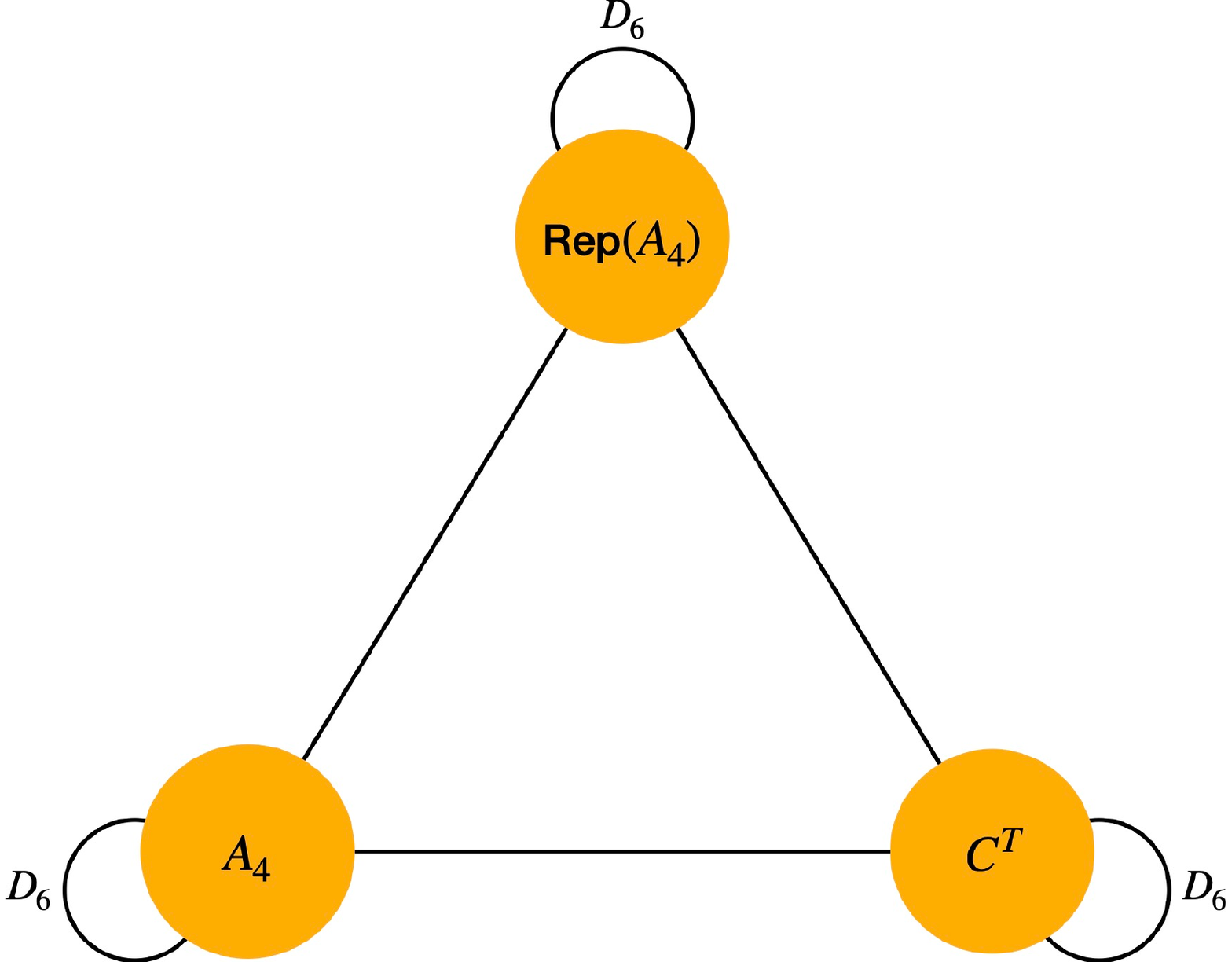}
    \caption{Brauer-Picard groupoid of Rep$(A_4)$ symmetry. We denote fusion categories as objects in the groupoid with nodes. We use straight links to denote different categorical symmetries that are connected via gauging, i.e., they are Morita equivalent to each other. The link attached to a single node denotes the Brauer-Picard group $\mathfrak{Brpic}(\text{Rep}(A_4))=D_6$.}
    \label{fig: a4 groupoid}
\end{figure}

Based on all possible seven gaugings in Table \ref{table:a4_duality}, we can build the generalized orbifold groupoid \cite{Gaiotto:2020iye, Diatlyk:2023fwf}, mathematically as Brauer-Picard groupoid, for Rep$(A_4)$, with the Brauer-Picard group reads \cite{nr2013}\footnote{We review some background of BrPic group(oid) and computation of $ \mathfrak{Brpic}(\text{Rep}(A_4))$ in Appendix \ref{app: BrPic}.}
\begin{equation}
    \mathfrak{Brpic}(\text{Rep}(A_4))=D_6,
\end{equation}
The $D_6$ above denotes the dihedral group of order 12. See Figure \ref{fig: a4 groupoid} for an illustration of the resulting groupoid\footnote{Strictly speaking, we present not the full Brauer-Picard 3-groupoid but its 1-truncation.}.


\section{Applications of Rep$(A_4)$ gaugings: Exceptional $c=1$ CFT and duality defects}\label{sec: c=1}

In our previous paper \cite{Perez-Lona:2023djo} (see also \cite{Thorngren:2021yso, Luo:2023ive}), we considered $c=1$ CFTs enjoying multiplicity-free noninvertible symmetries and discuss how the various gaugings can be identified as connecting different points in the $c=1$ moduli space. Here we consider the realization of Rep$(A_4)$ gaugings in $c=1$ CFTs.

Let us start with a compact boson whose target space is just a $S^1$ with radius $R$. Using left- and right-moving fields $X_L$ and $X_R$, we define $U(1)$-valued fields as momentum and winding variables
\begin{equation}
    \theta_m=\frac{X_L+X_R}{R},\quad \theta_w=R(X_L-X_R).
\end{equation}
Different values of $R$ correspond to theories on the circle branch of the $c=1$ moduli space, related by T-duality
\begin{equation}
\begin{split}
    R&\rightarrow \frac{1}{R},\\
    \theta_m &\leftrightarrow \theta_w,\\
    X_R&\rightarrow -X_R.
\end{split}
\end{equation}

At a generic $R$, the global symmetry enjoyed by the theory is 
\begin{equation}\label{eq: sym of circle branch}
    ( U(1)_m\times U(1)_w )\rtimes \mathbb{Z}_2^r,
\end{equation}
where subindices $m$ and $w$ denote the momentum and winding symmetries, respectively, and $\mathbb{Z}_2^r$ denotes the ``reflection'' symmetry:
\begin{equation}
   \mathbb{Z}_2^r: (\theta_m, \theta_w)\rightarrow (-\theta_m, -\theta_w).
\end{equation}
which sometimes is also referred to as ``charge conjugation.''

In order to realize a Rep$(G)$ symmetry, a general strategy is to think of $G$ as a subgroup of (\ref{eq: sym of circle branch}). Then, by gauging the $G$, one ends up with a $c=1$ theory enjoying a noninvertible Rep$(G)$ symmetry. In many cases, this leads to theories on the orbifold branch, which is defined by the $\Z_2^r$ orbifold at some radius $R$. For example, when $G=D_4=\mathbb{Z}_4\rtimes \mathbb{Z}_2^r$, gauging $G$ also includes gauging the $\mathbb{Z}_2^r$, which by definition results in a theory living on the orbifold branch. 

However, the finite subgroups of (\ref{eq: sym of circle branch}) do not include $A_4$. In order to start with an $A_4$ symmetric theory and gauge it to get a Rep$(A_4)$ symmetry, one needs a larger global symmetry group than (\ref{eq: sym of circle branch}) to start with. This leads us to consider the compact boson at the T-duality self-dual point $R=1$, known as the $SU(2)_1$ theory. This self-duality point of the $c=1$ circle branch enjoys a enhanced global symmetry compared to (\ref{eq: sym of circle branch}) as
\begin{equation}
   ( SU(2)\times SU(2) )/\mathbb{Z}_2\cong SO(4).
\end{equation}
It was found in \cite{Ginsparg:1987eb} that orbifolds of this self-dual theory admit an ADE classification:
\begin{itemize}
    \item A-type. This $\mathfrak{a}_n$ class corresponds to $\Z_n$ orbifolds, resulting in $SU(2)_1/\Z_n$ theories. These theories live at $R=n$ (or equivalently $R=1/n$ by T-duality) points in the circle branch. One special case is when $n=2$, where the resulting theory also belongs to the orbifold branch, i.e. it is the intersection point of the circle and the orbifold branch. 
    \item D-type. This $\mathfrak{d}_n$ class corresponds to  $D_n$ (order-$2n$ Dihedral group) orbifolds, resulting in $SU(2)_1/D_n$ theories. These theories live at $R=n$ (or equivalently $R=1/n$ by T-duality) points in the orbifold branch. These theories can be derived as $\Z_2^r$ orbifolds of $SU(2)_1/\Z_n$ theories.
    \item E-type. These three exceptional cases $\mathfrak{e_6, e_7}$ and $\mathfrak{e_8}$ correspond to $A_4, S_4$ and $A_5$ orbifolds, respectively, resulting in the following three theories
    \begin{equation}
    \frac{SU(2)_1}{A_4},\frac{SU(2)_1}{S_4}, \frac{SU(2)_1}{A_5}.
\end{equation}
These three theories are isolated in the $c=1$ moduli space. Namely, they belong to neither the circle nor the orbifold branch. 
\end{itemize}
Recall the fact that gauging a $G$ symmetry leads to a Rep$(G)$ quantum symmetry; thus, the three exceptional orbifold theories enjoy Rep$(A_4)$, Rep$(S_4)$ and Rep$(A_5)$ noninvertible symmetries. This result was observed in \cite{Thorngren:2021yso}.

\subsection{Rep$(A_4)$ gaugings of the $SU(2)_1/A_4$ theory}

\begin{figure}[h]
    \centering
    \includegraphics[width=16.5cm]{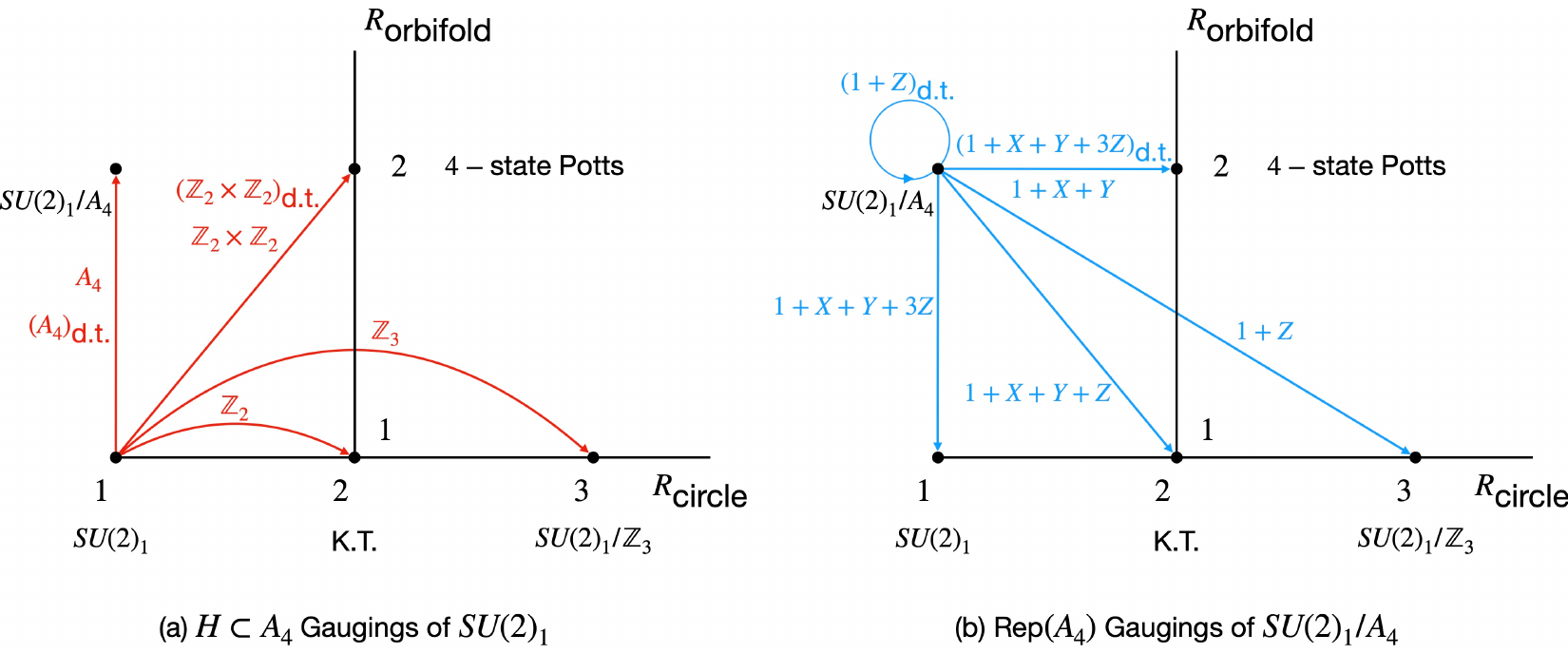}
    \caption{(a) Gauging various subgroups $H\subset A_4$ of the $SU(2)_1$ theory and (b) various Rep$(A_4)$ gaugings of the exceptional orbifold theory $SU(2)_1/A_4$ and the respectively resulting CFTs in $c=1$ moduli space.}
    \label{fig:c=1gauging}
\end{figure}

Utilizing the result in Table \ref{table:a4_duality}, it is straightforward to find various Rep$(A_4)$ gaugings of the $SU(2)_1/A_4$ theory from gauging $H\subset A_4$ subgroups of the $SU(2)_1$ theory. We present our results below and illustrate how these gaugings connect various theories in $c=1$ moduli space in Figure \ref{fig:c=1gauging}. 

\begin{figure}[h]
    \centering
    \includegraphics[width=16.5cm]{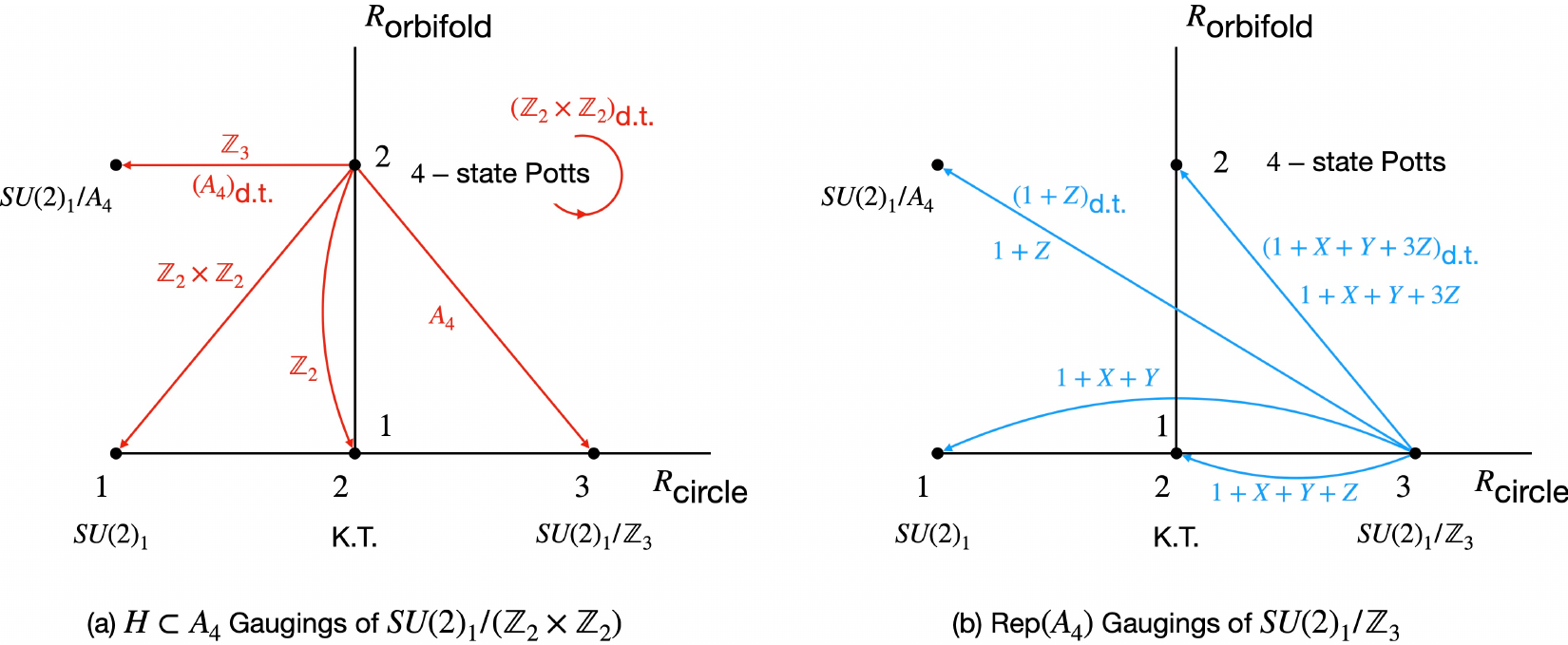}
    \caption{(a) Gauging various subgroups $H\subset A_4$ of the $SU(2)_1/(\Z_2\times \Z_2)$ theory and (b) various Rep$(A_4)$ gaugings of the exceptional orbifold theory $SU(2)_1/A_4$ and the respectively resulting CFTs in $c=1$ moduli space. Note that the self-duality of $SU(2)_1/(\Z_2\times \Z_2)$ theory under gauging $(\Z_2\times \Z_2)_{\text{d.t.}}$ matches the fact that the theory enjoys at least a Rep$(D_4)$ symmetry \cite{Perez-Lona:2023djo}.}
    \label{fig: more c=1 gaugings}
\end{figure}

\paragraph{$H\subset A_4$ gaugings of the $SU(2)_1$ theory.}
For $H\subset A_4$ gaugings of the $SU(2)_1$ theory, we can use the ADE classification of its orbifold theories to read off the following results:
\begin{itemize}
    \item $H=\mathbb{Z}_2$ and $H=\Z_3$ belong to the A-type orbifolds, and give rise to $R=2$ and $R=3$ theories on the circle branch respectively. In particular, $R=2$ on the circle branch is also $R=1$ on the orbifold branch. These correspond to two curved red lines in Figure \ref{fig:c=1gauging} (a).
    \item $H=\Z_2\times \Z_2 \cong D_2$ corresponds to the simplest D-type orbifold. The gauged theory is at $R=2$ on the orbifold branch. This corresponds to the diagonal red line in Figure \ref{fig:c=1gauging} (a).
    \item $H=A_4$ is the $\mathfrak{e}_6$ type orbifold, by definition leading to our interested $SU(2)_1/A_4$ theory. This corresponds to the vertical red line in Figure \ref{fig:c=1gauging} (a).
    \item $H=(\Z_2\times \Z_2)_\text{d.t.}$ and $H=(A_4)_\text{d.t.}$, in general are distinct from $\Z_2\times \Z_2$ and $A_4$ gaugings, respectively. However, in this case, the ’t Hooft anomaly for $SO(4)$ global symmetry\footnote{One way to understand this 't Hooft anomaly is it inherits from the well-known mixed anomaly between the $U(1)_m$ and the $U(1)_w$ symmetries for generic $R$ on the circle branch.} provides a chiral $\mathbb{Z}_2$ rotation which compensates this $H^2(\Z_2\times \Z_2; U(1))=H^2(A_4; U(1))=\Z_2$ choice of the discrete torsion. Therefore, they physically lead to the same orbifold theories as $H=\Z_2\times \Z_2$ and $H=(A_4)$ gaugings \cite{Thorngren:2021yso}.
    \item Since $A_4$ can be built as a group extension  $1\rightarrow \Z_2\times \Z_2 \rightarrow A_4 \rightarrow \Z_3 \rightarrow 1$, gauging $A_4$ can be performed in a two-step sequential gauging: first gauge $\Z_2\times \Z_2$ and then gauge $\Z_3$. This tells us that the exceptional $SU(2)_1/A_4$ CFT can also be derived from gauging a $\Z_3$ of the 4-state Potts model at $R=2$ on the orbifold branch. This corresponds to the horizontal red line in Figure \ref{fig: more c=1 gaugings}(a).
\end{itemize}

\paragraph{Rep$(A_4)$ gaugings of the $SU(2)_1/A_4$ theory.}
The above analysis of $H\subset A_4$ gaugings of the $SU(2)_1$ theory allows us to further discuss Rep$(A_4)$ gaugings of the $SU(2)_1/A_4$ theory. In particular, from the dual categorical symmetries listed in Table~\ref{table:a4_duality}. we know that when gauging $1+Z, 1+X+Y+Z,$ and $(1+Z)_{\text{d.t.}}$, the dual category is still Rep$(A_4)$. We thus present the following results: 
\begin{itemize}
    \item Gauging $1+X+Y+3Z$: This is gauging the full Rep$(A_4)$ without discrete torsion; thus, since it is the quantum symmetry, one recovers the $SU(2)_1$ theory. This corresponds to the vertical blue line in Figure~\ref{fig:c=1gauging}(b).
    
    \item Gauging $1+X+Y$ and $(1+X+Y+3Z)_\text{d.t.}$: These are associated to $\Z_2\times \Z_2$ and $(\Z_2\times \Z_2)_\text{d.t.}$ gaugings of $SU(2)_1$ according to Table \ref{table:a4_duality}. Furthermore, as we discussed above, the discrete torsion for $\Z_2\times \Z_2$ does not lead to a physically distinct orbifold theory; thus, these two gaugings both take the $SU(2)_1/A_4$ theory to $R=2$ on the orbifold branch. This corresponds to the horizontal blue line in Figure~\ref{fig:c=1gauging}(b).
    
	\item Gauging $1+X+Y+Z$: This is associated to the $\Z_2$ gauging of $SU(2)_1$. In particular, the resulting $c=1$ CFT at radius $R=2$ has a noninvertible Rep$(A_4)$ symmetry according to Table~\ref{table:a4_duality}. Note that this theory is the intersection point of the circle branch and the orbifold branch. This aligns with the general belief that $c=1$ CFTs on the orbifold branch enjoy a rich class of categorical symmetries. 
 
	\item Gauging $1+Z$: This is associated to $\Z_3$ gauging of $SU(2)_1$. In particular, the resulting $c=1$ CFT at radius $R=3$ enjoys noninvertible Rep$(A_4)$ symmetry according to Table \ref{table:a4_duality}. \emph{This is a remarkable result since this theory, i.e.~$SU(2)_1/\Z_3$, is on the circle branch}. To our knowledge, this is the first noninvertible symmetry other than the $\Z_N$ Tambara-Yamigami category found to appear on the circle branch.

	\item Gauging $(1+Z)_{\text{d.t.}}$: This is associated to $(A_4)_\text{d.t.}$ gauging of $SU(2)_1$. However, as we already pointed out, this discrete torsion for $A_4$ gauging does not lead to a physically distinct theory from the $A_4$ gauging with trivial action. Therefore, the resulting theory after gauging $(1+Z)_{\text{d.t.}}$ is still the exceptional CFT $SU(2)_1/A_4$, i.e.,
 \begin{equation*}
 \boxed{
     \text{$SU(2)_1/A_4$ is self-dual under gauging $(1+Z)_{\text{d.t.}}$}.
     }
 \end{equation*}
 We will investigate this self-duality in more detail in the following subsection.
\end{itemize}

Following similar steps, one can also build gauging patterns for the $SU(2)_1/(\Z_2\times \Z_2)$ (4-state Potts model) on the orbifold branch and those for the $SU(2)_1/\Z_3$ theories. The starting point is to notice that $SU(2)_1/(\Z_2\times \Z_2)$ enjoys a $A_4$ symmetry, according to Table \ref{table:a4_duality}, since it is obtained from gauging a $\Z_2\times \Z_2\subset A_4$ symmetry of $SU(2)_1$. Gauging its $A_4$ symmetry is equivalent to a two-step sequential gauging: first gauging $\Z_2\times \Z_2$ and then gauging $\Z_3$. This reproduces our result that the $SU(2)_1/\Z_3$ theory enjoys a Rep$(A_4)$ symmetry. We can then conclude the various $A_4$ and Rep$(A_4)$ gaugings for these two theories. See Figure \ref{fig: more c=1 gaugings} for an illustration.

Within these five theories connected via $A_4$ and Rep$(A_4)$ gaugings, the Kosterlitz-Thouless (K.T.) point is special since it enjoys neither $A_4$ nor Rep$(A_4)$ symmetry, but a $\mathcal{C}^T$ noninvertible triality symmetry (\ref{eq: triality}). Furthermore, this triality symmetry is anomalous, due to the fact that KT point does not admit a trivially gapped phase respecting this noninvertible symmetry \cite{Thorngren:2021yso}, thus we will not discuss its gaugings.

\subsection{Multiplicity of noninvertible defects from self-dual gauging} 
If a 2D QFT is self-dual under gauging a (zero-form) global symmetry, then a noninvertible duality defect constructed via the half-space gauging exists. Namely, performing the gauging in half of the spacetime, the resulting topological interface separating the original and the gauged theory is promoted to a defect due to the self-duality of the theory. The Kramers–Wannier duality defect of the Ising CFT is the simplest example of this type, implying the Ising CFT is self-dual under gauging $\Z_2$ symmetry \cite{Frohlich:2004ef}. Including this duality defect, one ends up with a symmetry larger than $\mathbb{Z}_2$, which in this case is $\Z_2$ Tambara-Yamagami fusion category.

In \cite{CLS23, Perez-Lona:2023djo, Diatlyk:2023fwf}, the construction of noninvertible defects is generalized to self-dualities under gauging noninvertible symmetries. This includes cases under gauging the full Rep$(\mathcal{H})$ fusion categorical symmetry as well as those under gauging an algebra object $\mathcal{A}$ as a subsymmetry of a fusion category \cite{Perez-Lona:2023djo}. For a theory with a fusion categorical symmetry $\mathcal{C}$, if it is self-dual under gauging the algebra object  $\mathcal{A}=\sum_ia_iA_i$, where $A_i$ are simple objects and $a_i$ are their multiplicities, the duality defect built from the half-space gauging enjoys the following self-fusion rules\footnote{We restrict to the case $\mathcal{D}=\bar{\mathcal{D}}$, where $\bar{\mathcal{D}}$ is the orientation reversal of $\mathcal{D}$, for simplicity.},
\begin{equation}\label{eq: fusion rules of duality defect}
\begin{split}
    &\mathcal{D}\times  \mathcal{D}=\mathcal{A}.  \\
\end{split}
\end{equation}
To complete the fusion algebra, one must specify fusions of simple objects in $\mathcal{A}$ with the duality defect $\mathcal{D}$. In the case of $\mathcal{A}=\text{Vec}_G$, i.e.~gauging invertible symmetries, the duality defect will simply absorb the simple objects in $\mathcal{A}$, namely
\begin{equation}\label{eq: absorb fusion}
        A_i\times \mathcal{D}=\mathcal{D}\times A_i=\langle A_i \rangle \mathcal{D}.
\end{equation}
Furthermore, this fusion rule could also be true for some special noninvertible cases. One simple example is the self-duality of gauging Rep$(\mathcal{H}_8)$ symmetry \cite{CLS23, Perez-Lona:2023djo, Diatlyk:2023fwf},
\begin{equation}\label{eq: fl for gauging reph8}
\begin{split}
    &\mathcal{D}\times \mathcal{D}=1+a+b+c+2m,\\
    &\mathcal{D}\times m=m\times \mathcal{D}=2\mathcal{D},\\
    &\mathcal{D}\times g=g\times\mathcal{D}=\mathcal{D},  \quad g\in \{ a,b,c \},
\end{split}
\end{equation}
where the new duality defect absorbs all $\Z_2$ lines as well as the noninvertible $m$ line when fused with them. 

However, in general, for self-dualities under gauging noninvertible symmetries, the fusion rule (\ref{eq: absorb fusion}) does not lead to associative fusion algebra and thus cannot be the correct result. Interestingly, in many cases, for a certain self-dual gauging, one can obtain a consistent fusion algebra only by introducing \emph{multiple} defects. We next demonstrate this multiplicity of duality defects through some explicit examples.

\subsubsection{Revisit self-dual gaugings of Rep$(S_3)$ and Rep$(D_4)$}
Before discussing our central example $SU(2)_1/A_4$, an exceptional CFT with self-dual Rep$(A_4)$ gaugings, let us revisit the self-dualities of gauging simpler noninvertible symmetries, namely Rep$(S_3)$ and Rep$(D_4)$ in $SU(2)_4/U(1)$ and $SU(2)_1/(\mathbb{Z}_2\times \mathbb{Z}_2)$, respectively, introduced in \cite{Perez-Lona:2023djo}. For ease of reading, we recall our notations for fusion rules of Rep$(S_3)$ as 
\begin{equation}
\begin{split}
    &Y\times Y=1+X+Y,\\
    &Y\times X=X\times Y=Y,\\
    &X\times X=1,
\end{split}
\end{equation}
and those of Rep$(D_4)$ are given by
\begin{equation}
\begin{split}
     &m\times m=1+a+b+c,\\
     &g\times m=m\times g=m, g\in \{a, b, c \},\\
     &a\times a=b\times b=c\times c=1,\\
     &a\times b=b\times a=c, a\times c=c\times a=b, b\times c=c\times b=a.
\end{split}
\end{equation}
The self-duality of gauging noninvertible symmetries gives rise to the following duality defects \cite{Perez-Lona:2023djo}
\begin{equation}
\begin{split}
    &\text{Rep}(S_3): \mathcal{D}\times \mathcal{D}=1+Y,\\
    &\text{Rep}(D_4): \mathcal{D}\times \mathcal{D}=1+b+m.
\end{split}
\end{equation}
Motivated by completing the fusion rules for $\mathcal{D}$'s, we will see each of these self-dual gaugings lead to multiple duality defects.

\paragraph{Two duality defects from gauging Rep$(S_3)$ in $SU(2)_4/U(1)$.}
Given the existence of the duality defect $\mathcal{D}$ in $SU(2)_4/U(1)$ theory via gauging the $1+Y$ algebra of Rep$(S_3)$,
\begin{equation}\label{eq: fl for subgauging reps3}
\begin{split}
     &\mathcal{D}\times \mathcal{D}=1 + Y,\\
\end{split}
\end{equation}
we next must specify the fusion rule for fusing $Y$ with $\mathcal{D}$. Following (\ref{eq: absorb fusion}), a naive guess would be that $\mathcal{D}$ absorbs lines from gauging invertible symmetries,  by multiplying their quantum dimensions, so that in this case $\mathcal{D}\times Y=Y\times \mathcal{D}=2\mathcal{D}$. However, the resulting fusion rule is not associative:
\begin{equation}
\begin{split}
&\mathcal{D}\times Y=Y\times \mathcal{D}=2\mathcal{D}\\
    \Rightarrow &(Y\times  \mathcal{D}) \times \mathcal{D}=2\mathcal{D}\times \mathcal{D}=2+2Y,\\
    &Y\times (\mathcal{D}\times \mathcal{D})=  1+X+2Y,\\
    \Rightarrow & (Y\times  \mathcal{D}) \times \mathcal{D}\neq Y\times (\mathcal{D}\times \mathcal{D}).
\end{split}
\end{equation}
The correct procedure is to add another duality defect (simple object) $\mathcal{D}'=X\times \mathcal{D}=\mathcal{D}\times X$, with the same self-fusion as $\mathcal{D}$, and with the following fusion rules
\begin{equation}\label{eq: comple fl for subgauging reps3}
\boxed{
\begin{split}
&Y\times Y=1+ X + Y, ~X\times X=1,\\
     &\mathcal{D}\times \mathcal{D}=\mathcal{D}'\times \mathcal{D}'=1 + Y,\\
     &X\times \mathcal{D}=\mathcal{D}\times X=\mathcal{D}',\\
      &Y\times \mathcal{D}=\mathcal{D}\times Y=\mathcal{D} + \mathcal{D}'.
\end{split}
}
\end{equation}
It is straightforward to check that the above fusion rules are associative. The Rep$(S_3)$ is realized as a subsymmetry (the first line in the above fusion rules) of this fusion categorical symmetry. 

This tells us the self-duality of gauging $1+Y$ of Rep$(S_3)$ implies \emph{two} duality defects, with the same self-fusion rule for condensing $1+Y$, and transformed to each other via fusion with $X$. Furthermore, the quantum dimensions of these two defects are both $\langle \mathcal{D}\rangle=\langle \mathcal{D}'\rangle =\sqrt{3}$. This non-integer quantum dimension implies the whole fusion category containing $\mathcal{D}$ and $\mathcal{D}'$ is anomalous, i.e. obstructs the trivially-gapped phase.

\paragraph{Two duality defects from gauging Rep$(D_4)$ in $SU(2)_1/(\mathbb{Z}_2\times \mathbb{Z}_2)$.}
Given the existence of the duality defect $\mathcal{D}$ in $SU(2)_1/(\mathbb{Z}_2\times \mathbb{Z}_2)$ theory via gauging the $1+b+m$ algebra of Rep$(D_4)$
\begin{equation}
    \mathcal{D}\times \mathcal{D}=1+b+m,
\end{equation}
we again want to determine the fusion rule for $m \times \mathcal{D}$.  Similarly to Rep$(S_3)$, the naive form (\ref{eq: absorb fusion}) is not associative: 
\begin{equation}
    \begin{split}
        &\mathcal{D}\times m=m\times \mathcal{D}=2\mathcal{D},\\
        \Rightarrow & (b\times \mathcal{D})\times \mathcal{D}=2\mathcal{D}\times \mathcal{D}=2+2b+2m,\\
        &b\times (\mathcal{D}\times \mathcal{D})=b\times (1+b+m)=1+b+m,\\
        \Rightarrow &(b\times \mathcal{D})\times \mathcal{D}\neq b\times (\mathcal{D}\times \mathcal{D}).
    \end{split}
\end{equation}
The correct thing to do is to add another duality defect $\mathcal{D}'$, 
\begin{equation}
    \mathcal{D}'=a\times \mathcal{D}=\mathcal{D}\times a=c\times \mathcal{C}=\mathcal{D}\times c,
\end{equation}
with the same self-fusion as $\mathcal{D}$. The full fusion rules including $\mathcal{D}$ and $\mathcal{D}'$ are given by 
\begin{equation}\label{eq: fl for rep(D8)}
\boxed{
    \begin{split}
     &m\times m=1+a+b+c,\\
     &g\times m=m\times g=m, g\in \{a, b, c \},\\
     &a\times a=b\times b=c\times c=1,\\
     &a\times b=b\times a=c, a\times c=c\times a=b, b\times c=c\times b=a,\\
          &\mathcal{D}\times \mathcal{D}=\mathcal{D}'\times \mathcal{D}'=1 + b + m,\\
    &\mathcal{D}\times \mathcal{D}'=\mathcal{D}'\times \mathcal{D}=a + c + m,\\
    &a\times \mathcal{D}=\mathcal{D}\times a=c\times \mathcal{D}=\mathcal{D}\times c=\mathcal{D}',\\
    &m\times \mathcal{D}=\mathcal{D}\times m=\mathcal{D} +\mathcal{D}'.
    \end{split}
    }
\end{equation}

In particular, the self-duality of gauing $1+b+m$ of Rep$(D_4)$ implies \emph{two} duality defects, transforming into one other under fusion with the $\Z_2$ generators $a$ and $c$. Furthermore, the quantum dimensions of these two new defects are both $\langle \mathcal{D} \rangle=\langle \mathcal{D}' \rangle=2$. 

This integral quantum dimension leads one to wonder whether this bigger noninvertible symmetry, with Rep$(D_4)$ as its subsymmetry, is non-anomalous. This is indeed the case. In fact, one can further identify this extended noninvertible symmetry with fusion rules in (\ref{eq: fl for rep(D8)}) as Rep$(D_8)$, where $D_8$ is the order 16 dihedral group. 

\subsubsection{Self-dual gauging of Rep$(A_4)$ and the resulting Rep$(SL(2,\Z_3))$ symmetry of $SU(2)_1/A_4$.}
Now, let us return to the exceptional $c=1$ CFT $SU(2)_1/A_4$. As we already discussed, it enjoys a $\mathcal{C}=\text{Rep}(A_4)$ symmetry, and in particular, it is self-dual under gauging the algebra object $\mathcal{A}=(1+Z)_\text{d.t.}$. Therefore, we can perform a half-space gauging of $\mathcal{A}$ and obtain at least one new noninvertible duality line defect $\mathcal{D}$ with the self-fusion in the form of (\ref{eq: fusion rules of duality defect}):
\begin{equation}
    \mathcal{D}\times \mathcal{D}=1+Z.
\end{equation}
If one assumes $\mathcal{D}$ is the only duality defect under this gauging, and it absorbs the $Z$ defect, then the fusion rules cannot be associative:
\begin{equation}
\begin{split}
    &\mathcal{D}\times Z=Z\times \mathcal{D}=3\mathcal{D},\\
    \Rightarrow&(Z\times \mathcal{D}) \times \mathcal{D}=3\mathcal{D}\times \mathcal{D}=3+3Z,\\
    &Z\times (\mathcal{D} \times \mathcal{D})=Z\times (1+Z)=1+X+Y+3Z,\\
\Rightarrow &(Z\times \mathcal{D}) \times \mathcal{D}\neq Z\times (\mathcal{D} \times \mathcal{D}).
\end{split}
\end{equation}
The correct result needs \emph{two more} defects $\mathcal{D}'$ and $\mathcal{D}''$, defined as 
\begin{equation}
    \mathcal{D}'=X\times \mathcal{D}, ~\mathcal{D}''=Y\times  \mathcal{D}.
\end{equation} 
The full fusion rules read
\begin{equation}\label{eq: fusion of rep(sl2z3)}
\boxed{
\begin{split}
&X\times X=Y, ~Y\times Y=X, ~X \times Y=Y \times X=1,\\
&Z\times Z=1+X+Y+2Z, ~X \times Z=Z \times X=Y \times Z=Z \times Y=Z,\\
    &\mathcal{D}\times \mathcal{D}=\mathcal{D}'\times \mathcal{D}''=\mathcal{D}''\times \mathcal{D}'=1 + Z, \\
    &\mathcal{D}'\times \mathcal{D}'=\mathcal{D}\times \mathcal{D}''=\mathcal{D}''\times \mathcal{D}=Y+Z,\\
    &\mathcal{D}''\times \mathcal{D}''=\mathcal{D}\times \mathcal{D}'=\mathcal{D}'\times \mathcal{D}=X+Z,\\
    &X\times \mathcal{D}=\mathcal{D}\times X=\mathcal{D}', ~Y\times \mathcal{D}=\mathcal{D}\times Y=\mathcal{D}'',\\
    &\mathcal{D}\times Z =Z \times \mathcal{D}=\mathcal{D}+\mathcal{D}'+\mathcal{D}''.
\end{split}
}
\end{equation}
We write the fusion rules so that from the first line to the last line, one can see how the symmetry of $SU(2)_1/A_4$ is extended.
\begin{itemize}
    \item The first line is just a $\mathbb{Z}_3$ invertible symmetry. If one does not know this theory is a $A_4$ orbifold, but only recognizes it as a $\Z_3$ orbifold of the 4-state Potts model as shown in Figure \ref{fig: more c=1 gaugings} (a), this will naively be the global symmetry as a quantum symmetry.
    \item Including the second line, one obtains a Rep$(A_4)$ symmetry. One can realize this global symmetry by noticing the theory is a $A_4$ orbifold.
    \item Including the third line, one finds out the non-trivial self-duality of this theory under gauging an algebra object of Rep$(A_4)$. Thus, the global symmetry cannot just be Rep$(A_4)$, but extended by the defect lines $\mathcal{D}, \mathcal{D}'$ and $\mathcal{D}''$.\footnote{This larger noninvertible symmetry can be regarded as a $\mathbb{Z}_2$-extended fusion category of the Rep$(A_4)$. We refer the reader to \cite{Perez-Lona:2023djo,CLS23} for more details on extending categorical symmetries from self-dualities under gauging noninvertible symmetries.}
\end{itemize}

The quantum dimension $\langle \mathcal{D} \rangle$ of $\mathcal{D}$ can be easily computed 
\begin{equation}
    \langle \mathcal{D} \rangle ^2=\langle 1 \rangle +\langle Z \rangle =4 \Rightarrow \langle \mathcal{D} \rangle =2,
\end{equation}
which is an integer. This satisfies the necessary condition for a noninvertible symmetry to admit a trivially gapped phase, i.e. to be non-anomalous. In fact, one can check further that the extended noninvertible symmetry given by (\ref{eq: fusion of rep(sl2z3)}) is a group-theoretical fusion category Rep$(SL(2,\Z_3))$. Namely,
\begin{equation*}
 \boxed{
     \text{$SU(2)_1/A_4$ theory enjoys a Rep$(SL(2,\Z_3))$ symmetry.}
     }
 \end{equation*}

This class of symmetry line defects for Rep$(SL(2,\Z_3))$ was found in \cite{Dijkgraaf:1989hb} as local operators associated with Verlinde lines of the $SU(2)_1/A_4$ theory\footnote{We thank Y. Choi and B. C. Rayhaun for pointing out the reference and valuable discussions on this point.}. The identification of our topological defect lines and local untwisted sector operators in \cite{Dijkgraaf:1989hb} are presented as follows, 
\begin{eqnarray}
    X & = & 1_1, \\
    Y & = & 1_2, \\
    Z & = & j, \\
    \mathcal{D} & = & \phi_0, \\
\mathcal{D}' & = & \phi_1, \\
   \mathcal{D}'' & = & \phi_2, \\
    1 & = & 1.
\end{eqnarray}
There are other local operators labeled by $\sigma$, $\tau$, $\omega_i^{\pm}$, $\theta_i^{\pm}$ in \cite{Dijkgraaf:1989hb}, but they belong to twisted sectors\footnote{We thank H.Y.~Zhang for pointing this out.} and so do not give rise to topological lines\footnote{For simplicity, considering the Ising CFT, there are three untwisted sector primary operators $1,\sigma,$ and $\epsilon$. There are three Verlinde topological lines $1, \mathcal{N}$ and $\eta$ corresponding to these three local fields. The fusion rules of these lines, which is TY$(\Z_2)$ in this case, can be reproduced by the OPEs of these local operators. There are also twisted sector fields such as $\mu$, i.e., disorder operator, but they do not give rise to extra topological lines.}. Therefore, what we have obtained from gauging Rep$(A_4)$ noninvertible symmetries are the complete set of Verlinde lines for this theory. Therefore, Rep$(SL(2,\Z_3))$ is the full categorical symmetry for the $SU(2)_1/A_4$ theory generated by topological lines.  Further natural questions include how to gauge this Rep$(SL(2,\Z_3))$ symmetry and what can be said about the resulting $c=1$ CFTs, which we leave for future work.

We close this section by remarking that including duality defects from self-dual gauging a noninvertible symmetry enjoys a elegant mathematical interpretation as a $\Z_2$-grading fusion category $\mathcal{C}'$ of the original fusion category $\mathcal{C}$ \cite{eno}. In the case of self-dual gaugings for Rep$(D_4)$ and Rep$(A_4)$, this $\Z_2$-extension for the fusion category can be regarded as 
\begin{equation}
\begin{split}
    &\text{Rep}(\Z_2)\leftarrow \text{Rep}(D_8) \leftarrow \text{Rep}(D_4),\\
    &\text{Rep}(\Z_2)\leftarrow \text{Rep}(SL(2,\Z_3)) \leftarrow \text{Rep}(A_4),\\
\end{split}
\end{equation} 
aligned with the fact $D_4=D_8/\Z_2$ and $A_4=SL(2,\Z_3)/\Z_2$, respectively. It would be interesting to investigate this $\Z_2$-grading picture and duality defects systematically in $c=1$ CFTs, which we leave for future work.

\section{Multiplicity-free example revisited: Rep$(D_4)$ gaugings from $D_4$ gaugings}\label{sec:repd4}

In \cite[section 3.2]{Perez-Lona:2023djo}, partition functions for eight gaugings of theories with Rep$(D_4)$ symmetry were constructed based on cosets from the eight (up to conjugation) subgroups of $D_4$.  This does not capture all gaugings of Rep$(D_4)$, however, because for each subgroup $G$ of $D_4$ we also have the possibility of turning on discrete torsion, classified by $H^2(G,U(1))$.  In the case at hand, the eight subgroups of $D_4$ are the trivial group, three copies of $\Z_2$, two copies of $Z_2\times\Z_2$, one copy of $\Z_4$ and the entire group.  Of these, we have the non-trivial cohomologies $H^2(D_4,U(1)) = H^2(\Z_2\times\Z_2,U(1))=\Z_2$, which means that there are three additional gaugings of $D_4$ which arise from turning on discrete torsion in the two $\Z_2\times\Z_2$ and the entire $D_4$, bringing us to eleven total gaugings of $D_4$.  Since gaugings of $G$ and Rep$(G)$ are known to be in one-to-one correspondence \cite{ostrik1}, we should also expect eleven gaugings of Rep$(D_4)$.

In this section, we will produce the remaining gaugings, based on the procedure outlined in \cite{FCDiag}.  The strategy here will be to read off the partial traces of the eleven $D_4$-symmetric topological phases from the eleven $D_4$ gaugings, map these to partial traces of the eleven Rep$(D_4)$-symmetric topological phases, and use the Rep$(D_4)\boxtimes$Rep$(D_4)$ diagonal gauging to construct the eleven gaugings of Rep$(D_4)$.

To begin with, we will establish notation for $D_4$ and write down the partition functions for its various gaugings.  Our notation will follow \cite{Perez-Lona:2023djo}.  We present the group as 
\begin{equation}
    \left\langle x,y|x^4=y^2=(xy)^2=1\right\rangle
\end{equation}
with conjugacy classes $[1]=\{1\}$, $[x]=\{x,x^3\}$, $[x^2]=\{x^2\}$, $[y]=\{y,x^2y\}$, and $[xy]=\{xy,x^3y\}$.  There are 40 commuting pairs of group elements, which naively would lead to 40 partial traces in a $D_4$ gauging.  However, Morita equivalence (which in this group-like case is simply conjugation equivalence) reduces this number to 22.  Making an arbitrary choice of representative in each equivalence class, the partition function for gauging all of $D_4$ can be written as
\begin{multline}
\label{d4_gauging}
\frac{1}{8}\big[Z_{1,1}+(Z_{1,x^2}+Z_{x^2,1}+Z_{x^2,x^2})+2(Z_{1,x}+Z_{x,1}+Z_{x,x}+Z_{x,x^2}+Z_{x^2,x}+Z_{x,x^3})+2(Z_{1,y}+Z_{y,1}+Z_{y,y})\\
+2(Z_{1,xy}+Z_{xy,1}+Z_{xy,xy})\pm2(Z_{x^2,y}+Z_{y,x^2}+Z_{y,x^2y})\pm2(Z_{x^2,xy}+Z_{xy,x^2}+Z_{xy,x^3y})\big]
\end{multline}
where terms in the same modular orbit have been collected in parentheses, and $\pm$ provides the $\Z_2$ discrete torsion.  With this notation, the remaining nine gaugings of $D_4$ are the trivial gauging
\be
Z_{1,1},
\ee
the three $\Z_2$ gaugings
\be
\frac{1}{2}\big[Z_{1,1}+(Z_{1,x^2}+Z_{x^2,1}+Z_{x^2,x^2})\big],
\ee
\be
\frac{1}{2}\big[Z_{1,1}+(Z_{1,y}+Z_{y,1}+Z_{y,y})\big],
\ee
\be
\frac{1}{2}\big[Z_{1,1}+(Z_{1,xy}+Z_{xy,1}+Z_{xy,xy})\big],
\ee
the two $\Z_2\times\Z_2$ gaugings (each with $\Z_2$ discrete torsion)
\be
\frac{1}{4}\big[Z_{1,1}+(Z_{1,x^2}+Z_{x^2,1}+Z_{x^2,x^2})+2(Z_{1,y}+Z_{y,1}+Z_{y,y})\pm2(Z_{x^2,y}+Z_{y,x^2}+Z_{y,x^2y})\big],
\ee
\be
\frac{1}{4}\big[Z_{1,1}+(Z_{1,x^2}+Z_{x^2,1}+Z_{x^2,x^2})+2(Z_{1,xy}+Z_{xy,1}+Z_{xy,xy})\pm2(Z_{x^2,xy}+Z_{xy,x^2}+Z_{xy,x^3y})\big],
\ee
and the $\Z_4$ gauging
\be
\frac{1}{4}\big[Z_{1,1}+(Z_{1,x^2}+Z_{x^2,1}+Z_{x^2,x^2})+2(Z_{1,x}+Z_{x,1}+Z_{x,x}+Z_{x,x^2}+Z_{x^2,x}+Z_{x,x^3})\big].
\ee
Noting that these eleven gaugings are a result of taking the product of a $D_4$-symmetric theory with the eleven $D_4$-symmetric topological phases and gauging the diagonal $D_4$, we can deduce the partial traces of these phases under $D_4$ gauging.  Noting that the diagonal $D_4$ will have the form (\ref{d4_gauging}) where now each $Z_{g,h}$ is the product of partial traces from the two individual theories, we simply read off what the partial traces of the topological phases must be to produce each of the eleven gaugings.  The result of this exercise is given in Table~\ref{table:d4}.

\begin{table}
    \centering
    \begin{tabular}{c | c | c | c | c | c | c | c | c | c | c | c}
         & $(D_4)_+$ & $(D_4)_-$ & $(\Z_2^2)_+^{(x^2,y)}$ & $(\Z_2^2)_-^{(x^2,y)}$ & $(\Z_2^2)_+^{(x^2,xy)}$ & $(\Z_2^2)_-^{(x^2,xy)}$ & $\Z_4$ & $\Z_2^{x^2}$ & $\Z_2^{y}$ & $\Z_2^{xy}$ & $\id$\\\hline
         $Z_{1,1}$ & 1 & 1 & 2 & 2 & 2 & 2 & 2 & 4 & 4 & 4 & 8\\
         $Z_{1,x^2}$ & 1 & 1 & 2 & 2 & 2 & 2 & 2 & 4 & 0 & 0 & 0\\
         $Z_{x^2,1}$ & 1 & 1 & 2 & 2 & 2 & 2 & 2 & 4 & 0 & 0 & 0\\
         $Z_{x^2,x^2}$ & 1 & 1 & 2 & 2 & 2 & 2 & 2 & 4 & 0 & 0 & 0\\
         $Z_{1,x}$ & 1 & 1 & 0 & 0 & 0 & 0 & 2 & 0 & 0 & 0 & 0\\
         $Z_{x,1}$ & 1 & 1 & 0 & 0 & 0 & 0 & 2 & 0 & 0 & 0 & 0\\
         $Z_{x,x}$ & 1 & 1 & 0 & 0 & 0 & 0 & 2 & 0 & 0 & 0 & 0\\
         $Z_{x,x^2}$ & 1 & 1 & 0 & 0 & 0 & 0 & 2 & 0 & 0 & 0 & 0\\
         $Z_{x^2,x}$ & 1 & 1 & 0 & 0 & 0 & 0 & 2 & 0 & 0 & 0 & 0\\
         $Z_{x,x^3}$ & 1 & 1 & 0 & 0 & 0 & 0 & 2 & 0 & 0 & 0 & 0\\
         $Z_{1,y}$ & 1 & 1 & 2 & 2 & 0 & 0 & 0 & 0 & 2 & 0 & 0\\
         $Z_{y,1}$ & 1 & 1 & 2 & 2 & 0 & 0 & 0 & 0 & 2 & 0 & 0\\
         $Z_{y,y}$ & 1 & 1 & 2 & 2 & 0 & 0 & 0 & 0 & 2 & 0 & 0\\
         $Z_{1,xy}$ & 1 & 1 & 0 & 0 & 2 & 2 & 0 & 0 & 0 & 2 & 0\\
         $Z_{xy,1}$ & 1 & 1 & 0 & 0 & 2 & 2 & 0 & 0 & 0 & 2 & 0\\
         $Z_{xy,xy}$ & 1 & 1 & 0 & 0 & 2 & 2 & 0 & 0 & 0 & 2 & 0\\
         $Z_{x^2,y}$ & 1 & $-1$ & 2 & $-2$ & 0 & 0 & 0 & 0 & 0 & 0 & 0\\
         $Z_{y,x^2}$ & 1 & $-1$ & 2 & $-2$ & 0 & 0 & 0 & 0 & 0 & 0 & 0\\
         $Z_{y,x^2y}$ & 1 & $-1$ & 2 & $-2$ & 0 & 0 & 0 & 0 & 0 & 0 & 0\\
         $Z_{x^2,xy}$ & 1 & $-1$ & 0 & 0 & 2 & $-2$ & 0 & 0 & 0 & 0 & 0\\
         $Z_{xy,x^2}$ & 1 & $-1$ & 0 & 0 & 2 & $-2$ & 0 & 0 & 0 & 0 & 0\\
         $Z_{xy,x^3y}$ & 1 & $-1$ & 0 & 0 & 2 & $-2$ & 0 & 0 & 0 & 0 & 0\\
    \end{tabular}
    \caption{Partial traces obtained when gauging $D_4$-symmetric topological phases.}
    \label{table:d4}
\end{table}

The next step is to relate the $D_4$ partial traces to Rep$(D_4)$ ones.  Rep$(D_4)$ has five simple objects which we will write as $1$, $a$, $b$, $c$ and $m$, corresponding to the five irreps of $D_4$.  The characters for these irreps act as
\be
\label{d4_chars}
\begin{tabular}{c|c|c|c|c|c|}
& $[1]$ & $[x^2]$ & $[x]$ & $[y]$ & $[xy]$ \\
\hline
$\chi_1$ & $1$ & $1$ & $1$ & $1$ & $1$ \\
\hline
$\chi_a$ & $1$ & $1$ & $1$ & $-1$ & $-1$ \\
\hline
$\chi_b$ & $1$ & $1$ & $-1$ & $1$ & $-1$ \\
\hline
$\chi_c$ & $1$ & $1$ & $-1$ & $-1$ & $1$ \\
\hline
$\chi_m$ & $2$ & $-2$ & $0$ & $0$ & $0$ \\
\hline
\end{tabular},
\ee
from which one can deduce the fusion rules
\be
\begin{tabular}{c | c c c c c}
$\otimes$ & 1 & $a$ & $b$ & $c$ & $m$\\\hline
1 & 1 & $a$ & $b$ & $c$ & $m$\\
$a$ & $a$ & 1 & $c$ & $b$ & $m$\\
$b$ & $b$ & $c$ & 1 & $a$ & $m$\\
$c$ & $c$ & $b$ & $a$ & 1 & $m$\\
$m$ & $m$ & $m$ & $m$ & $m$ & $1+a+b+c$
\end{tabular}.
\ee
Recall \cite{Perez-Lona:2023djo} that the partition function for gauging the regular representation of Rep$(D_4)$ is naturally expressed in terms of 28 partial traces as
\begin{align}
\label{d4_orb_1abc2m}
    \frac{1}{8}\bigg[&Z_{1,1}^1+(Z_{1,a}^a+Z_{a,1}^a+Z_{a,a}^1)+(Z_{1,b}^b+Z_{b,1}^b+Z_{b,b}^1)+(Z_{1,c}^c+Z_{c,1}^c+Z_{c,c}^1) \non\\
    &+\left(\frac{\beta_1}{\beta_2\beta_9}Z_{a,b}^c+\frac{\beta_1}{\beta_3\beta_5}Z_{a,c}^b+\frac{\beta_6}{\beta_5\beta_{10}}Z_{b,a}^c+\frac{\beta_6}{\beta_2\beta_7}Z_{b,c}^a+\frac{\beta_{11}}{\beta_7\beta_9}Z_{c,a}^b+\frac{\beta_{11}}{\beta_3\beta_{10}}Z_{c,b}^a\right)\non\\
    &-2\left(Z_{1,m}^m+Z_{m,1}^m+Z_{m,m}^1-\frac{\beta_1}{\beta_4\beta_{13}}Z_{a,m}^m-\frac{\beta_{18}}{\beta_{13}\beta_{19}}Z_{m,a}^m+\frac{\beta_{18}}{\beta_4\beta_{19}}Z_{m,m}^a\right)\non\\
    &+2\left(Z_{1,m}^m+Z_{m,1}^m+Z_{m,m}^1+\frac{\beta_6}{\beta_8\beta_{14}}Z_{b,m}^m+\frac{\beta_{18}}{\beta_{14}\beta_{16}}Z_{m,b}^m+\frac{\beta_{18}}{\beta_8\beta_{16}}Z_{m,m}^b\right)\non\\
    &+2\left(Z_{1,m}^m+Z_{m,1}^m+Z_{m,m}^1-\frac{\beta_{11}}{\beta_{12}\beta_{15}}Z_{c,m}^m-\frac{\beta_{18}}{\beta_{15}\beta_{17}}Z_{m,c}^m-\frac{\beta_{18}}{\beta_{12}\beta_{17}}Z_{m,m}^c\right)\bigg]
\end{align}
where the complex parameters $\beta_i$ correspond to gauge freedom in the Rep$(D_4)$ associator.\footnote{These 28 partial traces are not independent, but satisfy non-trivial relations, at least some of which can be obtained by `nucleating' bubbles involving noninvertible lines (see \cite{FCDiag}, appendix C).  While we will not do so explicitly, taking all such relations into account should cause the number of degrees of freedom in the $D_4$ and Rep$(D_4)$ partial traces to match.}  In order to express the Rep$(D_4)$ partial traces in terms of the $D_4$ ones, we assume that we have obtained our Rep$(D_4)$-symmetric theory by (fully) gauging some theory with non-anomalous $D_4$ symmetry.  We know, then, that the partial trace $Z_{1,1}^1$, which is just the resulting partition function, must satisfy
\begin{multline}
\label{pt_rel1}
Z_{1,1}^1=\frac{1}{8}\big[Z_{1,1}+(Z_{1,x^2}+Z_{x^2,1}+Z_{x^2,x^2})+2(Z_{1,x}+Z_{x,1}+Z_{x,x}+Z_{x,x^2}+Z_{x^2,x}+Z_{x,x^3})\\
+2(Z_{1,y}+Z_{y,1}+Z_{y,y})+2(Z_{1,xy}+Z_{xy,1}+Z_{xy,xy})\\
+2(Z_{x^2,y}+Z_{y,x^2}+Z_{y,x^2y})+2(Z_{x^2,xy}+Z_{xy,x^2}+Z_{xy,x^3y})\big].
\end{multline}
We can obtain many other such relations by noting that the resulting Rep$(D_4)$ is the quantum dual to the original $D_4$, and as such we know its action on the Hilbert space of the $D_4$-symmetric theory -- it simply acts via characters.  Take, for instance, the Rep$(D_4)$ partial trace $Z_{1,a}^a$.  This corresponds to a Hilbert space of states which are obtained by wrapping the torus with an $a$ line, which acts on the original $D_4$ states by applying $\chi_a$ to their twisted sector label (the first subscript).  Using the character table (\ref{d4_chars}) we can conclude that $Z_{1,a}^a$ must be expressible as
\begin{multline}
\label{pt_rel2}
Z_{1,a}^a=\frac{1}{8}\big[Z_{1,1}+(Z_{1,x^2}+Z_{x^2,1}+Z_{x^2,x^2})+2(Z_{1,x}+Z_{x,1}+Z_{x,x}+Z_{x,x^2}+Z_{x^2,x}+Z_{x,x^3})\\
+2(Z_{1,y}-Z_{y,1}-Z_{y,y})+2(Z_{1,xy}-Z_{xy,1}-Z_{xy,xy})\\
+2(Z_{x^2,y}-Z_{y,x^2}-Z_{y,x^2y})+2(Z_{x^2,xy}-Z_{xy,x^2}-Z_{xy,x^3y})\big].
\end{multline}
By this method we can obtain any partial trace where one of the indices is the identity:
\begin{multline}
\label{pt_rel3}
Z_{a,1}^a=\frac{1}{8}\big[Z_{1,1}+(Z_{1,x^2}+Z_{x^2,1}+Z_{x^2,x^2})+2(Z_{1,x}+Z_{x,1}+Z_{x,x}+Z_{x,x^2}+Z_{x^2,x}+Z_{x,x^3})\\
+2(-Z_{1,y}+Z_{y,1}-Z_{y,y})+2(-Z_{1,xy}+Z_{xy,1}-Z_{xy,xy})\\
+2(-Z_{x^2,y}+Z_{y,x^2}-Z_{y,x^2y})+2(-Z_{x^2,xy}+Z_{xy,x^2}-Z_{xy,x^3y})\big],
\end{multline}
\begin{multline}
\label{pt_rel4}
Z_{a,a}^1=\frac{1}{8}\big[Z_{1,1}+(Z_{1,x^2}+Z_{x^2,1}+Z_{x^2,x^2})+2(Z_{1,x}+Z_{x,1}+Z_{x,x}+Z_{x,x^2}+Z_{x^2,x}+Z_{x,x^3})\\
+2(-Z_{1,y}-Z_{y,1}+Z_{y,y})+2(-Z_{1,xy}-Z_{xy,1}+Z_{xy,xy})\\
+2(-Z_{x^2,y}-Z_{y,x^2}+Z_{y,x^2y})+2(-Z_{x^2,xy}-Z_{xy,x^2}+Z_{xy,x^3y})\big],
\end{multline}
\begin{multline}
\label{pt_rel5}
Z_{1,b}^b=\frac{1}{8}\big[Z_{1,1}+(Z_{1,x^2}+Z_{x^2,1}+Z_{x^2,x^2})+2(Z_{1,x}-Z_{x,1}-Z_{x,x}-Z_{x,x^2}+Z_{x^2,x}-Z_{x,x^3})\\
+2(Z_{1,y}+Z_{y,1}+Z_{y,y})+2(Z_{1,xy}-Z_{xy,1}-Z_{xy,xy})\\
+2(Z_{x^2,y}+Z_{y,x^2}+Z_{y,x^2y})+2(Z_{x^2,xy}-Z_{xy,x^2}-Z_{xy,x^3y})\big],
\end{multline}
\begin{multline}
\label{pt_rel6}
Z_{b,1}^b=\frac{1}{8}\big[Z_{1,1}+(Z_{1,x^2}+Z_{x^2,1}+Z_{x^2,x^2})+2(-Z_{1,x}+Z_{x,1}-Z_{x,x}+Z_{x,x^2}-Z_{x^2,x}-Z_{x,x^3})\\
+2(Z_{1,y}+Z_{y,1}+Z_{y,y})+2(-Z_{1,xy}+Z_{xy,1}-Z_{xy,xy})\\
+2(Z_{x^2,y}+Z_{y,x^2}+Z_{y,x^2y})+2(-Z_{x^2,xy}+Z_{xy,x^2}-Z_{xy,x^3y})\big],
\end{multline}
\begin{multline}
\label{pt_rel7}
Z_{b,b}^1=\frac{1}{8}\big[Z_{1,1}+(Z_{1,x^2}+Z_{x^2,1}+Z_{x^2,x^2})+2(-Z_{1,x}-Z_{x,1}+Z_{x,x}-Z_{x,x^2}-Z_{x^2,x}+Z_{x,x^3})\\
+2(Z_{1,y}+Z_{y,1}+Z_{y,y})+2(-Z_{1,xy}-Z_{xy,1}+Z_{xy,xy})\\
+2(Z_{x^2,y}+Z_{y,x^2}+Z_{y,x^2y})+2(-Z_{x^2,xy}-Z_{xy,x^2}+Z_{xy,x^3y})\big],
\end{multline}
\begin{multline}
\label{pt_rel8}
Z_{1,c}^c=\frac{1}{8}\big[Z_{1,1}+(Z_{1,x^2}+Z_{x^2,1}+Z_{x^2,x^2})+2(Z_{1,x}-Z_{x,1}-Z_{x,x}-Z_{x,x^2}+Z_{x^2,x}-Z_{x,x^3})\\
+2(Z_{1,y}-Z_{y,1}-Z_{y,y})+2(Z_{1,xy}+Z_{xy,1}+Z_{xy,xy})\\
+2(Z_{x^2,y}-Z_{y,x^2}-Z_{y,x^2y})+2(Z_{x^2,xy}+Z_{xy,x^2}+Z_{xy,x^3y})\big],
\end{multline}
\begin{multline}
\label{pt_rel9}
Z_{c,1}^c=\frac{1}{8}\big[Z_{1,1}+(Z_{1,x^2}+Z_{x^2,1}+Z_{x^2,x^2})+2(-Z_{1,x}+Z_{x,1}-Z_{x,x}+Z_{x,x^2}-Z_{x^2,x}-Z_{x,x^3})\\
+2(-Z_{1,y}+Z_{y,1}-Z_{y,y})+2(Z_{1,xy}+Z_{xy,1}+Z_{xy,xy})\\
+2(-Z_{x^2,y}+Z_{y,x^2}-Z_{y,x^2y})+2(Z_{x^2,xy}+Z_{xy,x^2}+Z_{xy,x^3y})\big],
\end{multline}
\begin{multline}
\label{pt_rel10}
Z_{c,c}^1=\frac{1}{8}\big[Z_{1,1}+(Z_{1,x^2}+Z_{x^2,1}+Z_{x^2,x^2})+2(-Z_{1,x}-Z_{x,1}+Z_{x,x}-Z_{x,x^2}-Z_{x^2,x}+Z_{x,x^3})\\
+2(-Z_{1,y}-Z_{y,1}+Z_{y,y})+2(Z_{1,xy}+Z_{xy,1}+Z_{xy,xy})\\
+2(-Z_{x^2,y}-Z_{y,x^2}+Z_{y,x^2y})+2(Z_{x^2,xy}+Z_{xy,x^2}+Z_{xy,x^3y})\big],
\end{multline}
\be
\label{pt_rel11}
Z_{1,m}^m=\frac{1}{4}\big[Z_{1,1}+(Z_{1,x^2}-Z_{x^2,1}-Z_{x^2,x^2})+2(Z_{1,x}-Z_{x^2,x})+2(Z_{1,y})+2(Z_{1,xy})-2(Z_{x^2,y})-2(Z_{x^2,xy})\big],
\ee
\be
\label{pt_rel12}
Z_{m,1}^m=\frac{1}{4}\big[Z_{1,1}+(-Z_{1,x^2}+Z_{x^2,1}-Z_{x^2,x^2})+2(Z_{x,1}-Z_{x,x^2})+2(Z_{y,1})+2(Z_{xy,1})-2(Z_{y,x^2})-2(Z_{xy,x^2})\big],
\ee
\be
\label{pt_rel13}
Z_{m,m}^1=\frac{1}{4}\big[Z_{1,1}+(-Z_{1,x^2}-Z_{x^2,1}+Z_{x^2,x^2})+2(-Z_{x,x}+Z_{x,x^3})+2(Z_{y,y})+2(Z_{xy,xy})-2(Z_{y,x^2y})-2(Z_{xy,x^3y})\big].
\ee
Now we require similar relations for the remaining 15 partial traces with mixed (non-trivial) indices.  A sensible guess is that we can get e.g.~$Z_{a,m}^m$ by taking $Z_{1,m}^m$ and acting on the second index in each $D_4$ partial trace with $\chi_a$.  The factors of $\beta$ out front can be fixed by comparing to the full Rep$(D_4)$ partition function.  This gives
\be
\label{pt_rel14}
Z_{a,m}^m=\frac{\beta_4\beta_{13}}{4\beta_1}\big[Z_{1,1}+(Z_{1,x^2}-Z_{x^2,1}-Z_{x^2,x^2})+2(Z_{1,x}-Z_{x^2,x})-2(Z_{1,y})-2(Z_{1,xy})+2(Z_{x^2,y})+2(Z_{x^2,xy})\big],
\ee
\be
\label{pt_rel15}
Z_{m,a}^m=\frac{\beta_{13}\beta_{19}}{4\beta_{18}}\big[Z_{1,1}+(-Z_{1,x^2}+Z_{x^2,1}-Z_{x^2,x^2})+2(Z_{x,1}-Z_{x,x^2})-2(Z_{y,1})-2(Z_{xy,1})+2(Z_{y,x^2})+2(Z_{xy,x^2})\big],
\ee
\be
\label{pt_rel16}
Z_{m,m}^a=-\frac{\beta_4\beta_{19}}{4\beta_{18}}\big[Z_{1,1}+(-Z_{1,x^2}-Z_{x^2,1}+Z_{x^2,x^2})+2(-Z_{x,x}+Z_{x,x^3})-2(Z_{y,y})-2(Z_{xy,xy})+2(Z_{y,x^2y})+2(Z_{xy,x^3y})\big],
\ee
\be
\label{pt_rel17}
Z_{b,m}^m=\frac{\beta_8\beta_{14}}{4\beta_6}\big[Z_{1,1}+(Z_{1,x^2}-Z_{x^2,1}-Z_{x^2,x^2})+2(-Z_{1,x}+Z_{x^2,x})+2(Z_{1,y})-2(Z_{1,xy})-2(Z_{x^2,y})+2(Z_{x^2,xy})\big],
\ee
\be
\label{pt_rel18}
Z_{m,b}^m=\frac{\beta_{14}\beta_{16}}{4\beta_{18}}\big[Z_{1,1}+(-Z_{1,x^2}+Z_{x^2,1}-Z_{x^2,x^2})+2(-Z_{x,1}+Z_{x,x^2})+2(Z_{y,1})-2(Z_{xy,1})-2(Z_{y,x^2})+2(Z_{xy,x^2})\big],
\ee
\be
\label{pt_rel19}
Z_{m,m}^b=\frac{\beta_8\beta_{16}}{4\beta_{18}}\big[Z_{1,1}+(-Z_{1,x^2}-Z_{x^2,1}+Z_{x^2,x^2})+2(Z_{x,x}-Z_{x,x^3})+2(Z_{y,y})-2(Z_{xy,xy})-2(Z_{y,x^2y})+2(Z_{xy,x^3y})\big],
\ee
\be
\label{pt_rel20}
Z_{c,m}^m=-\frac{\beta_{12}\beta_{15}}{4\beta_{11}}\big[Z_{1,1}+(Z_{1,x^2}-Z_{x^2,1}-Z_{x^2,x^2})+2(-Z_{1,x}+Z_{x^2,x})-2(Z_{1,y})+2(Z_{1,xy})+2(Z_{x^2,y})-2(Z_{x^2,xy})\big],
\ee
\be
\label{pt_rel21}
Z_{m,c}^m=-\frac{\beta_{15}\beta_{17}}{4\beta_{18}}\big[Z_{1,1}+(-Z_{1,x^2}+Z_{x^2,1}-Z_{x^2,x^2})+2(-Z_{x,1}+Z_{x,x^2})-2(Z_{y,1})+2(Z_{xy,1})+2(Z_{y,x^2})-2(Z_{xy,x^2})\big],
\ee
\be
\label{pt_rel22}
Z_{m,m}^c=-\frac{\beta_{12}\beta_{17}}{4\beta_{18}}\big[Z_{1,1}+(-Z_{1,x^2}-Z_{x^2,1}+Z_{x^2,x^2})+2(Z_{x,x}-Z_{x,x^3})-2(Z_{y,y})+2(Z_{xy,xy})+2(Z_{y,x^2y})-2(Z_{xy,x^3y})\big]
\ee
for the partial traces involving $m$ and
\begin{multline}
\label{pt_rel23}
Z_{a,b}^c=\frac{\beta_2\beta_9}{8\beta_1}\big[Z_{1,1}+(Z_{1,x^2}+Z_{x^2,1}+Z_{x^2,x^2})+2(Z_{1,x}-Z_{x,1}-Z_{x,x}-Z_{x,x^2}+Z_{x^2,x}-Z_{x,x^3})\\
+2(-Z_{1,y}+Z_{y,1}-Z_{y,y})+2(-Z_{1,xy}-Z_{xy,1}+Z_{xy,xy})\\
+2(-Z_{x^2,y}+Z_{y,x^2}-Z_{y,x^2y})+2(-Z_{x^2,xy}-Z_{xy,x^2}+Z_{xy,x^3y})\big],
\end{multline}
\begin{multline}
\label{pt_rel24}
Z_{a,c}^b=\frac{\beta_3\beta_5}{8\beta_1}\big[Z_{1,1}+(Z_{1,x^2}+Z_{x^2,1}+Z_{x^2,x^2})+2(Z_{1,x}-Z_{x,1}-Z_{x,x}-Z_{x,x^2}+Z_{x^2,x}-Z_{x,x^3})\\
+2(-Z_{1,y}-Z_{y,1}+Z_{y,y})+2(-Z_{1,xy}+Z_{xy,1}-Z_{xy,xy})\\
+2(-Z_{x^2,y}-Z_{y,x^2}+Z_{y,x^2y})+2(-Z_{x^2,xy}+Z_{xy,x^2}-Z_{xy,x^3y})\big],
\end{multline}
\begin{multline}
\label{pt_rel25}
Z_{b,a}^c=\frac{\beta_{5}\beta_{10}}{8\beta_6}\big[Z_{1,1}+(Z_{1,x^2}+Z_{x^2,1}+Z_{x^2,x^2})+2(-Z_{1,x}+Z_{x,1}-Z_{x,x}+Z_{x,x^2}-Z_{x^2,x}-Z_{x,x^3})\\
+2(Z_{1,y}-Z_{y,1}-Z_{y,y})+2(-Z_{1,xy}-Z_{xy,1}+Z_{xy,xy})\\
+2(Z_{x^2,y}-Z_{y,x^2}-Z_{y,x^2y})+2(-Z_{x^2,xy}-Z_{xy,x^2}+Z_{xy,x^3y})\big],
\end{multline}
\begin{multline}
\label{pt_rel26}
Z_{b,c}^a=\frac{\beta_2\beta_7}{8\beta_6}\big[Z_{1,1}+(Z_{1,x^2}+Z_{x^2,1}+Z_{x^2,x^2})+2(-Z_{1,x}-Z_{x,1}+Z_{x,x}-Z_{x,x^2}-Z_{x^2,x}+Z_{x,x^3})\\
+2(Z_{1,y}-Z_{y,1}-Z_{y,y})+2(-Z_{1,xy}+Z_{xy,1}-Z_{xy,xy})\\
+2(Z_{x^2,y}-Z_{y,x^2}-Z_{y,x^2y})+2(-Z_{x^2,xy}+Z_{xy,x^2}-Z_{xy,x^3y})\big],
\end{multline}
\begin{multline}
\label{pt_rel27}
Z_{c,a}^b=\frac{\beta_7\beta_9}{8\beta_{11}}\big[Z_{1,1}+(Z_{1,x^2}+Z_{x^2,1}+Z_{x^2,x^2})+2(-Z_{1,x}+Z_{x,1}-Z_{x,x}+Z_{x,x^2}-Z_{x^2,x}-Z_{x,x^3})\\
+2(-Z_{1,y}-Z_{y,1}+Z_{y,y})+2(Z_{1,xy}-Z_{xy,1}-Z_{xy,xy})\\
+2(-Z_{x^2,y}-Z_{y,x^2}+Z_{y,x^2y})+2(Z_{x^2,xy}-Z_{xy,x^2}-Z_{xy,x^3y})\big],
\end{multline}
\begin{multline}
\label{pt_rel28}
Z_{c,b}^a=\frac{\beta_3\beta_{10}}{8\beta_{11}}\big[Z_{1,1}+(Z_{1,x^2}+Z_{x^2,1}+Z_{x^2,x^2})+2(-Z_{1,x}-Z_{x,1}+Z_{x,x}-Z_{x,x^2}-Z_{x^2,x}+Z_{x,x^3})\\
+2(-Z_{1,y}+Z_{y,1}-Z_{y,y})+2(Z_{1,xy}-Z_{xy,1}-Z_{xy,xy})\\
+2(-Z_{x^2,y}+Z_{y,x^2}-Z_{y,x^2y})+2(Z_{x^2,xy}-Z_{xy,x^2}-Z_{xy,x^3y})\big]
\end{multline}
for the remainder.  Using the 28 above relations, we can transform the $D_4$ partial traces of Table~\ref{table:d4} to their Rep$(D_4)$ equivalents, given in Table~\ref{table:repd4}.

\begin{table}
    \centering
    \begin{tabular}{c | c | c | c | c | c | c | c | c | c | c | c}
         & $(D_4)_+$ & $(D_4)_-$ & $(\Z_2^2)_+^{(x^2,y)}$ & $(\Z_2^2)_-^{(x^2,y)}$ & $(\Z_2^2)_+^{(x^2,xy)}$ & $(\Z_2^2)_-^{(x^2,xy)}$ & $\Z_4$ & $\Z_2^{x^2}$ & $\Z_2^{y}$ & $\Z_2^{xy}$ & $\id$\\\hline
         $Z_{1,1}^1$ & 5 & 2 & 4 & 1 & 4 & 1 & 4 & 2 & 2 & 2 & 1\\
         $Z_{1,a}^a$ & 1 & 2 & 0 & 1 & 0 & 1 & 4 & 2 & 0 & 0 & 1\\
         $Z_{a,1}^a$ & 1 & 2 & 0 & 1 & 0 & 1 & 4 & 2 & 0 & 0 & 1\\
         $Z_{a,a}^1$ & 1 & 2 & 0 & 1 & 0 & 1 & 4 & 2 & 0 & 0 & 1\\
         $Z_{1,b}^b$ & 1 & 0 & 4 & 1 & 0 & 1 & 0 & 2 & 2 & 0 & 1\\
         $Z_{b,1}^b$ & 1 & 0 & 4 & 1 & 0 & 1 & 0 & 2 & 2 & 0 & 1\\
         $Z_{b,b}^1$ & 1 & 0 & 4 & 1 & 0 & 1 & 0 & 2 & 2 & 0 & 1\\
         $Z_{1,c}^c$ & 1 & 0 & 0 & 1 & 4 & 1 & 0 & 2 & 0 & 2 & 1\\
         $Z_{c,1}^c$ & 1 & 0 & 0 & 1 & 4 & 1 & 0 & 2 & 0 & 2 & 1\\
         $Z_{c,c}^1$ & 1 & 0 & 0 & 1 & 4 & 1 & 0 & 2 & 0 & 2 & 1\\
         $Z_{1,m}^m$ & 0 & 2 & 0 & 2 & 0 & 2 & 0 & 0 & 2 & 2 & 2\\
         $Z_{m,1}^m$ & 0 & 2 & 0 & 2 & 0 & 2 & 0 & 0 & 2 & 2 & 2\\
         $Z_{m,m}^1$ & 0 & 2 & 0 & 2 & 0 & 2 & 0 & 0 & 2 & 2 & 2\\
         $\frac{\beta_1}{\beta_4\beta_{13}}Z_{a,m}^m$ & 0 & $-2$ & 0 & $-2$ & 0 & $-2$ & 0 & 0 & 0 & 0 & 2\\
         $\frac{\beta_{18}}{\beta_{13}\beta_{19}}Z_{m,a}^m$ & 0 & $-2$ & 0 & $-2$ & 0 & $-2$ & 0 & 0 & 0 & 0 & 2\\
         $\frac{\beta_{18}}{\beta_4\beta_{19}}Z_{m,m}^a$ & 0 & 2 & 0 & 2 & 0 & 2 & 0 & 0 & 0 & 0 & $-2$\\
         $\frac{\beta_6}{\beta_8\beta_{14}}Z_{b,m}^m$ & 0 & 0 & 0 & 2 & 0 & $-2$ & 0 & 0 & 2 & 0 & 2\\
         $\frac{\beta_{18}}{\beta_{14}\beta_{16}}Z_{m,b}^m$ & 0 & 0 & 0 & 2 & 0 & $-2$ & 0 & 0 & 2 & 0 & $2$\\
         $\frac{\beta_{18}}{\beta_8\beta_{16}}Z_{m,m}^b$ & 0 & 0 & 0 & 2 & 0 & $-2$ & 0 & 0 & 2 & 0 & $2$\\
         $\frac{\beta_{11}}{\beta_{12}\beta_{15}}Z_{c,m}^m$ & 0 & 0 & 0 & 2 & 0 & $-2$ & 0 & 0 & 0 & $-2$ & $-2$\\
         $\frac{\beta_{18}}{\beta_{15}\beta_{17}}Z_{m,c}^m$ & 0 & 0 & 0 & 2 & 0 & $-2$ & 0 & 0 & 0 & $-2$ & $-2$\\
         $\frac{\beta_{18}}{\beta_{12}\beta_{17}}Z_{m,m}^c$ & 0 & 0 & 0 & 2 & 0 & $-2$ & 0 & 0 & 0 & $-2$ & $-2$\\
         $\frac{\beta_1}{\beta_2\beta_9}Z_{a,b}^c$ & $-1$ & 0 & 0 & 1 & 0 & 1 & 0 & 2 & 0 & 0 & 1\\
         $\frac{\beta_1}{\beta_3\beta_5}Z_{a,c}^b$ & $-1$ & 0 & 0 & 1 & 0 & 1 & 0 & 2 & 0 & 0 & 1\\
         $\frac{\beta_6}{\beta_5\beta_{10}}Z_{b,a}^c$ & $-1$ & 0 & 0 & 1 & 0 & 1 & 0 & 2 & 0 & 0 & 1\\
         $\frac{\beta_6}{\beta_2\beta_7}Z_{b,c}^a$ & $-1$ & 0 & 0 & 1 & 0 & 1 & 0 & 2 & 0 & 0 & 1\\
         $\frac{\beta_{11}}{\beta_7\beta_9}Z_{c,a}^b$ & $-1$ & 0 & 0 & 1 & 0 & 1 & 0 & 2 & 0 & 0 & 1\\
         $\frac{\beta_{11}}{\beta_3\beta_{10}}Z_{c,b}^a$ & $-1$ & 0 & 0 & 1 & 0 & 1 & 0 & 2 & 0 & 0 & 1\\
    \end{tabular}
    \caption{Partial traces obtained when gauging Rep$(D_4)$-symmetric topological phases, obtained by duality from $D_4$.  Each phase is labeled by its corresponding $D_4$ gauging.}
    \label{table:repd4}
\end{table}

Knowing that the diagonal $\Rep(D_4)\boxtimes\Rep(D_4)$ gauging must reduce to the gauging (\ref{d4_orb_1abc2m}) of the regular representation of Rep$(D_4)$ when one of the theories is the Rep$(D_4)$ SPT phase which corresponds to the trivial $D_4$ gauging, we can infer the form of the diagonal gauging.  Specifically, assume $L$ and $R$ are Rep$(D_4)$-symmetric theories, with partition functions $Z_L$ and $Z_R$ and whose junction vector spaces have coefficients $\beta_i^L$ and $\beta_i^R$.  Then, the diagonal gauging of $L\otimes R$ has partition function
\begin{align}
\label{rd4rd4_diag}
    Z_{\text{diag.}}=\ & \frac{1}{8}\ls \mathcal{Z}_{1,1}^1+\mathcal{Z}_{1,a}^a+\mathcal{Z}_{1,b}^b+\mathcal{Z}_{1,c}^c+\mathcal{Z}_{1,m}^m+\mathcal{Z}_{a,1}^a+\mathcal{Z}_{a,a}^1+\frac{\mathcal{B}_1}{\mathcal{B}_2\mathcal{B}_9}\mathcal{Z}_{a,b}^c+\frac{\mathcal{B}_1}{\mathcal{B}_3\mathcal{B}_5}\mathcal{Z}_{a,c}^b\right.\non\\
    & \qquad\left. +\frac{\mathcal{B}_1}{\mathcal{B}_4\mathcal{B}_{13}}\mathcal{Z}_{a,m}^m+\mathcal{Z}_{b,1}^b+\frac{\mathcal{B}_6}{\mathcal{B}_5\mathcal{B}_{10}}\mathcal{Z}_{b,a}^c+\mathcal{Z}_{b,b}^1+\frac{\mathcal{B}_6}{\mathcal{B}_2\mathcal{B}_7}\mathcal{Z}_{b,c}^a+\frac{\mathcal{B}_6}{\mathcal{B}_8\mathcal{B}_{14}}\mathcal{Z}_{b,m}^m\right.\non\\
    & \qquad\left. +\mathcal{Z}_{c,1}^c+\frac{\mathcal{B}_{11}}{\mathcal{B}_7\mathcal{B}_9}\mathcal{Z}_{c,a}^b + \frac{\mathcal{B}_{11}}{\mathcal{B}_3\mathcal{B}_{10}}\mathcal{Z}_{c,b}^a + \mathcal{Z}_{c,c}^1+\frac{\mathcal{B}_{11}}{\mathcal{B}_{12}\mathcal{B}_{15}}\mathcal{Z}_{c,m}^m\right.\non\\
    & \qquad\left. +\mathcal{Z}_{m,1}^m+\frac{\mathcal{B}_{18}}{\mathcal{B}_{13}\mathcal{B}_{19}}\mathcal{Z}_{m,a}^m+\frac{\mathcal{B}_{18}}{\mathcal{B}_{14}\mathcal{B}_{16}}\mathcal{Z}_{m,b}^m+\frac{\mathcal{B}_{18}}{\mathcal{B}_{15}\mathcal{B}_{17}}\mathcal{Z}_{m,c}^m\right.\non\\
    & \qquad\left. +\mathcal{Z}_{m,m}^1+\frac{\mathcal{B}_{18}}{\mathcal{B}_4\mathcal{B}_{19}}\mathcal{Z}_{m,m}^a +\frac{\mathcal{B}_{18}}{\mathcal{B}_8\mathcal{B}_{16}}\mathcal{Z}_{m,m}^b+\frac{\mathcal{B}_{18}}{\mathcal{B}_{12}\mathcal{B}_{17}}\mathcal{Z}_{m,m}^c\rs
\end{align}
where $\mathcal{Z}_{s,p}^q=(Z_{s,p}^q)_L(Z_{s,p}^q)_R$ are products of the partial traces we would have found from gauging the individual Rep$(D_4)$ symmetries and $\mathcal{B}_i=\beta^L_i\beta^R_i$ are the products of the individual beta coefficients.  Now, plugging the values from Table~\ref{table:repd4} into (\ref{rd4rd4_diag}) produces the partition functions for the eleven physically distinct gaugings of Rep$(D_4)$.

We find that there are three gaugings for which the regular representation is the algebra object, the partition functions for which can be written as
\begin{align}
    \frac{1}{8}\bigg[&Z_{1,1}^1+(Z_{1,a}^a+Z_{a,1}^a+Z_{a,a}^1)+(Z_{1,b}^b+Z_{b,1}^b+Z_{b,b}^1)+(Z_{1,c}^c+Z_{c,1}^c+Z_{c,c}^1) \non\\
    &+\left(\frac{\beta_1}{\beta_2\beta_9}Z_{a,b}^c+\frac{\beta_1}{\beta_3\beta_5}Z_{a,c}^b+\frac{\beta_6}{\beta_5\beta_{10}}Z_{b,a}^c+\frac{\beta_6}{\beta_2\beta_7}Z_{b,c}^a+\frac{\beta_{11}}{\beta_7\beta_9}Z_{c,a}^b+\frac{\beta_{11}}{\beta_3\beta_{10}}Z_{c,b}^a\right)\non\\
    &+2\gamma_a\left(Z_{1,m}^m+Z_{m,1}^m+Z_{m,m}^1-\frac{\beta_1}{\beta_4\beta_{13}}Z_{a,m}^m-\frac{\beta_{18}}{\beta_{13}\beta_{19}}Z_{m,a}^m+\frac{\beta_{18}}{\beta_4\beta_{19}}Z_{m,m}^a\right)\non\\
    &+2\gamma_b\left(Z_{1,m}^m+Z_{m,1}^m+Z_{m,m}^1+\frac{\beta_6}{\beta_8\beta_{14}}Z_{b,m}^m+\frac{\beta_{18}}{\beta_{14}\beta_{16}}Z_{m,b}^m+\frac{\beta_{18}}{\beta_8\beta_{16}}Z_{m,m}^b\right)\non\\
    &+2\gamma_c\left(Z_{1,m}^m+Z_{m,1}^m+Z_{m,m}^1-\frac{\beta_{11}}{\beta_{12}\beta_{15}}Z_{c,m}^m-\frac{\beta_{18}}{\beta_{15}\beta_{17}}Z_{m,c}^m-\frac{\beta_{18}}{\beta_{12}\beta_{17}}Z_{m,m}^c\right)\bigg]
\end{align}
where $(\gamma_a,\gamma_b,\gamma_c)$ can take the values $(-1,1,1)$, $(1,-1,1)$ or $(1,1-1)$.  The remaining eight gaugings have unique algebra objects.  We have
\begin{itemize}
    \item $1+a+m$:
    \be
    \frac{1}{4}\left[Z_{1,1}^1+(Z_{1,a}^a+Z_{a,1}^a+Z_{a,a}^1)+\left(Z_{1,m}^m+Z_{m,1}^m+Z_{m,m}^1-\frac{\beta_1}{\beta_4\beta_{13}}Z_{a,m}^m-\frac{\beta_{18}}{\beta_{13}\beta_{19}}Z_{m,a}^m+\frac{\beta_{18}}{\beta_4\beta_{19}}Z_{m,m}^a\right)\right]
    \ee
    \item $1+b+m$:
    \be
    \frac{1}{4}\left[Z_{1,1}^1+(Z_{1,b}^b+Z_{b,1}^b+Z_{b,b}^1)+\left(Z_{1,m}^m+Z_{m,1}^m+Z_{m,m}^1+\frac{\beta_6}{\beta_8\beta_{14}}Z_{b,m}^m+\frac{\beta_{18}}{\beta_{14}\beta_{16}}Z_{m,b}^m+\frac{\beta_{18}}{\beta_8\beta_{16}}Z_{m,m}^b\right)\right]
    \ee
    \item $1+c+m$:
    \be
    \frac{1}{4}\left[Z_{1,1}^1+(Z_{1,c}^c+Z_{c,1}^c+Z_{c,c}^1)+\left(Z_{1,m}^m+Z_{m,1}^m+Z_{m,m}^1-\frac{\beta_{11}}{\beta_{12}\beta_{15}}Z_{c,m}^m-\frac{\beta_{18}}{\beta_{15}\beta_{17}}Z_{m,c}^m-\frac{\beta_{18}}{\beta_{12}\beta_{17}}Z_{m,m}^c\right)\right]
    \ee
    \item $1+a+b+c$:
    \begin{align}
    \frac{1}{4}\bigg[&Z_{1,1}^1+(Z_{1,a}^a+Z_{a,1}^a+Z_{a,a}^1)+(Z_{1,b}^b+Z_{b,1}^b+Z_{b,b}^1)+(Z_{1,c}^c+Z_{c,1}^c+Z_{c,c}^1) \non\\
    &+\left(\frac{\beta_1}{\beta_2\beta_9}Z_{a,b}^c+\frac{\beta_1}{\beta_3\beta_5}Z_{a,c}^b+\frac{\beta_6}{\beta_5\beta_{10}}Z_{b,a}^c+\frac{\beta_6}{\beta_2\beta_7}Z_{b,c}^a+\frac{\beta_{11}}{\beta_7\beta_9}Z_{c,a}^b+\frac{\beta_{11}}{\beta_3\beta_{10}}Z_{c,b}^a\right)\bigg]
    \end{align}
    \item $1+a$:
    \be
    \frac{1}{2}\left[Z_{1,1}^1+(Z_{1,a}^a+Z_{a,1}^a+Z_{a,a}^1)\right]
    \ee
    \item $1+b$:
    \be
    \frac{1}{2}\left[Z_{1,1}^1+(Z_{1,b}^b+Z_{b,1}^b+Z_{b,b}^1)\right]
    \ee
    \item $1+c$:
    \be
    \frac{1}{2}\left[Z_{1,1}^1+(Z_{1,c}^c+Z_{c,1}^c+Z_{c,c}^1)\right]
    \ee
    \item $1$:
    \begin{align}
    \label{rd4_triv_me}
    \frac{1}{8}\bigg[&5Z_{1,1}^1+(Z_{1,a}^a+Z_{a,1}^a+Z_{a,a}^1)+(Z_{1,b}^b+Z_{b,1}^b+Z_{b,b}^1)+(Z_{1,c}^c+Z_{c,1}^c+Z_{c,c}^1) \non\\
    &-\left(\frac{\beta_1}{\beta_2\beta_9}Z_{a,b}^c+\frac{\beta_1}{\beta_3\beta_5}Z_{a,c}^b+\frac{\beta_6}{\beta_5\beta_{10}}Z_{b,a}^c+\frac{\beta_6}{\beta_2\beta_7}Z_{b,c}^a+\frac{\beta_{11}}{\beta_7\beta_9}Z_{c,a}^b+\frac{\beta_{11}}{\beta_3\beta_{10}}Z_{c,b}^a\right)\bigg]
    \end{align}
\end{itemize}
These eleven gaugings are collected along with their corresponding gaugings of $D_4$ in Table~\ref{table:d4_duality}.  The results match those found in \cite[Table 2]{Diatlyk:2023fwf}.  We also list the Rep$(D_4)$-symmetric topological phase, using the naming convention of \cite[appendix C.3]{Bhardwaj:2024qrf}.

\begin{table}
    \centering
    \begin{tabular}{c | c | c}
    $D_4$ Gauging & Rep$(D_4)$-symmetric Phase & Rep$(D_4)$ Algebra Object \\\hline
    $\id$ & SPT & $1+a+b+c+2m$ \\
    $(\Z_2\times\Z_2)^{(x^2,y)}_-$ & SPT & $1+a+b+c+2m$ \\
    $(\Z_2\times\Z_2)^{(x^2,xy)}_-$ & SPT & $1+a+b+c+2m$ \\
    $(D_4)_-$ & $\Z_2$ SSB & $1+a+m$ \\
    $\Z_2^y$ & $\Z_2$ SSB & $1+b+m$ \\
    $\Z_2^{xy}$ & $\Z_2$ SSB & $1+c+m$ \\
    $\Z_2^{x^2}$ & Rep$(D_4)/(\Z_2\times\Z_2)$ SSB & $1+a+b+c$ \\
    $\Z_4$ & $\Z_2\times\Z_2$ SSB & $1+a$ \\
    $(\Z_2\times\Z_2)^{(x^2,y)}_+$ & $\Z_2\times\Z_2$ SSB & $1+b$ \\
    $(\Z_2\times\Z_2)^{(x^2,xy)}_+$ & $\Z_2\times\Z_2$ SSB & $1+c$ \\
    $(D_4)_+$ & Rep$(D_4)$ SSB & 1 \\
    \end{tabular}
    \caption{Relation between gaugings of $D_4$, topological phases carrying Rep$(D_4)$ symmetry and the algebra objects of the corresponding gaugings of Rep$(D_4)$.}
    \label{table:d4_duality}
\end{table}

The partition function (\ref{rd4_triv_me}) for the trivial Rep$(D_4)$ gauging deserves further comment.  One would likely expect the trivial gauging to have trivial partition function $Z_{1,1}^1$, rather than the expression above.  As explained in \cite{FCDiag}, one can reconcile these two expressions by noting that (\ref{rd4_triv_me}) is the partition function for gauging an algebra which is Morita equivalent to the trivial one, so physically (\ref{rd4_triv_me}) and $Z_{1,1}^1$ lie in the same equivalence class.  In fact, the partial trace constraints imposed by Morita equivalence explain why there is no choice of discrete torsion when gauging the $\Z_2\times\Z_2$ subcategory of Rep$(D_4)$.  Following the construction in \cite[appendix D]{FCDiag}, the Morita trivial algebra with object $m\otimes 1\otimes\overline{m}=1+a+b+c$ has a partition function that would correspond to one of the two naive choice of discrete torsion in $\Z_2\times\Z_2\subset\Rep(D_4)$.

The results of this section have been derived in a general gauge, parameterized by the $\beta$ coefficients.  To match with the choice more commonly made in the literature, we can set
\be
\beta_1=\beta_4^2,\quad \beta_3=\frac{\beta_4^2}{\beta_2},\quad \beta_5=-\beta_2,\quad \beta_6=\beta_4^2,\quad \beta_7=-\frac{\beta_4^2}{\beta_2},\quad \beta_8=\pm \beta_4,\quad \beta_9=-\frac{\beta_4^2}{\beta_2},\non
\ee
\be
\beta_{10}=\frac{\beta_4^2}{\beta_2},\quad \beta_{11}=-\frac{\beta_4^4}{\beta_2^2},\quad \beta_{12}=\mp\frac{\beta_4^2}{\beta_2},\quad \beta_{13}=-\beta_4,\quad \beta_{14}=\pm \beta_4,\quad \beta_{15}=\mp\frac{\beta_4^2}{\beta_2},\non
\ee
\be
\label{eq:StandardD4Coeffs}
\beta_{17}=\frac{\beta_2\beta_{16}}{\beta_4},\quad \beta_{18}=\pm \beta_4 \beta_{16},\quad \beta_{19}=\pm \beta_{16}.
\ee
Then all components of the associator are equal to $+1$ except for
\be
\tilde{K}^{m,i}_{j,m}(m,m) = \tilde{K}^{i,m}_{m,j}(m,m) = \chi(i,j),
\qquad\tilde{K}^{m,m}_{m,m}(i,j)=\hlf\chi(i,j),
\ee
where $i$ and $j$ run over $1,a,b,c$ and
\be
\chi(i,j)=\lp\begin{matrix} 1 & 1 & 1 & 1 \\ 1 & 1 & -1 & -1 \\ 1 & -1 & 1 & -1 \\ 1 & -1 & -1 & 1 \end{matrix}\rp,
\ee
is the non-trivial bi-character for $\Z_2^2$.  In such a gauge, the partition function for gauging the regular representation (with choice of discrete torsion) of Rep$(D_4)$ takes the form
\begin{align}
    \frac{1}{8}\bigg[&Z_{1,1}^1+(Z_{1,a}^a+Z_{a,1}^a+Z_{a,a}^1)+(Z_{1,b}^b+Z_{b,1}^b+Z_{b,b}^1)+(Z_{1,c}^c+Z_{c,1}^c+Z_{c,c}^1) \non\\
    &-\left(Z_{a,b}^c+Z_{a,c}^b+Z_{b,a}^c+Z_{b,c}^a+Z_{c,a}^b+Z_{c,b}^a\right)\non\\
    &+2\gamma_a\left(Z_{1,m}^m+Z_{m,1}^m+Z_{m,m}^1+Z_{a,m}^m+Z_{m,a}^m+Z_{m,m}^a\right)\non\\
    &+2\gamma_b\left(Z_{1,m}^m+Z_{m,1}^m+Z_{m,m}^1+Z_{b,m}^m+Z_{m,b}^m+Z_{m,m}^b\right)\non\\
    &+2\gamma_c\left(Z_{1,m}^m+Z_{m,1}^m+Z_{m,m}^1+Z_{c,m}^m+Z_{m,c}^m+Z_{m,m}^c\right)\bigg]
\end{align}
where again we have $(\gamma_a,\gamma_b,\gamma_c)$ equal to $(-1,1,1)$, $(1,-1,1)$ or $(1,1-1)$.

\section{Decomposition}   \label{sect:decomp}

\subsection{General framework}

Recall that decomposition \cite{Hellerman:2006zs,Sharpe:2022ene} is the statement that a local $d$-dimensional
quantum field theory with a global $(d-1)$-form symmetry is equivalent to a disjoint union of quantum field theories.
One realization is in two-dimensional gauge theories in which a subgroup of the (zero-form) gauge group acts trivially.  Such theories have a global one-form symmetry, and so decompose.

In our previous paper \cite{Perez-Lona:2023djo}, we considered decomposition in the special case that the entire gauged noninvertible symmetry acted trivially.  For ordinary finite groups, this correspoinds to the case that the entire finite group acts trivially -- known as
Dijkgraaf-Witten theory.  More formally, this corresponds to 
a fiber functor
\begin{equation*}
    F:\mathcal{C}\to \text{Vec}
\end{equation*}
or more suggestively a sequence 
\begin{equation*}
\mathcal{C}\xrightarrow{\text{id}}\mathcal{C}\xrightarrow{F}\text{Vec}
\end{equation*}
resembling the setup of a trivially acting group
\begin{equation*}
    G\xrightarrow{\text{id}} G\xrightarrow{} 1
\end{equation*}

In this section, we generalize to cases in which only part of the gauged symmetry acts trivially.  In other words, in terms of ordinary finite groups, in this section we consider the generalization of the case that only a normal subgroup of the gauge group acts trivially.

To that end, we need to define the noninvertible generalization of ``normal subgroup.''
This should be captured by a similar sequence of fusion categories
\begin{equation*}
    \mathcal{N}\hookrightarrow \mathcal{C}\xrightarrow{\pi} \mathcal{Q} 
\end{equation*}
where $\pi:\mathcal{C}\to\mathcal{Q}$ is a fiber functor to some other fusion category, such that $\pi(\mathcal{N})\subset \langle \mathbbm{1}_{\mathcal{Q}}\rangle\cong \text{Vec}$. This is to be thought of as the analogue of a trivially-acting normal subgroup
\begin{equation*}
    N\hookrightarrow G\to G/N
\end{equation*}
where one thinks of the fusion category $\mathcal{Q}$ as a "quotient" $\mathcal{C}/\mathcal{N}$ (much as we think of Vec, the trivial linear category, as $\mathcal{C}/\mathcal{C}$). 

In our work, we only consider gauging fusion categories of the form
Rep(${\cal H})$, for ${\cal H}$ some Hopf algebra.  One way to define analogues of normal subgroups
is via Hopf ideals ${\cal I}$, see for example \cite[prop. 4.2]{natale},
\cite{meir}.  One way to describe an analogue of a normal subgroup is to start with a Hopf ideal
${\cal I}$ defining a short exact sequence\footnote{
See e.g.~\cite{nlabhopf} for more information on short exact sequences of Hopf algebras.
}
\begin{equation*}
\mathcal{I}\hookrightarrow\mathcal{H}\to\mathcal{H}/\mathcal{I}
\end{equation*}
from which we get an appropriate sequence of representation categories
\begin{equation*}
    \text{Rep}(\mathcal{H}/\mathcal{I})\to \text{Rep}(\mathcal{H})\to \text{Rep}(\mathcal{I})
\end{equation*}
so that we can take only $\text{Rep}(\mathcal{H}/\mathcal{I})$ to act trivially.

Now, to be clear, this is not the most general case since $\text{Rep}(\mathcal{H})$ still admits a fiber functor (see for example \cite{zuoliu} for quotients in greater generality).  That being said, we do need the trivially-acting subcategory to admit a fiber functor, since by definition of exact sequence of functors one gets a fiber functor, see aroung \cite[Def. 4.2]{natale}. We will consider such a more general example in section~\ref{sect:decomp:1px}.

As one computes partition functions, an important point is that a monoidal functor Rep$({\cal H}) \rightarrow{\rm Rep}({\cal I})$ will relate the associators in Rep$({\cal H})$ and Rep$({\cal I})$.
Specifically, recall that a
monoidal functor \cite[Definition 2.4.1]{EGNO} from $C$ to $C'$ is a pair $(F, J)$ where $F : C \to C'$
is a functor, and 
\begin{equation}
    J = \{J_{X,Y}  : \: F(X) \otimes F(Y) \: \stackrel{\sim}{\longrightarrow} \: F(X\otimes Y ) \: \vert \: X, Y \in C\}
    \end{equation}
is a natural isomorphism (a set of intertwiners), such that $F(1)$ is isomorphic to $1'$, and the following diagram holds
\begin{equation}   \label{eq:relate-assoc}
\begin{tikzcd}
(F(X)\otimes F(Y))\otimes F(Z) \arrow[rrr, "{a_{F(X),F(Y),F(Z)}}"] \arrow[d, "{J_{X,Y}\otimes \text{id}_{F(Z)}}"'] &  &  & F(X)\otimes (F(Y)\otimes F(Z)) \arrow[d, "{\text{id}_{F(X)}\otimes J_{Y,Z}}"] \\
F(X\otimes Y)\otimes F(Z) \arrow[d, "{J_{X\otimes Y,Z}}"']                                                         &  &  & F(X)\otimes F(Y\otimes Z) \arrow[d, "{J_{X,Y\otimes Z}}"]                     \\
F((X\otimes Y)\otimes Z) \arrow[rrr, "{F(a_{X,Y,Z})}"']                                                            &  &  & F(X\otimes (Y\otimes Z))                                                     
\end{tikzcd}
\end{equation}

For example, consider gauging Rep$(G)$ for $G$ a finite group.  Let $N$ be a normal subgroup of $G$.  Then,
Rep$(G/N)$ is a subcategory of Rep$(G)$, which for example we could take to act trivially.
Then, Rep$(N)$ is the quotient.  Schematically:
\begin{equation}
        1 \: \longrightarrow \: {\mathbb C}[N] \: \longrightarrow \: {\mathbb C}[G] \: \longrightarrow \: {\mathbb C}[G/N] \: \longrightarrow \: 1,
\end{equation}
    hence, using Rep$(G) = {\rm Rep}({\mathbb C}[G])$,
    \begin{equation}
        {\rm Rep}({\mathbb C}[G/N]) \: \longrightarrow \: {\rm Rep}({\mathbb C}[G]) \: \longrightarrow \: {\rm Rep}({\mathbb C}[N])
    \end{equation}

Next, we consider computing a partition function when a subalgebra of the gauged Frobenius algebra
acts trivially.  Formally, given a partition function
\begin{equation}
    Z \: = \: \sum_{a,b,c} \mu_{a,b}^c \Delta^{b,a}_c Z_{a,b}^c,
\end{equation}
the monoidal functor $(F,J)$ maps it to\footnote{
Roughly, we can think of
\begin{equation}
    \Delta \circ \mu \in {\rm Hom}(L_1 \otimes L_2, L_2 \otimes L_1),
\end{equation}
so 
\begin{equation}
    F(\Delta \circ \mu) \in {\rm Hom}( F(L_1 \otimes L_2), F(L_2 \otimes L_1) ).
\end{equation}
However, we want something in Hom$(F(L_1) \otimes F(L_2), F(L_2) \otimes F(L_1))$, and the $J$'s perform that conversion.
}
\begin{equation}
    \sum_{a,b,c} \left( J^{-1}_{b,a} \Delta_c^{b,a} 
    \mu_{a,b}^c  J_{a,b} \right)
    Z_{F(a),F(b)}^{F(c)},
\end{equation}
which is of the form
\begin{equation}
    \sum_{a,b,c} \mu_{F(a),F(b)}^{F(c)} \Delta_{F(c)}^{F(b),F(a)} Z_{F(a),F(b)}^{F(c)}.
\end{equation}
The partial traces $Z_{F(a),F(b)}^{F(c)}$ are computed using standard methods;
applying $J$'s as above specifies how to map the coefficients.

We consider some examples in the next subsections.

\subsection{Trivial example: Trivially-acting subcategories of Vec$(G)$}

Let us apply this reasoning to ordinary orbifolds by a finite group
$G$.  As discussed in \cite{Perez-Lona:2023djo},
in the current language, this means we consider
gauging Vec$(G) = {\rm Rep}( {\mathbb C}[G]^*)$.  
For a normal subgroup of $G$, a subcategory is Vec$(N)$, with coset Vec$(G/N)$.
In a little more detail, we have the exact sequences
\begin{equation}
    1 \: \longrightarrow \: {\mathbb C}[N] \: \longrightarrow \:
    {\mathbb C}[G] \: \longrightarrow \: {\mathbb C}[G/N] \: \longrightarrow \: 1,
\end{equation}
\begin{equation}
    1 \: \longrightarrow \: {\mathbb C}[G/N]^* \: \longrightarrow \:
    {\mathbb C}[G]^* \: \longrightarrow \: {\mathbb C}[N]^* \: \longrightarrow \: 1,
\end{equation}
hence
\begin{equation}
    {\rm Rep}\left( {\mathbb C}[N]^* \right)
    \: \longrightarrow \: {\rm Rep}\left( {\mathbb C}[G]^* \right)
    \: \longrightarrow \: {\rm Rep}\left( {\mathbb C}[G/N]^* \right)
    .
\end{equation}

In this case, if $N$ acts trivially, then the monoidal functor is generated by the map that
sends $g \in G$ (identified with a representation of ${\mathbb C}[G]^*$) to $g N \in G/N$,
and we can\footnote{
One can make other choices, but in this case, that choice exists and is clearly preferred.
} take the natural transformations $J$ to all be trivial.
Since the $J$'s are trivial, the coefficients $\mu_{a,b}^c \Delta^{b,a}_c$ are unmodified (and in any event,
also trivial, as discussed in \cite{Perez-Lona:2023djo}), and so describing partition functions of orbifolds in which subgroups act trivially immediately reduces to standard old computations,
see for example \cite{Pantev:2005rh,Pantev:2005wj,Pantev:2005zs}.

\subsection{Example: Trivially-acting Rep$({\mathbb Z}_Q)$ of Rep$({\mathbb Z}_{PQ})$}

In this section, we will consider a family of examples involving a Rep$({\mathbb Z}_N)$ symmetry
with a trivially-acting Rep$({\mathbb Z}_Q)$ symmetry.  Now, Rep$({\mathbb Z}_N) \cong \text{Vec}({\mathbb Z}_N)$,
so in principle these examples are equivalent to ordinary orbifolds; however, we will use nontrivial
intertwiners in our description of Rep$({\mathbb Z}_N)$, so as to give a nontrivial demonstration
of the underlying technology.

Because these examples are, in principle, equivalent to ordinary orbifolds, we will be able to compare
the results of our computations to known results for decomposition in ordinary cyclic orbifolds, which will provide a solid consistency check on our methods.

\subsubsection{Setup}

We begin by considering Rep$({\mathbb Z}_N)$ with nontrivial intertwiners.
In this section we will establish basics, then compute partition functions for gauged
Rep$({\mathbb Z}_N)$ theories for various Frobenius algebras, then turn to the analysis of the
case that a Rep$({\mathbb Z}_Q)$ subcategory acts trivially.

Let $a$ denote the generator of ${\mathbb Z}_N$.
Write Rep(${\mathbb Z}_N) = \{ 1, X, \cdots, X^{N-1} \}$,
where $X$ generates the irreducible representations of ${\mathbb Z}_N$, and
where
\begin{equation}
    X(a^k) \: = \: \xi^k,
\end{equation}
for $\xi = \exp(2 \pi i / N)$. 

We will work with a nontrivial
associator in Rep$({\mathbb Z}_N)$, defined by
\begin{equation}
    e_{X^k} \otimes e_{X^m} \: = \: \beta_{k,m} e_{X^{k+m}},
\end{equation}
where
\begin{equation}
    \beta_{0,k} \: = \: \beta_{k,0} \: = \: 1,
\end{equation}
and the sum $k+m$ is mod $N$.

\subsubsection{Frobenius algebras}

Now, since gaugings are determined by Frobenius algebras,
let us compute possible Frobenius algebra structures that can be gauged in
Rep$({\mathbb Z}_N)$, in order to eventually see the effect of the trivially-acting subalgebra in various examples of gaugings.

As discussed in \cite{Perez-Lona:2023djo},
to gauge Rep$({\cal H})$, we put a Frobenius algebra structure on ${\cal H}^*$.
Here, we wish to gauge Rep$({\mathbb Z}_N) = {\rm Rep}( {\mathbb C}[{\mathbb Z}_N])$, so we put a 
Frobenius algebra stucture on ${\mathbb C}[{\mathbb Z}_N]^*$.

We begin with the Frobenius algebra for the regular representation of ${\mathbb Z}_N$, and then discuss other choices.

Following the conventions of  \cite{Perez-Lona:2023djo}, for $g, h \in {\mathbb Z}_N$,
we let $v_g$ denote elements of a basis for ${\mathbb C}[{\mathbb Z}_N]^*$,
defined by
\begin{equation}
    v_g(h) \: = \: \delta_{g,h}.
\end{equation}
The Frobenius algebra structure is defined by
\begin{equation}
    \mu_*( v_g \otimes v_h) \: = \: \delta_{g,h} v_g,
    \: \: \:
    \Delta_F( v_g) \: = \: v_g \otimes v_g.
\end{equation}
We relate this to the representation basis via characters, as
\begin{equation}
    e_{X^r} \: = \: \sum_{k=0}^{N-1} \chi_{X^r}(a^k) v_{a^k}, 
\end{equation}
where
\begin{equation}
    \chi_{X^r}(a^k) \: = \: \exp\left( 2 \pi i \frac{k r}{N} \right).
\end{equation}
Then,
\begin{eqnarray}
    \mu_*\left( e_{X^p} \otimes e_{X^q} \right)
    & = &
    \sum_{k,\ell=0}^{N-1} \chi_{X^p}(a^k) \chi_{X^q}(a^{\ell}) \mu_*\left( v_{a^k} \otimes v_{a^{\ell}} \right),
    \\
    & = & \sum_{k=0}^{N-1} \chi_{X^p}(a^k) \chi_{X^q}(a^k) v_{a^k},
    \\
    & = & \sum_{k=0}^{N-1} \chi_{X^{p+q}}(a^k) v_{a^k},
    \\
    & = & e_{X^{p+q}}.
\end{eqnarray}
In components, if we use a basis for Hom$(X^p \otimes X^q, X^r)$ given by the intertwiners, namely
\begin{equation}  \label{eq:mult-basis}
    e_{X^p} \otimes e_{X^q} \: = \: \beta_{p,q} \, e_{X^{p+q}},
\end{equation}
we have
\begin{equation}   \label{eq:zN:mult1}
    \mu_{X^p, X^q}^{X^r} \: = \: \mu_{p,q}^r \: = \: \frac{\delta_{r,p+q}}{ \beta_{p,q}}
\end{equation}

Similarly, the comultiplications are given by
\begin{eqnarray}
    \Delta_F\left( e_{X^p} \right) & = &
    \sum_{k=0}^{N-1} \chi_{X^p}( a^k) \Delta_F( v_{a^k} ),
    \\
    & = &
    \sum_{k=0}^{N-1} \chi_{X^p}(a^k) v_{a^k} \otimes v_{a^k},
    \\
    & = &
    \sum_{k=0}^{N-1} \exp\left( 2 \pi i \frac{kp}{N} \right) v_{a^k}\otimes v_{a^k},
    \\
    & = &
    \frac{1}{N} \sum_{k,\ell=0}^{N-1} \exp\left( 2 \pi i \frac{kp}{N} \right)
    \left[ \sum_{q=0}^{N-1} \exp\left( 2 \pi i \frac{q (\ell-k)}{N} \right) \right] v_{a^k} \otimes v_{a^{\ell}},
    \\
    & = &
    \frac{1}{N} \sum_{q=0}^{N-1} \sum_{k,\ell=0}^{N-1} \chi_{X^{p-q}}(a^k) \chi_{X^q}(a^{\ell}) v_{a^k} \otimes v_{a^{\ell}},
    \\
    & = &
    \frac{1}{N} \sum_{q=0}^{N-1} e_{X^{p-q}} \otimes e_{X^q},
\end{eqnarray}
where we have used the fact that
\begin{equation}
    \sum_{q=0}^{N-1} \exp\left( 2 \pi i \frac{q (\ell-k)}{N} \right) \: = \: 
    N \delta_{\ell,k}.
\end{equation}

To compute the corresponding components $\Delta_{L_1,L_2}^{L_3}$, we need to pick a basis for Hom$(X^p, X^r \otimes X^q)$.  We pick a basis\footnote{
In the computations in our previous paper \cite{Perez-Lona:2023djo}, we frequently constructed such a basis by rewriting the comultiplication as the product of a coevaluation and an ordinary multiplication. 
For example, 
the comultiplication component $\Delta_{L_1}^{L_2,L_3}$ is computed using
\begin{center}
        \begin{tikzpicture}
			\draw[->] (0,0) [partial ellipse=-90:-180:0.5cm and 0.5cm];
			\draw[->] (0,0) [partial ellipse=-90:0:0.5cm and 0.5cm];   
        	\draw (-0.5,0) -- (-0.5,1.5);
            \node at (-0.9,0) {$L_2$};
            \node at (0.1,0) {$\overline{L}_2$};
			\draw (1,0) [partial ellipse=180:0:0.5cm and 0.5cm];
			\draw[->] (1.5,-1.5) -- (1.5,0);
            \node at (1.9,0) {$L_1$};
			\draw[->] (1,0.5) -- (1,1);
            \node at (1.4,1) {$L_3$};
			\draw (1,1) -- (1,1.5);
			\filldraw[black] (1,0.5) circle (2pt);
			\filldraw[black] (0,-0.5) circle (2pt);
    \end{tikzpicture}
\end{center}
Here, we pick a different basis, for simplicity.
} by rewriting~(\ref{eq:mult-basis}) as
\begin{equation}
    e_{X^p} \: = \: \frac{1}{\beta_{p-q,q}} e_{X^{p-q}} \otimes e_{X^q}.
\end{equation}
In terms of that basis, the comultiplication components are then
\begin{equation}  \label{eq:zN:comult1}
    \Delta_{X^p}^{X^r,X^q} \: = \: \Delta_p^{r,q} \: = \:
  \delta_{r,p-q}\, \frac{ \beta_{p-q,q} }{N}.
\end{equation}

Next, let us discuss other choices of subalgebras.  Suppose $N$ is divisible by some integer $m < N$,
so that we have
\begin{equation}
    1 \: \longrightarrow \: {\mathbb Z}_m \: \longrightarrow \: {\mathbb Z}_N \: \longrightarrow \: {\mathbb Z}_{N/m} \: \longrightarrow \: 1,
\end{equation}
hence
\begin{equation}
    {\rm Rep}( {\mathbb Z}_{N/m}) \: \longrightarrow \:
    {\rm Rep}( {\mathbb Z}_N) \: \longrightarrow \:
    {\rm Rep}( {\mathbb Z}_m).
\end{equation}
Following \cite[section 2.5]{Perez-Lona:2023djo}, let $K = {\mathbb Z}_m$, and we use a basis
$v_{gK}$ for ${\mathbb C}[{\mathbb Z}_N/{\mathbb Z}_m]^*$,
The Frobenius algebra structure is defined by
\begin{equation}
    \mu_*\left( v_{g K} \otimes v_{h K} \right) \: = \: \delta_{gK, hK} v_{gK}, \: \: \:
    \Delta_F \left( v_{gK} \right) \: = \: v_{gK} \otimes v_{gK}.
\end{equation}
The action of $a \in {\mathbb Z}_N$ is defined by
\begin{equation}
    a: \: v_{gK} \: \mapsto \: v_{ag K},
\end{equation}
and it is straightforward to check that the $\{ v_{gK} \}$ form a basis for the regular representation of
${\mathbb Z}_{N/m} \cong {\mathbb Z}_N / {\mathbb Z}_m$.
We relate this to the representation basis via characters, as
\begin{equation}
    e_{X^{rm}} \: = \: \sum_{k=0}^{N/m - 1} \chi_{X^{rm}}(a^k) v_{a^k K},
\end{equation}
where for simplicity we have chosen $\chi$ to be a character of the ${\mathbb Z}_N$ representation:
\begin{equation}
    \chi_{X^{rm}}(a^k) \: = \: \exp\left( 2 \pi i \frac{k m r}{N} \right).
\end{equation}
Then, more or less repeating the computations for the previous case, it is straightforward to compute
\begin{equation}
    \mu_*\left( e_{X^{mp}} \otimes e_{X^{mq}} \right) \: = \: e_{X^{m(p+q)}},
\end{equation}
hence, using the same basis as before,
\begin{equation}  \label{eq:zNm:mult1}
\mu_{X^{mp} , X^{mq}}^{ X^{mr}} \: = \: \mu_{mp,mp}^{mr} \: = \: \frac{ \delta_{r, p+q} }{\beta_{mp,mq}}.
\end{equation}
Similarly, the comultiplications are computed to be
\begin{equation}
    \Delta_F\left( e_{X^{mp}} \right) \: = \: \frac{m}{N} \sum_{q=0}^{N/m-1} e_{X^{m(p-q)} } \otimes e_{X^{mq}},
\end{equation}
with components
\begin{equation}  \label{eq:zNm:comult1}
    \Delta_{X^{mp}}^{X^{mr}, X^{mq}} \: = \: \Delta_{mp}^{mr,mq} \: = \: \delta_{r,p-q} \frac{ \beta_{mp-mq, mq}}{ N/m}.
\end{equation}

\subsubsection{Partition functions}  \label{sect:zNzQ:partfn}

Next, we compute partition functions, for each of the two Frobenius algebras.
\begin{itemize}
    \item $A = 1 + X + X^2 + \cdots$, the regular representation of ${\mathbb Z}_N$.

Here, the nonzero multiplication~(\ref{eq:zN:mult1}) and comultiplication~(\ref{eq:zN:comult1}) components are
\begin{equation}
    \mu_{p,q}^{p+q} \: = \: \frac{1}{\beta_{p,q}},
    \: \: \:
    \Delta_p^{p-q,q} \: = \: \frac{ \beta_{p-q,q} }{N}.
\end{equation}
hence the $T^2$ partition function of an $A$-gauged theory has the form~\cite[equ'n (2.91)]{Perez-Lona:2023djo}
\begin{eqnarray}
    Z & = & \sum_{p,q = 0}^{N-1} \mu_{p,q}^{p+q} \Delta_{p+q}^{q,p} \, Z_{p,q}^{p+q},
    \\
    & = & \frac{1}{N} \sum_{p,q=0}^{N-1} \frac{ \beta_{q,p} }{ \beta_{p,q} } \, Z_{p,q}^{p+q}.
\end{eqnarray}
(We remind the reader that although this is in principle equivalent to a ${\mathbb Z}_N$ orbifold,
we are taking into account possible nontrivial intertwiners, defined by the $\beta_{p,q}$.)

\item $A_m = 1 + X^m + X^{2m} + \cdots$, the regular representation of ${\mathbb Z}_{N/m}$.

Here, the nonzero multiplication~(\ref{eq:zNm:mult1}) and comultiplication~(\ref{eq:zNm:comult1}) components are
\begin{equation}
    \mu_{mp,mq}^{mp+mq} \: = \: \frac{1}{\beta_{mp,mq}}, \: \: \:
    \Delta_{mp}^{mp-mq,mq} \: = \: \frac{ \beta_{mp-mq,mq} }{ N/m },
\end{equation}
hence the $T^2$ partition function of an $A_m$-gauged theory has the form~\cite[equ'n (2.91)]{Perez-Lona:2023djo}
\begin{eqnarray}
    Z & = & \sum_{p,q = 0}^{N/m-1} \mu_{mp,mq}^{mp+mq} \Delta_{mp+mq}^{mq,mp} \, Z_{mp,mq}^{mp+mq},
    \\
    & = & \frac{m}{N} \sum_{p,q = 0}^{N/m-1} \frac{ \beta_{mq,mp} }{ \beta_{mp,mq}} \, Z_{mp,mq}^{mp+mq}.
\end{eqnarray}

\end{itemize}

Next, we shall take into account the fact that Rep$({\mathbb Z}_Q)$ acts trivially,
first by constructing a monoidal functor Rep$({\mathbb Z}_N) \rightarrow {\rm Rep}( {\mathbb Z}_P)$,
then by simplifying the partition functions in the various Frobenius algebras, comparing to standard
results for decomposition in cyclic orbifolds.

\subsubsection{Monoidal functor Rep$({\mathbb Z}_N) \rightarrow {\rm Rep}({\mathbb Z}_P)$}

Now, suppose that 
Rep$({\mathbb Z}_N)$ has a trivially-acting
Rep$({\mathbb Z}_Q)$, where $N = PQ$.  Given
\begin{equation}
    1 \: \longrightarrow \: {\mathbb Z}_P = \langle b \rangle \: \longrightarrow \:
    {\mathbb Z}_N = \langle a \rangle \: \longrightarrow \: {\mathbb Z}_Q \: \longrightarrow \: 1,
\end{equation}
where $a^P = b$, we have
\begin{equation}
    {\rm Rep}({\mathbb Z}_Q) \: \longrightarrow \: {\rm Rep} ( {\mathbb Z}_N) \: \longrightarrow \: 
    {\rm Rep}({\mathbb Z}_P) .
\end{equation}
Recall Rep(${\mathbb Z}_N) = \{ 1, X, \cdots, X^{N-1} \}$,
where $X$ generates the irreducible representations of ${\mathbb Z}_N$, and
where
\begin{equation}
    X(a^k) \: = \: \xi^k,
\end{equation}
for $\xi = \exp(2 \pi i / N)$.  Let $Z$ generate the irreducible ${\mathbb Z}_P$ representations, so
that $Z^P = 1$.  

To describe the decomposition of a theory in which a Frobenius algebra associated to Rep$({\mathbb Z}_N)$ has been gauged, we must construct a monoidal functor $(F,J): {\rm Rep}({\mathbb Z}_N) \rightarrow {\rm Rep}({\mathbb Z}_P)$, which we do next.  This functor has the property that $F(X^Q) = 1$.  For simplicity, we assume that $P$ divides $Q$, then two simple possibilities for $F$ are
determined by
\begin{itemize}
    \item $F(X) = Z$, or
    \item $F(X) = 1$.
\end{itemize}
We assume the former option, so that $F$ is nontrivial.

Now, we assume that the associator in Rep$({\mathbb Z}_P)$ is trivial, but recall there is a nontrivial
associator in Rep$({\mathbb Z}_N)$, defined by
\begin{equation}
    e_{X^k} \otimes e_{X^m} \: = \: \beta_{k,m} e_{X^{k+m}},
\end{equation}
where
\begin{equation}
    \beta_{0,k} \: = \: \beta_{k,0} \: = \: 1.
\end{equation}
Then, for example,
\begin{eqnarray}
    e_{X^k} \otimes \left( e_{X^{\ell}} \otimes e_{X^m} \right) & = &
    \beta_{\ell,m} \, e_{X^k} \otimes e_{X^{\ell+m}},   \label{eq:assoc:z4z2:1}
    \\
    & = & \beta_{\ell,m}\, \beta_{k,\ell+m} \, e_{X^{k+\ell+m}},
    \\
    \left( e_{X^k} \otimes e_{X^{\ell}} \right) \otimes e_{X^m} & = & 
    \beta_{k,\ell}\, e_{X^{k+\ell}} \otimes e_{X^m}, 
    \\
    & = & \beta_{k,\ell}\, \beta_{k+\ell,m} \,e_{X^{k+\ell+m}}.  \label{eq:assoc:z4z2:2}
\end{eqnarray}
The natural transformations $J_{X,Y}: F(X) \otimes F(Y) \rightarrow F(X \otimes Y)$ are taken to be
\begin{equation}
\begin{array}{cl}
    J_{e_{X^k},e_{X^{\ell}}} = \beta_{k,\ell}: & F\left(e_{X^k}\right) \otimes F\left(e_{X^{\ell}}\right) 
    \rightarrow F\left( e_{X^k} \otimes e_{X^{\ell}} \right),
    \label{eq:zNzQ:jpq}
    \\
    J_{e_{X^k} \otimes e_{X^{\ell}}, e_{X^m}} = \beta_{k+\ell,m}: &
    F\left( e_{X^k} \otimes e_{X^{\ell}} \right) \otimes F\left( e_{X^m} \right) \rightarrow
    F\left( (e_{X^k} \otimes e_{X^{\ell}}) \otimes e_{X^m} \right),
    \\
    J_{e_{X^k}, e_{X^{\ell}} \otimes e_{X^m}} = \beta_{k, \ell+m}: &
    F\left( e_{X^k} \right) \otimes F\left( e_{X^{\ell}} \otimes e_{X^m} \right) \rightarrow
    F\left( e_{X^k} \otimes ( e_{X^{\ell}} \otimes e_{X^m} ) \right).
\end{array}
\end{equation}
Then, using the fact that
\begin{equation}
    F(\alpha): F\left( ( e_{X^k} \otimes e_{X^{\ell}} ) \otimes e_{X^m} \right) \: \longrightarrow \:
    F\left( e_{X^k} \otimes ( e_{X^{\ell}} \otimes e_{X^m} ) \right)
\end{equation}
is given by
\begin{equation}
    \frac{ \beta_{\ell,m} \,\beta_{k,\ell+m} }{ \beta_{k,\ell} \, \beta_{k+\ell,m}},
\end{equation}
from equations~(\ref{eq:assoc:z4z2:1}), (\ref{eq:assoc:z4z2:2}),
and also using the fact that the associator in Rep$({\mathbb Z}_P)$ is trivial,
it is straightforward to check that~(\ref{eq:relate-assoc}) commutes, and so $(F,J)$ define a monoidal functor.

\subsubsection{Decomposition}

To understand the decomposition of the gauging of a Frobenius algebra in Rep$({\mathbb Z}_N)$
with trivially-acting Rep$({\mathbb Z}_Q)$, we turn to the partition functions on $T^2$.
Recall that to interpret the results, we map a partition function
\begin{equation}
    Z \: = \: \sum_{a,b,c} \mu_{a,b}^c \Delta^{b,a}_c Z_{a,b}^c,
\end{equation}
to
\begin{equation}
    \sum_{a,b,c} \left( J^{-1}_{b,a} \Delta_c^{b,a} 
    \mu_{a,b}^c  J_{a,b} \right)
    Z_{F(a),F(b)}^{F(c)},
\end{equation}
using the monoidal functor $(F,J)$ constructed previously.

We will do this for each of the Frobenius algebras in turn.
\begin{itemize}
    \item $A = 1 + X + X^2 + \cdots$, the regular representation of ${\mathbb Z}_N$.

    Recall from section~\ref{sect:zNzQ:partfn} that here the $T^2$ partition function is
    \begin{equation}
        Z_{T^2}\left( [X/A] \right) \: = \: \frac{1}{N} \sum_{p,q=0}^{N-1} \frac{ \beta_{q,p} }{ \beta_{p,q} } \, Z_{X^p, X^q}^{X^{p+q}}.
    \end{equation}
    Further, from~(\ref{eq:zNzQ:jpq}), we have
    \begin{equation}
        J_{p,q}: F\left(e_{X^p}\right) \otimes F\left(e_{X^q} \right) \: \longrightarrow \:
        J\left( e_{X^p} \otimes e_{X^q} \right)
     \end{equation}
     is given by $J_{p,q} = \beta_{p,q}$.
     Since $F(X) = Z$, we have that, with the trivial Rep$({\mathbb Z}_Q)$ action, the partition
     function should be reinterpreted as
     \begin{eqnarray}
         Z_{T^2}\left( [X/A] \right) & = & \frac{1}{N}  \sum_{p,q=0}^{N-1} J_{q,p}^{-1} \frac{ \beta_{q,p} }{ \beta_{p,q} } J_{p,q} \, Z_{Z^p, Z^q}^{Z^{p+q} } ,
         \\
         & = & \frac{1}{N}  \sum_{p,q=0}^{N-1} Z_{Z^p, Z^q}^{Z^{p+q} } ,
         \\
         & = & \frac{Q^2}{N} \sum_{a,b = 0}^{P-1} Z_{Z^a, Z^b}^{Z^{a+b}}
         \\
         & = & Q \, Z_{T^2}\left( [X/{\mathbb Z}_P] \right),
     \end{eqnarray}
     where we have used the fact that $Z^P = 1$.
     This suggests that the original theory, the $A$-gauged theory with trivially-acting
     Rep$({\mathbb Z}_Q)$, is equivalent to $Q$ copies of a ${\mathbb Z}_P$ orbifold.

Identifying Rep$({\mathbb Z}_k)$ with ${\mathbb Z}_k$, this result is exactly as expected
from \cite{Hellerman:2006zs} for a cyclic orbifold.  In the language of ordinary orbifolds,
this is a ${\mathbb Z}_N$ orbifold with a trivially-acting subgroup ${\mathbb Z}_Q$,
where $N=PQ$.  From \cite{Hellerman:2006zs}, this should be equivalent to a disjoint union of $Q$ copies of a ${\mathbb Z}_{N/Q = P}$ orbifold, as we computed above.

    \item $A_m = 1 + X^m + X^{2m} + \cdots$, the regular representation of ${\mathbb Z}_{N/m}$.

    Recall from section~\ref{sect:zNzQ:partfn} that here the $T^2$ partition function is
    \begin{equation}
        Z_{T^2}\left( [X/A_m] \right) \: = \:  \frac{m}{N} \sum_{p,q = 0}^{N/m-1} \frac{ \beta_{mq,mp} }{ \beta_{mp,mq}} \, Z_{X^{mp},X^{mq}}^{X^{mp+mq}}.
    \end{equation}
    Proceeding as before, we have that, with the trivial Rep$({\mathbb Z}_Q)$ action, the partition function should be reinterpreted as
    \begin{eqnarray}
        Z_{T^2}\left( [X/A_m] \right) & = &  \frac{m}{N} \sum_{p,q = 0}^{N/m-1} J_{mq,mp}^{-1} \frac{ \beta_{mq,mp} }{ \beta_{mp,mq}} 
        J_{mp,mq} \, Z_{Z^{mp}, Z^{mq}}^{Z^{mp+mq}},
        \\
        & = & \frac{m}{N} \sum_{a,b = 0}^{N/m-1} Z_{Z^{ma}, Z^{mb}}^{Z^{ma+mb}},
    \end{eqnarray}

Now, let us interpret this result in special cases, to compare against expectations.  For simplicity, assume that $Q$ and $P$ have no common divisors greater than one.
\begin{itemize}
    \item If $m=Q$, so that gauging $A_m$ corresponds to gauging $Z_{N/m = P}$, then the partition function above becomes
    \begin{equation}
        Z_{T^2}\left( [X/A_m] \right) \: = \: \frac{1}{P} \sum_{a,b = 0}^{P-1} Z_{ma, mb} \: = \: Z\left( [X/{\mathbb Z}_P] \right),
    \end{equation}
    consistent with the statement that the QFT of $[X/A_m]$ matches that of a single
    ${\mathbb Z}_P$ orbifold, which is consistent with expectations from
    \cite{Hellerman:2006zs}, since no part of the ${\mathbb Z}_P$ need act trivially.
    \item If $m=P$, so that gauging $A_m$ corresponds to gauging ${\mathbb Z}_{N/m = Q}$, then the partition function above becomes
    \begin{equation}
        Z_{T^2}\left( [X/A_m] \right) \: = \: \frac{1}{Q} \sum_{a,b = 0}^{Q-1} Z_{ma,mb} \: = \: \frac{Q^2}{Q} Z_{1,1} \: = \: Q Z(X) \: = \: Z\left( \coprod_Q X \right),
    \end{equation}
    consistent with the statement that the QFT of $[X/A_m]$ matches that of a disjoint union of $Q$ copies of $X$, which is consistent with expectations from
    \cite{Hellerman:2006zs}, since all of the orbifold group acts trivially.
\end{itemize}

\end{itemize}

\subsection{Examples: trivially-acting subcategories of Rep$(S_3)$}

In this section, we will outline examples of decomposition under trivially-acting
noninvertible symmetry groups, for subcategories of Rep$(S_3)$.  In these examples, we will only compute part the pertinent monoidal
functor directly to compute the result.  

There is a procedure, given in \cite{Bhardwaj:2024qrf}, to calculate the possible ways in which noninvertible symmetries can act non-faithfully.  The method used there involves identifying gapless phases carrying the symmetry in question.  These exist in addition to the more thoroughly studied gapped topological phases.  In particular, symmetric phases with a single ground state exist for each consistent action of that symmetry, i.e.~there is one in which the symmetry acts trivially, one in which it acts completely faithfully, and all possibilities in between.

When applied to Rep$(S_3)$-symmetric theories, which is done in \cite[section VI]{Bhardwaj:2024qrf}, we learn that there are four (gapped plus gapless) Rep$(S_3)$-symmetric phases with a single ground state, thus four possible actions of the Rep$(S_3)$ symmetry.  These include the two cases in which Rep$(S_3)$ acts totally trivially and totally effectively.  In one of the remaining intermediate cases, the group-like $\Z_2$ subcategory generated by $X$ acts trivially.  The remaining effective symmetry is a $\Z_3$ for which $Y$ acts as the sum of the two non-trivial group elements.  The final case is one in which the $Y$ line is not independent of the other two simple lines, but instead acts as $1+X$.  This is a type of non-faithful action that is unique to the noninvertible case, since in the group-like case every simple object is weight one and the only option is for certain group elements to act as the identity.  Below we examine the two intermediate cases using the technology developed in this section.

\subsubsection{Trivially-acting subcategory Rep$({\mathbb Z}_2)$ of Rep$(S_3)$}    
\label{sect:triv-repz2-reps3}
 
 Consider gauging Rep$(S_3)$, and above, and take $N = {\mathbb Z}_3$, then as $S_3/{\mathbb Z}_3 = {\mathbb Z}_2$, we have
    \begin{equation}
        {\rm Rep}( {\mathbb Z}_2 ) \: \longrightarrow \: {\rm Rep}( S_3 ) \: \longrightarrow \: {\rm Rep}( {\mathbb Z}_3 ).
    \end{equation}

    Label the simple objects of Rep$({\mathbb Z}_2)$ by $\{ 1, X\}$, the simple objects of Rep$(S_3)$ by $\{1, X, Y \}$,
    and the simple objects of Rep$({\mathbb Z}_3)$ by $\{1, Z, Z^2 \}$.  The functor 
    \begin{equation}
        {\rm Rep}( {\mathbb Z}_2) \: \longrightarrow \: {\rm Rep}(S_3)
    \end{equation}
    sends $X$ to $X$.  We need to construct a symmetric monoidal functor $(F, J): {\rm Rep}(S_3) \rightarrow {\rm Rep}({\mathbb Z}_3)$.  If we make the ansatz $F(X) = 1$, $F(Y) = \alpha Z + \beta Z^2$,
    then from requiring compatibility with the fusion products
    \begin{equation}
        X \otimes X \: \cong \: 1, \: \: \:
        X \otimes Y \: \cong \: Y, \: \: \:
        Y \otimes Y \: \cong \: 1 + X + Y,
    \end{equation}
    we find the consistency condition
    \begin{equation}
        F(Y \otimes Y) \: = \: F(Y) \otimes F(Y) \: \cong \: F(1) + F(X) + F(Y) \: = \: 2 + F(Y), 
    \end{equation}
    which implies, after evaluating on the ansatz, $\alpha = \beta = 1$, so that $F(Y) = Z + Z^2$.

    Now, to finish constructing the monoidal functor, we also have to specify natural transformations $J$.  However, in this example, working out all of the $J$'s is an extremely laborious exercise.
    Instead, below we list some of the $J$'s which can be determined, which will be sufficient for our computations.
\begin{eqnarray}
    J_{1,1}:
&F(e)\otimes F(e)\mapsto\ F(e \otimes e),  \label{eq:z2s3:j11} \\
J_{1,X}: &
F(e)\otimes F(e_X)\mapsto\ F(e \otimes e_X),  \label{eq:z2s3:j1x}\\
J_{1,Y}:&
F(e)\otimes F(e_{Y1})\mapsto\  F(e \otimes e_{Y1}),  \label{eq:z2s3:jey1}\\
& F(e)\otimes F(e_{Y2})\mapsto\  F(e \otimes e_{Y2}),\\
J_{X,1}:&
F(e_X) \otimes F(e)\mapsto\ F(e_X \otimes e),  \label{eq:z2s3:jx1}\\
J_{X,X}:&
F(e_X)\otimes F(e_X) \mapsto\  \beta_1 \, F(e_X\otimes e_X),  \label{eq:z2s3:jxx} \\
J_{X,Y}:&
F(e_X)\otimes F(e_{Y1})\mapsto\  \beta_2 \, F(e_X\otimes e_{Y1}),\\ 
& F(e_X)\otimes F(e_{Y2})\mapsto\  - \beta_2 \, F(e_X\otimes e_{Y2}),\\
J_{Y,1}:& 
F(e_{Y1})\otimes F(e)\mapsto\ F(e_{Y1} \otimes e),\\
& F(e_{Y2})\otimes F(e)\mapsto\ F(e_{Y2} \otimes e),\\
J_{Y,X}:& F(e_{Y1})\otimes F(e_X)\mapsto\ \beta_3 \, F(e_{Y1}\otimes e_X),\\
&F(e_{Y2})\otimes F(e_X)\mapsto\  - \beta_3 \, F(e_{Y2}\otimes e_X),
\label{eq:z2s3:jyx}
\end{eqnarray}
Natural transformations such as $J_{X \otimes Y,X}$ (which should be distinguished
from $J_{Y,X}$ in general) are determined by the $J$'s above.
We can safely denote $F(\tilde{K}(\vec{\beta}))=\tilde{K} (\vec{\beta})$ since $F$ is faithful. Then, for example, the $(X,X,Y)$ diagram is
\begin{center}
    \begin{tikzcd}
(F(e_X)\otimes F(e_X))\otimes F(e_{Y1}) \arrow[rr, "="] \arrow[d, "{J_{X,X}\otimes\text{id}_{F(Y)}}"'] &  & F(e_X)\otimes (F(e_X)\otimes F(e_{Y1})) \arrow[d, "{\text{id}_{F(X)}\otimes J_{X,Y}}"] \\
\beta_1 F(e_X\otimes e_X)\otimes F(e_{Y1}) \arrow[d, "{J_{X\otimes Y,Y}}"']                            &  & \beta_2 F(e_X)\otimes F(e_X\otimes e_{Y1}) \arrow[d]                                   \\
\beta_1 F((e_X\otimes e_X)\otimes e_{Y1}) \arrow[rr, "F(\tilde{K})=\frac{-\beta_2^2}{\beta_1}{}"]      &  & -\beta_2^2 F(e_X\otimes (e_X\otimes e_{Y1}))                            
\end{tikzcd}
\end{center}

Next, let us compare $T^2$ partition functions.  We will do this for each of the Frobenius algebras one can gauge, as listed in \cite[section 3.1]{Perez-Lona:2023djo}.  In each case, in the partition function, we will replace $Z_{A,B}^C$ by $Z_{F(A),F(B)}^{F(C)}$.
Specifically, this means we set $X$ to $1$ (since $X$ acts trivially, so the corresponding line is equivalent to the identity defect), and set $Y$ to $F(Y) = Z + Z^2$.  Since each partial trace is a correlation function, the partial traces are linear in their indices, meaning\footnote{
As a consistency check, this is also consistent with behavior of 
hemisphere partition functions.
The hemisphere partition function $Z$ for D-branes $A$, $B$ obeys
\begin{equation}
    Z(A \oplus B) \: = \: Z(A) + Z(B).
\end{equation}
Furthermore, if $A$ and $B$ are bound via e.g.~the cone construction, the same result holds.
If $C$ is such a cone, a condensation of some tachyon $A \rightarrow B$, then 
\begin{equation}
    Z(C) \: = \: Z(A) + Z(B).
\end{equation}
Ultimately this is because the hemisphere partition functions only depend upon topological K theory
classes.  (We would like to thank J.~Knapp for an explanation of this point.)
For this reason, in the partial traces, we take $Z_{A+B,C+D}^{E+F}$ to expand linearly across the
terms, as assumed above.
}
\begin{equation}
    Z_{A+B,C}^{D} \: = \: Z_{A,C}^D + Z_{B,C}^D.
\end{equation}

In addition, in the present case, since Rep$({\mathbb Z}_3) \cong \text{Vec}({\mathbb Z}_3)$, we can interpret the image as an ordinary orbifold, and simplify using the fact that in an ordinary orbifold, $Z_{a,b}^c \neq 0$ only when
$c=ab$.   Putting this together, we find, for example,
\begin{eqnarray}
    Z_{1, Z+Z^2}^{Z + Z^2} & = &
    Z_{1,Z}^Z + Z_{1,Z^2}^Z + Z_{1,Z}^{Z^2} + Z_{1,Z^2}^{Z^2},
    \\
    & = & Z_{1,Z}^Z + Z_{1,Z^2}^{Z^2},
    \\
    Z_{Z+Z^2,Z+Z^2}^{Z+Z^2} & = &
    Z_{Z,Z}^Z + Z_{Z,Z^2}^Z + Z_{Z^2,Z}^Z + Z_{Z^2,Z^2}^Z 
    + Z_{Z,Z}^{Z^2} + Z_{Z,Z^2}^{Z^2} + Z_{Z^2,Z}^{Z^2} + 
    Z_{Z^2,Z^2}^{Z^2},
    \\
    & = & Z_{Z^2,Z^2}^Z + Z_{Z,Z}^{Z^2},
\end{eqnarray}
In the Rep$({\mathbb Z}_3) \cong \text{Vec}({\mathbb Z}_3)$ orbifold, we will often omit superscripts, as they are redundant. 

With these simplifications in mind, we can now simplify $T^2$ partition function expressions for each of the gauged Frobenius algebras considered in \cite[section 3.1]{Perez-Lona:2023djo}.
\begin{itemize}
\item $A = 1+X$.  First, we consider gauging the Frobenius algebra $A = 1+X$.
As discussed in \cite[section 3.1.5]{Perez-Lona:2023djo}, the partition function is
\begin{equation}
\label{rs3_1+x_pf}
Z_{1+X} = \frac{1}{2} \left[  Z_{1,1}^1+Z_{1,X}^X+Z_{X,1}^X+Z_{X,X}^1\right].
\end{equation}
If we take $X$ to act trivially, then each of the partial traces $Z_{a,b}^c$ reduces to $Z_{1,1}^1$.
Furthermore, all but one of the relevant $J$ natural transformations~(\ref{eq:z2s3:j11}), (\ref{eq:z2s3:j1x}), (\ref{eq:z2s3:jx1}), (\ref{eq:z2s3:jxx}) is trivial.  The only nontrivial one,
(\ref{eq:z2s3:jxx}), appears together with its inverse, and so cancels out.
As a result, when reinterpreted, the partition function reduces to reduces to
\begin{equation}
    Z_{1+X} \: = \: \frac{1}{2} \left[ 4 Z_{1,1}^1 \right] \: = \: 2 Z_{1,1}^1,
\end{equation}
consistent with a disjoint union of 2 copies of the original theory:
\begin{equation}
    \left[ {\cal T} / A \right] \: = \: \coprod_2 {\cal T}.
\end{equation}
Since the gauged Rep$({\mathbb Z}_2) \cong {\mathbb Z}_2$, and all of it acts trivially,
this decomposition is consistent with expectations.
This also matches results from our previous work \cite[section 5.4.1]{Perez-Lona:2023djo}, where we computed $[ {\cal T}/(1+X)]$ in the case that both $X$ and $Y$ act trivially, and there also, recovered a disjoint union of two copies of ${\cal T}$, matching the result above.

\item $A = 1+Y$.  As discussed in \cite[section 3.1.6]{Perez-Lona:2023djo}, the partition function is
\begin{equation}
\label{rs3_1+y_pf}
Z_{1+Y} = \frac{1}{3}\left[ Z_{1,1}^1+Z_{1,Y}^Y+Z_{Y,1}^Y+Z_{Y,Y}^1+\frac{\beta_4}{2\beta_5^2}Z_{Y,Y}^Y\right],
\end{equation}
Taking $X$ to act trivially, we can expand as above.  As we do not have the $J_{Y,Y}$ natural transformations,
we will leave the coefficients of the (images of the) $Z_{Y,Y}^1$ and $Z_{Y,Y}^Y$ terms undetermined,
denoting those coefficients merely with the symbol $\#$.  As the $J_{1,1}$, $J_{1,Y}$, and $J_{Y,1}$ natural
transformations are trivial, the partition function reduces to
\begin{eqnarray}
    Z_{1+Y} & = & \frac{1}{3} \left[ 
    Z_{1,1}^1 + \left( Z_{1,Z}^Z + Z_{1,Z^2}^{Z^2} \right) + \left( Z_{Z,1}^Z + Z_{Z^2,1}^{Z^2} \right) 
    + \# \left( Z_{Z,Z^2}^1 + Z_{Z^2,Z}^1 \right) + \# \left( Z_{Z,Z}^{Z^2} + Z_{Z^2,Z^2}^Z \right) \right].
    \nonumber
\end{eqnarray}
Modular invariance constrains the result and determines the coefficients $\#$ to be $1$.
This is one copy of the $T^2$ partition function of an ordinary ${\mathbb Z}_3 \cong {\rm Rep}({\mathbb Z}_3)$ orbifold, suggesting that for $A = 1+Y$,
\begin{equation} \label{eq:xtrivial:1+y}
    \left[ {\cal T} / A \right] \: = \: \left[ {\cal T} / {\rm Rep}({\mathbb Z}_3) \right].
\end{equation}
Since the Frobenius algebra $A = 1+Y$ does not involve the trivially-acting Rep$({\mathbb Z}_2)$, this is
consistent with expectations -- we are simply reinterpreting the Rep$(S_3)$ orbifold in this special case.

This is also consistent with results from our previous work
\cite[section 5.4.1]{Perez-Lona:2023djo}, where it was argued that if everything acts trivially, then 
\begin{equation}
    [ {\cal T} / (1+Y) ] \: = \: \coprod_3 {\cal T}.
\end{equation}
Here, if everything acts trivially, then Rep$({\mathbb Z}_3) \cong {\mathbb Z}_3$
acts trivially, and a trivially-acting ${\cal Z}_3$ orbifold should just return a disjoint union of three copies of the original theory.  Thus, our result above in equation~(\ref{eq:xtrivial:1+y}) correctly specializes to the results in \cite[section 5.4.1]{Perez-Lona:2023djo}.

\item $A = 1+X+2Y$:
Recall that the general expression for a gauged Rep$(S_3)$ correlation
function, for the full Frobenius algebra $A = 1 + X + 2Y$, is
\cite[equ'n (3.211)]{Perez-Lona:2023djo}
\begin{eqnarray}
\label{rs3_rr_pf}
    Z_{1+X+2Y} & = & \frac{1}{6} \biggl[  Z_{1,1}^1 \: + \: \left( Z_{1,X}^X + Z_{X,1}^X + Z^1_{X,X} \right) \: + \: 2 \left( Z_{1,Y}^Y + Z_{Y,1}^Y + Z_{Y,Y}^1 +  \frac{\beta_4}{2 \beta_5^2} Z_{Y,Y}^Y \right)
    \nonumber \\
    & & \hspace*{0.5in}
    \: - \:
    \frac{2 \beta_1}{\beta_2 \beta_3} \left(  Z^Y_{X,Y} +
    \frac{\beta_2 \beta_4}{\beta_1 \beta_6} Z^Y_{Y,X} - \frac{\beta_3 \beta_4}{\beta_1 \beta_6} Z_{Y,Y}^X - \frac{\beta_2 \beta_3 \beta_4}{2 \beta_1 \beta_5^2} Z_{Y,Y}^Y \right) \biggr]
\end{eqnarray}

For trivially-acting Rep$({\mathbb Z}_2) \subset {\rm Rep}(S_3)$, we can simplify this as before.
We use $\#$ to denote coefficients that depend upon $J_{Y,Y}$, and which we cannot therefore describe.
\begin{eqnarray} 
    Z_{1+X+2Y} & = & \frac{1}{6} \biggl[ 
    4 Z_{1,1} \: + \: 2 \left( Z_{1,Z} + Z_{1,Z^2} + Z_{Z,1} + Z_{Z^2,1} + \# Z_{Z,Z^2} +  \# Z_{Z^2,Z} \right)
    \nonumber \\
    & & \hspace*{0.25in} 
    \: + \: \# \left( Z_{Z^2,Z^2} + Z_{Z,Z} \right)
    \: - \: J_{Y,X}^{-1} \left( 2 \frac{\beta_1}{\beta_2 \beta_3} \right) J_{X,Y} \left( Z_{1,Z} + Z_{1,Z^2} \right)
    \nonumber \\
    & & \hspace*{0.25in} 
    \: - \: J_{X,Y}^{-1} \left( 2 \frac{ \beta_4}{\beta_3 \beta_6} \right) J_{Y,X} \left( Z_{Z,1} + Z_{Z^2,1} \right)
    \: + \: \# \left( Z_{Z,Z^2} + Z_{Z^2,Z} \right)
        \nonumber \\
    & & \hspace*{0.25in} 
    \: + \:  \# \left( Z_{Z^2,Z^2} + Z_{Z,Z} \right)
    \biggr]
    \label{eq:z2s3-partfn-1}
\end{eqnarray}

Next, let us compute the coefficients involving $J$'s.  Since $Y$ is two-dimensional,
we should check two possibilities for each.  For the functor $J_{Y,X}^{-1} \circ \alpha \circ J_{X,Y}$, we have
\begin{equation}
    F(e_X) \otimes F(e_{Yi}) \: \stackrel{J_{X,Y}}{\longrightarrow} \:
    F(e_X \otimes e_{Yi}) \: \stackrel{\alpha}{\longrightarrow} \:
    F(e_{Yi} \otimes e_X) \: \stackrel{ J_{Y,X}^{-1}}{\longrightarrow} \:
    F(e_{Yi})  \otimes F(e_X)
\end{equation}
for $i \in \{1,2\}$.  Depending upon the value of $i$, each $J$ will differ by a sign;
however, the signs cancel out so that
$J_{Y,X}^{-1} \circ \alpha \circ J_{X,Y}$ involves multiplication by $\beta_2/\beta_3$.
In particular, 
\begin{equation}
    J_{Y,X}^{-1} \left( 2 \frac{\beta_1}{\beta_2 \beta_3} \right) J_{X,Y} \: = \:
    2 \frac{ \beta_1 }{ \beta_3^2 }.
\end{equation}
The remaining $\beta$-dependence is a consequence of the fact that a basis was
chosen in \cite{Perez-Lona:2023djo} arising from composition with coevaluation.
In \cite[section 3.5.1]{Perez-Lona:2023djo} it was observed that the standard choices of
$\beta$ are $\beta_1 = \beta_3 = -1$, for which
\begin{equation}
    J_{Y,X}^{-1} \left( 2 \frac{\beta_1}{\beta_2 \beta_3} \right) J_{X,Y} \: = \:
    2 \frac{ \beta_1 }{ \beta_3^2 } \: = \: -2.
\end{equation}
Similarly,
\begin{equation}
    J_{X,Y}^{-1} \left( 2 \frac{ \beta_4}{\beta_3 \beta_6} \right) J_{Y,X} \: = \:
    2 \frac{ \beta_4}{\beta_2 \beta_6}.
\end{equation}
Proceeding as before, with the choices $\beta_2 = \beta_4 = +1$, $\beta_6 = -1$ yields
\begin{equation}
    J_{X,Y}^{-1} \left( 2 \frac{ \beta_4}{\beta_3 \beta_6} \right) J_{Y,X} \: = \:
    2 \frac{ \beta_4}{\beta_2 \beta_6} \: = \: -2.
\end{equation}
Assembling the pieces we then have
\begin{eqnarray} 
    Z_{1+X+2Y} & = & \frac{1}{6} \biggl[ 
    4 Z_{1,1} \: + \: 2 \left( Z_{1,Z} + Z_{1,Z^2} + Z_{Z,1} + Z_{Z^2,1} + \# Z_{Z,Z^2} +  \# Z_{Z^2,Z} \right)
    \nonumber \\
    & & \hspace*{0.25in} 
    \: + \: \# \left( Z_{Z^2,Z^2} + Z_{Z,Z} \right)
    \: - \: (-2) \left( Z_{1,Z} + Z_{1,Z^2} \right)
    \nonumber \\
    & & \hspace*{0.25in} 
    \: - \: (-2) \left( Z_{Z,1} + Z_{Z^2,1} \right)
    \: + \: \# \left( Z_{Z,Z^2} + Z_{Z^2,Z} \right)
        \nonumber \\
    & & \hspace*{0.25in} 
    \: + \:  \# \left( Z_{Z^2,Z^2} + Z_{Z,Z} \right)
    \biggr],
    \\
    & = & \frac{2}{3}\bigl[  Z_{1,1} + Z_{1,Z} + Z_{1,Z^2} + Z_{Z,1} + Z_{Z^2,1} 
    \nonumber \\
    & & \hspace*{0.25in} + \# Z_{Z,Z^2} + \# Z_{Z^2,Z} + \# Z_{Z^2,Z^2} + \# Z_{Z,Z} \bigr].
    \label{eq:z2s3-partfn-2}
\end{eqnarray}
(Note that for readability we have absorbed normalization factors into the ``$\#$,'' so that their values have changed
between the two equations above.)
As before, modular invariance uniquely determines the values of $\#$,
so we see that from the $T^2$ partition function $Z_{1+X+2Y}$, gauging $A = 1 + X + 2Y$ with trivially-acting $X$, is equivalent to a disjoint union of two copies of an ordinary ${\mathbb Z}_3 \cong {\rm Rep}( {\mathbb Z}_3 )$ orbifold:
\begin{equation} \label{eq:xtrivial:1+x+2y}
    \left[ {\cal T} / A \right] \: = \: \coprod_2 \left[ {\cal T} / 
    {\rm Rep}({\mathbb Z}_3) \right].
\end{equation}

This also is consistent with expectations.  After all, we have in effect gauged
both Rep$({\mathbb Z}_2) \cong {\mathbb Z}_2$, and also Rep$({\mathbb Z}_3) \cong {\mathbb Z}_3$.  The former acts trivially, and so should result into a decomposition into two universes.  The latter does not act trivially, so each universe should involve a Rep$({\mathbb Z}_3)$ orbifold.

As another consistency check, let us also compare to results from our previous
paper \cite[section 5.4.1]{Perez-Lona:2023djo}.
There, we argued that if everything acts trivially, then $[ {\cal T}/(1+X+2Y)]$ is equivalent to a disjoint union of six copies of ${\cal T}$.  Here, if the Rep$({\mathbb Z}_3) \cong {\mathbb Z}_3$ orbifold acts trivially, then it should be equivalent to a disjoint union of three copies of ${\cal T}$, so that a disjoint union of two copies of the Rep$({\mathbb Z}_3)$ orbifold is equivalent to a disjoint union of a total of six copies of ${\cal T}$.  Thus, our result in equation~(\ref{eq:xtrivial:1+x+2y}) correctly specializes to results in
\cite[section 5.4.1]{Perez-Lona:2023djo}.
\end{itemize}

\subsubsection{$Y = 1+X$ in Rep$(S_3)$}   \label{sect:decomp:1px}

So far we have studied examples where the action of some of the simples is identical to the action of (sums of) the identity operator. Here, motivated by 
\cite{Bhardwaj:2024qrf}, we analyze a slight generalization of this situation, where the action of a particular simple object can be expressed as the action of a sum of other simple objects.  In particular, in this example, the trivially-acting subsymmetry is not of the form Rep$({\cal H}/{\cal I})$ for any Hopf ideal
${\cal I}$.

Recall that an endomorphism of Hopf algebras $f:{\mathcal H}\to {\mathcal H}$ induces a tensor endofunctor of representation categories
\begin{equation*}
    F:{\rm Rep}({\mathcal H})\to {\rm Rep}({\mathcal H}),
\end{equation*}
which at the level of objects is simply given by precomposition with $f$. That is, given a representation $(V,\rho:\mathcal{H}\to \text{End}(V))$, the functor $F$ maps this to $(V,\rho\circ f)$.

Depending on the endomorphism $f$, it can be the case that the induced functor $F$ factors through a subcategory $\imath:\mathcal{C}\hookrightarrow {\rm Rep}({\mathcal H})$, meaning there exists a tensor functor
\begin{equation*}
    G: {\rm Rep}({\mathcal H}) \to {\mathcal C},
\end{equation*}
such that $F\cong \imath\circ G$ as tensor functors.

In particular, an irreducible representation $R$ of $\mathcal{H}$ that is not fixed by $f$ will map to a sum of other irreducible representations. This has the interpretation of $R$ acting non-faithfully.

Let us now specialize to $\mathcal{H}=\C[S_3]$. A simple example is given by the trivial endomorphism
\begin{eqnarray*}
    f:& S_3\to S_3\\
    & g\mapsto 1.
\end{eqnarray*}
Clearly, the induced tensor endofunctor acts at the level of objects as
\begin{equation*}
    F(R)\cong \text{dim}(R) 1
\end{equation*}
for $1$ the trivial representation of $S_3$. Hence we obtain a tensor functor
\begin{equation*}
    G: \text{Rep}(S_3)\to \text{Vec},
\end{equation*}
where $\imath:\text{Vec}\hookrightarrow \text{Rep}(S_3)$ is the subcategory of $\text{Rep}(S_3)$ generated by the trivial representation. The functor $G$ is simply the fiber functor that witnesses the whole category acting trivially.

Now we analyze the case where $f$ raises every element of $S_3$ to its third power:
\begin{eqnarray*}
    f:& S_3\to S_3,\\
    & a\mapsto a^3=a,\\
    & b\mapsto b^3=1.
\end{eqnarray*}
The one-dimensional irrep $X$ is characterized by
\begin{eqnarray*}
    \rho_X(a) =-1, & \rho_X(b) = 1,
\end{eqnarray*}
so that its image under the induced endofunctor $F$ is
\begin{eqnarray*}
    (\rho\circ f)(a) = \rho(a) = -1, &  (\rho\circ f)(b) = \rho (1) = 1,
\end{eqnarray*}
meaning $X$ is fixed by $F$.

On the other hand, the two-dimensional irrep $Y$ is described by
\begin{eqnarray*}
    \rho_Y(a)=\lp\begin{matrix} -1 & 0 \\ 0 & 1 \end{matrix}\rp,\qquad\rho_Y(b)=\lp\begin{matrix} -\hlf & -\frac{\sqrt{3}}{2} \\ \frac{\sqrt{3}}{2} & -\hlf \end{matrix}\rp.
\end{eqnarray*}
so that its image is
\begin{eqnarray*}
    (\rho_Y\circ f)(a)=\rho_Y(a)=\lp\begin{matrix} -1 & 0 \\ 0 & 1 \end{matrix}\rp,\qquad(\rho_Y\circ f)(b)=\rho_Y(1)= \lp\begin{matrix} 1 & 0 \\ 0 & 1 \end{matrix}\rp.
\end{eqnarray*}
This identifies the image of $Y$ as
\begin{equation}
    F(Y)\cong X+1.
\end{equation}
Noting that the image of all simples, and hence of the whole category, factors through the $\imath:\text{Vec}(\mathbb{Z}_2)\hookrightarrow \text{Rep}(S_3)$ subcategory spanned by the simples $(1,X)$, the restriction of $F$ to such a subcategory gives a tensor functor
\begin{eqnarray}
    G:& \text{Rep}(S_3)\to \text{Vec}(\Z_2),\\
    &(1,X,Y)\mapsto (1,X,1+X),
\end{eqnarray}
which we interpret as $Y$ having a non-faithful action.

The fact that $Y$ could in principle admit an action identical to that of $1+X$ is suggested by this replacement being consistent with the fusion rules
\begin{eqnarray*}
    Y\otimes Y\cong 1+X+Y\Rightarrow (1+X)\otimes (1+X)\cong 1+X+ (1+X).
\end{eqnarray*}
However, the perspective described above ensures that this can be done consistently at the level of fusion categories. In particular, the existence of a tensor functor $G$ allows us to map partition functions:

As before, to specify the modular functor, we need to specify more than just
$F$, we need to specify natural isomorphism $J$.  As we take the target to
have trivial associators, we can use the same natural isomorphism $J$ (for
cases listed) in equations~(\ref{eq:z2s3:j11}) through (\ref{eq:z2s3:jyx}).
This list is not complete, but will suffice for our computations here.

With these simplifications in mind, we can now simplify $T^2$ partition function expressions for each of the gauged Frobenius algebras considered in \cite[section 3.1]{Perez-Lona:2023djo}.
\begin{itemize}
\item $A = 1+X$.  First, we consider gauging the Frobenius algebra $A = 1+X$.
As discussed in \cite[section 3.1.5]{Perez-Lona:2023djo}, the partition function is~(\ref{rs3_1+x_pf}), which we repeat below:
\begin{equation}
Z_{1+X} = \frac{1}{2} \left[  Z_{1,1}^1+Z_{1,X}^X+Z_{X,1}^X+Z_{X,X}^1\right].
\end{equation}
Mapping $Y$ to $1+X$ has no effect on this partition function.
We see that in this case,
\begin{equation}  \label{eq:yeq1+x:1+x}
    [ {\cal T} / (1+X) ] \: = \: [ {\cal T}/ {\mathbb Z}_2].
\end{equation}

Now, let us compare to results from our previous work \cite[section 5.4.1]{Perez-Lona:2023djo}.  There, when both $X$ and $Y$ act trivially,
we found 
\begin{equation}
    [ {\cal T} / (1+X) ] \: = \: \coprod_2 {\cal T}.
\end{equation}
Comparing our result above, when the ${\mathbb Z}_2$ acts trivially,
the ${\mathbb Z}_2$ orbifold should be equivalent to a disjoint union of two copies of ${\cal T}$, and so we see that the result above, equation~(\ref{eq:yeq1+x:1+x}),
correctly specializes to the results in \cite[section 5.4.1]{Perez-Lona:2023djo}.

\item $A = 1+Y$.  As discussed in \cite[section 3.1.6]{Perez-Lona:2023djo}, the partition function is~(\ref{rs3_1+y_pf}), which we repeat below:
\begin{equation}
Z_{1+Y} = \frac{1}{3}\left[ Z_{1,1}^1+Z_{1,Y}^Y+Z_{Y,1}^Y+Z_{Y,Y}^1+\frac{\beta_4}{2\beta_5^2}Z_{Y,Y}^Y\right],
\end{equation}
As the $J_{1,1}$, $J_{1,Y}$, and $J_{Y,1}$ natural transformations are trivial,
the partition function reduces to
\begin{equation}
    Z_{1+Y} \: = \: \frac{1}{3} \left[ Z_{1,1}^1 + Z_{1,1+X}^{1+X} + Z_{1+X,1}^{1+X} + \alpha_1 Z_{1+X,1+X}^1 + \alpha_2 Z_{1+X,1+X}^{1+X} \right],
\end{equation}
where $\alpha_{1,2}$ are constants whose determination requires knowledge of the
remaining $J$'s.  Simplifying, this becomes
\begin{eqnarray}
    Z_{1+Y} & = & \frac{1}{3} \Bigl[ Z_{1,1}^1 + \left( Z_{1,1}^1 + Z_{1,X}^X \right) + \left( Z_{1,1}^1 + Z_{X,1}^X \right) + \alpha_1 \left( Z_{1,1}^1 + Z_{X,X}^1 \right)
    \nonumber \\
    & & \hspace*{0.5in}
    + \alpha_2 \left( Z_{1,1}^1 + Z_{1,X}^X + Z_{X,1}^X + Z_{X,X}^1 \right) \Bigr].
\end{eqnarray}

Since the image is an ordinary ${\mathbb Z}_2$ orbifold, modular invariance relates $Z_{1,X}^X$, $Z_{X,1}^X$, and $Z_{X,X}^1$, which must appear with the same coefficient.  This implies that $\alpha_1 = 1$, so we can write the partition function as
\begin{eqnarray}
    Z_{1+Y} & = & \frac{1}{3} \Bigl[ (3) Z_{1,1}^1 + \left( Z_{1,1}^1 + Z_{1,X}^X + Z_{X,1}^X + Z_{X,X}^1 \right)
    \nonumber \\
    & & \hspace*{0.5in}
   + \alpha_2 \left( Z_{1,1}^1 + Z_{1,X}^X + Z_{X,1}^X + Z_{X,X}^1 \right) \Bigr],
   \\
   & = &
   Z_{1,1}^1 + (2) \frac{\alpha_2 + 1}{3} Z\left( [ {\cal T}/{\mathbb Z}_2] \right).
\end{eqnarray}
Now, the coefficients of each term need to be integers, so we require that
\begin{equation}
    \alpha_2 \: = \: \frac{3k + 1}{2}
\end{equation}
for some integer $k$.
This implies that
\begin{equation}
    Z_{1+Y} \: = \: Z\left( {\cal T} \coprod_{k+1}  [{\cal T}/{\mathbb Z}_2] \right),
\end{equation}
consistent with the statement that
\begin{equation}  \label{eq:yeq1+x:step1}
[ {\cal T}/ A] \: = \: {\cal T} \coprod_{k+1} [{\cal T}/{\mathbb Z}_2],
\end{equation}
for some integer $k$.

Now, let us compare to results from our previous work 
\cite[section 5.4.1]{Perez-Lona:2023djo}.  There, when both $X$ and $Y$ act trivially,
we found 
\begin{equation}
    [{\cal T}/(1+Y)] = \coprod_3 {\cal T}.
\end{equation}
Here, if the ${\mathbb Z}_2$ acts trivially, then $[ {\cal T} / {\mathbb Z}_2 ]$ is equivalent to a disjoint union of two copies of ${\cal T}$, so equation~(\ref{eq:yeq1+x:step1}) correctly specializes to
our previous results \cite[section 5.4.1]{Perez-Lona:2023djo} if $k=0$.

Thus, after comparing to previous results, we conclude that here,
\begin{equation}
    [ {\cal T} / (1+Y) ] \: = \: {\cal T} \coprod [ {\cal T} / {\mathbb Z}_2].
\end{equation}

\item $A = 1+X+2Y$:
Recall that the general expression for a gauged Rep$(S_3)$ correlation
function, for the full Frobenius algebra $A = 1 + X + 2Y$, is~(\ref{rs3_rr_pf}),
\cite[equ'n (3.211)]{Perez-Lona:2023djo}, which we repeat below:
\begin{eqnarray}
    Z_{1+X+2Y} & = & \frac{1}{6} \biggl[  Z_{1,1}^1 \: + \: \left( Z_{1,X}^X + Z_{X,1}^X + Z^1_{X,X} \right) \: + \: 2 \left( Z_{1,Y}^Y + Z_{Y,1}^Y + Z_{Y,Y}^1 +  \frac{\beta_4}{2 \beta_5^2} Z_{Y,Y}^Y \right)
    \nonumber \\
    & & \hspace*{0.5in}
    \: - \:
    \frac{2 \beta_1}{\beta_2 \beta_3} \left(  Z^Y_{X,Y} +
    \frac{\beta_2 \beta_4}{\beta_1 \beta_6} Z^Y_{Y,X} - \frac{\beta_3 \beta_4}{\beta_1 \beta_6} Z_{Y,Y}^X - \frac{\beta_2 \beta_3 \beta_4}{2 \beta_1 \beta_5^2} Z_{Y,Y}^Y \right) \biggr]
\end{eqnarray}
As the $J_{1,1}$, $J_{1,X}$, $J_{1,Y}$, $J_{X,1}$, and $J_{Y,1}$ natural transformations are trivial, and the $J_{X,X}$ natural transformation is just a multiplication by a scalar, this expression reduces to
\begin{eqnarray}
    Z_{1+X+2Y} & = & \frac{1}{6} \biggl[ Z_{1,1}^1 + \left( Z_{1,X}^X + Z_{X,1}^X + Z_{X,X}^1 \right) + 2 Z_{1,1+X}^{1+X} + 2 Z_{1+X,1}^{1+X} 
    \nonumber \\
    & & \hspace*{0.25in}
    + \alpha_1 Z_{1+X,1+X}^1 + \alpha_2 Z_{1+X,1+X}^{1+X} + \alpha_3 Z_{X,1+X}^{1+X} + \alpha_4 Z_{1+X,X}^{1+X}
    \nonumber \\
    & & \hspace*{0.25in}
    + \alpha_5 Z_{1+X,1+X}^X + \alpha_6 Z_{1+X,1+X}^{1+X}  \biggr],
    \\
    & = &
    \frac{1}{6} \biggl[ Z_{1,1}^1 + \left( Z_{1,X}^X + Z_{X,1}^X + Z_{X,X}^1 \right) + 2 \left( Z_{1,1}^1 + Z_{1,X}^X \right)
    + 2 \left( Z_{1,1}^1 + Z_{X,1}^X \right)
    \nonumber \\
    & & \hspace*{0.25in}
    + \alpha_1 \left( Z_{1,1}^1 + Z_{X,X}^1 \right) 
    + \alpha_2 \left( Z_{1,1}^1 + Z_{1,X}^X + Z_{X,1}^X + Z_{X,X}^1 \right)
    \nonumber \\
    & & \hspace*{0.25in}
    + \alpha_3 \left( Z_{X,1}^X + Z_{X,X}^1 \right)
    + \alpha_4 \left( Z_{1,X}^X + Z_{X,X}^1 \right)
    \nonumber \\
    & & \hspace*{0.25in}
    + \alpha_5 \left( Z_{1,X}^X + Z_{X,1}^X \right)
    + \alpha_6 \left( Z_{1,1}^1 + Z_{1,X}^X + Z_{X,1}^X + Z_{X,X}^1 \right) \biggr],
\end{eqnarray}
where $\alpha_{1-6}$ are constants whose determination requires knowledge of the remaining natural transformations $J$.

As before, as this reduces to a ${\mathbb Z}_2$ orbifold, we can apply modular invariance:  the same number of $Z_{1,X}^X$, $Z_{X,1}^X$, and $Z_{X,X}^1$ should appear.  Counting terms above, there are
\begin{itemize}
    \item $5 + \alpha_1 + \alpha_2 + \alpha_6$ copies of $Z_{1,1}^1$,
    \item $3 + \alpha_2 + \alpha_4 + \alpha_5 + \alpha_6$ copies of $Z_{1,X}^X$,
    \item $3 + \alpha_2 + \alpha_3 + \alpha_5 + \alpha_6$ copies of $Z_{X,1}^X$,
    \item $1 + \alpha_1  + \alpha_2 + \alpha_3 + \alpha_4 + \alpha_6$ copies of $Z_{X,X}^1$.
\end{itemize}
Given that the number of copies of $Z_{1,X}^X$, $Z_{X,1}^X$, and $Z_{X,X}^1$ should match, we find the constraints
\begin{itemize}
    \item $\alpha_3 = \alpha_4$,
    \item $\alpha_1 + \alpha_3 + \alpha_4 = 2 + \alpha_3 + \alpha_5 = 2 + \alpha_4 + \alpha_5$, or more simply,
    $\alpha_1 + \alpha_3 = 2 + \alpha_5$.
\end{itemize}
Altogether, this means that the partition function $Z_{1+X+2Y}$ consists of 
\begin{equation}
    \frac{2}{6} \left( 1 + \alpha_1  + \alpha_2 + \alpha_3 + \alpha_4 + \alpha_6 \right) \: = \: \frac{1}{3} \left( 3 + \alpha_2 + \alpha_3 + \alpha_5 + \alpha_6 \right)
\end{equation}
copies of the partition function of $[ {\cal T}/{\mathbb Z}_2]$, and
\begin{equation}
    \frac{1}{6} \left[ \left(5 + \alpha_1 + \alpha_2 + \alpha_6\right) - \left(1 + \alpha_1  + \alpha_2 + \alpha_3 + \alpha_4 + \alpha_6\right)  \right]
    \: = \:
    \frac{1}{3} \left( 2 - \alpha_3 \right)
\end{equation}
copies of the partition function of ${\cal T}$, or more simply,
\begin{equation}   
    [ {\cal T}/ (1+X+2Y)] \: = \: \coprod_{(1/3)(2 - \alpha_3)} {\cal T} \, 
    \coprod_{(1/3)(3 + \alpha_2 + \alpha_3 + \alpha_5 + \alpha_6} [ {\cal T}/{\mathbb Z}_2].
\end{equation}

Since $(2-\alpha_3)/3$ must be a positive integer, we see
\begin{equation}
    \alpha_3 \: = \: \alpha_4 \: = \: 2 - 3k
\end{equation}
for some integer $k \geq 0$.
Similarly, we require
\begin{equation}
    \frac{1}{3} \left( 3 + \alpha_2 + \alpha_3 + \alpha_5 + \alpha_6 \right) \: = \: 
    \ell \: \in {\mathbb Z}_{\geq 0}, 
\end{equation}
so that altogether
\begin{equation}   \label{eq:yeq1+x:1+x+2y:part1}
    [ {\cal T}/ (1+X+2Y)] \: = \: \coprod_k {\cal T} \,
    \coprod_{\ell} [{\cal T}/{\mathbb Z}_2].
\end{equation}

Now, let us compare to previous results.  
From \cite[section 5.4.1]{Perez-Lona:2023djo}, if both $X$ and $Y$ act trivially,
then 
\begin{equation}
    [ {\cal T} / (1+X+2Y)] \: = \: \coprod_6 {\cal T}.
\end{equation}
Specializing equation~(\ref{eq:yeq1+x:1+x+2y:part1}) to this case, and using the fact that a trivially-acting ${\mathbb Z}_2$ orbifold is equivalent to a disjoint union of two copies of the original theory, we find
that $[ {\cal T}/ (1+X+2Y)]$ specializes to a disjoint union of 
\begin{equation}
    k + 2 \ell
\end{equation}
copies of ${\cal T}$.  Thus, we find that equation~(\ref{eq:yeq1+x:1+x+2y:part1}) will correctly specialize to results in 
\cite[section 5.4.1]{Perez-Lona:2023djo} if $k$, $\ell$ take any of the values listed
in the table below:
\begin{center}
    \begin{tabular}{cc}
    $k$ & $\ell$ \\ \hline
    $0$ & $3$ \\
    $2$ & $2$ \\
    $4$ & $1$ \\
    $6$ & $0$
    \end{tabular}
\end{center}
Of course, the correct choice is ultimately dictated by the correct natural transformations $J$, but as the complete computation is extremely laborious, we leave it for future work.

We close this section with an intuitive argument justifying a conjecture for the result.
The functor Rep$(S_3) \rightarrow {\rm Rep}({\mathbb Z}_2)$ is obtained by precomposing every representation with the power map, which sends the order-two generator of $S_3$ to itself, and the order-three generator to 1. 
Now, gauging Rep($S_3$) should give a theory with $S_3$ symmetry, but the fact that the action $Y=1+X$ comes from sending the order-three generator to 1, suggests that the action of the quantum symmetry $S_3$ also factors through that power map, so that only the order-two generator acts nontrivially and the order-three generator acts trivially. The only way for the order-three generator to act trivially is to have three copies of a single theory, which in terms of the structure above suggests that $k=0$ and $\ell=3$:
\begin{equation}
[ {\cal T}/ (1+X+2Y)] \: = \: \coprod_3 [ {\cal T}/{\mathbb Z}_2].    
\end{equation}
In other words, the order-two generator acts nontrivially on each orbifold theory $[{\cal T}/{\mathbb Z}_2]$ as a ${\mathbb Z}_2$ quantum symmetry, whereas the order-three generator just permutes the three different copies of $[{\cal T}/{\mathbb Z}_2]$.

\end{itemize}

\subsubsection{Completely-trivially-acting Rep$(S_3)$}

For completeness, in this subsection we review the case that all of the Rep$(S_3)$ acts trivially.
This case was previously discussed in \cite{Perez-Lona:2023djo}; we briefly outline the results and 
discuss relations to the results of this section.

Briefly, in the case that all of Rep$(S_3)$ acts trivially, we previously argued that
\cite[section 5.4.1]{Perez-Lona:2023djo}
\begin{eqnarray}
        \left[ {\cal T} / (1+X) \right] & = & \coprod_2 {\cal T}, 
        \\
        \left[ {\cal T} / (1+Y) \right] & = & \coprod_3 {\cal T},
        \\
        \left[ {\cal T} / (1+X+2Y) \right] & = & \coprod_6 {\cal T}.
\end{eqnarray}

In section~\ref{sect:triv-repz2-reps3}, we computed decompositions when $X$ acts trivially, but not necessarily $Y$,
and we checked that the results obtained were consistent with the results above when the rest of
Rep$(S_3)$ also acts trivially.

In section~\ref{sect:decomp:1px}, we computed decompositions when $Y = 1+X$, another special case.
Here, we were not able to uniquely determine decompositions using our (limited) knowledge of the $J$ natural transformations and modular invariance alone, but we were able to use the results above to determine a prediction for the results for decomposition.

\section{Conclusions}

In this paper, we have extended our previous work \cite{Perez-Lona:2023djo} on gauging noninvertible symmetries in two dimensions to include non-multiplicity-free cases, and have studied explicitly the example of Rep$(A_4)$.  We apply these gaugings to $c=1$ CFTs and find theories enjoying Rep$(A_4)$ symmetry, even on the circle branch. We find that for a self-duality under gauging noninvertible symmetries, one generally introduces not only a single duality duality but multiple new defects. We obtain the self-duality under gauging Rep$(A_4)$ in the exceptional $SU(2)_1/A_4$ CFT implies a larger symmetry than Rep$(A_4)$, namely Rep$(SL(2,\Z_3))$.  We have also further discussed decomposition in theories with trivially-acting gauged noninvertible (sub)symmetries.

One matter to which we hope to return in the future is to compute the analogues of twist fields in ordinary
orbifolds. For noninvertible symmetries, this needs an explicit computation for the corresponding defect Hilbert spaces.

Another matter to which we hope to return is to check explicitly at the level of partition functions that orbifolding by a quantum symmetry returns the original theory.  This is well-known for abelian orbifolds, and has been argued abstractly \cite{Bhardwaj:2017xup} for nonabelian orbifolds (where the quantum symmetry is a noninvertible symmetry of the form Rep$(G)$ if the original orbifold was a $G$ orbifold); however, we would also like to compute this explicitly at the level of partition functions.

Based on the Rep$(A_4)$ symmetry we find on the circle branch (which previously are not expected to enjoy a rich noninvertible symmetry structrue), as well as the multiplicity of defects built via self-dual gauging noninvertible symmetries, we hope to perform a systematic investigation of $\Z_2$-graded fusion categories and its interplay with other noninvertible defects in the context of $c=1$ CFTs.

\section*{Acknowledgements}

We would like to thank Y.~Choi, J.~Knapp, R.~Radhakrishnan, B.~Rayhaun, and H.~Y.~Zhang for useful conversations.
E.S.~and X.Y.~were partially supported by NSF grant PHY-2310588.
D.R.~thanks the Kavli Institute for Theoretical Physics (KITP) and the 21st Simons Physics Summer Workshop for hospitality during part of this work.
X.Y.~thanks the ICTP String-Math 2024, the Institute of Theoretical Physics (IFT), the 2024 IHES Summer School -- Symmetries and Anomalies: A Modern Take, and 21st Simons Physics Summer Workshop for their hospitality during part of this work. 

\appendix

\section{Details of $A_4$ calculations}  \label{app:a4-details}

\subsection{Fusion intertwiners}
\label{appsub:A4FusionIntertwiners}

For any fusion involving the trivial irrep we have a canonical choice,
\be
\la_{1,R}^R(ev)=\la_{R,1}^R(ve)=v.
\ee

For the fusions of the one-dimensional irreps, our task is easy
\begin{align}
    \la_{X,X}^Y(e_Xe_X)=\ & \beta_1e_Y,\\
    \la_{X,Y}^1(e_Xe_Y)=\ & \beta_2e,\\
    \la_{Y,X}(e_Ye_X)=\ & \beta_3e,\\
    \la_{Y,Y}^X(e_Ye_Y)=\ & \beta_4e_X.
\end{align}

For fusions involving the $Z$ irrep, it will be useful to specify two alternative bases in which $\rho_Z(a)$ and $\rho_Z(b)$ are diagonalized.  For the former, we take $\{u_1,u_2,e_3\}$, where
\begin{align}
    u_1=\ & \frac{1}{\sqrt{2}}\lp e_1-ie_2\rp,\\
    u_2=\ & \frac{1}{\sqrt{2}}\lp e_1+ie_2\rp,
\end{align}
or inversely,
\begin{align}
    e_1=\ & \frac{1}{\sqrt{2}}\lp u_1+u_2\rp,\\
    e_2=\ & \frac{i}{\sqrt{2}}\lp u_1-u_2\rp.
\end{align}
The eigenvalues of $u_1$, $u_2$, and $e_3$ under $\rho_Z(a)$ are respectively $\zeta$, $\zeta^2$, and $1$ (in particular they are all distinct).  To diagonalize $\rho_Z(b)$ we take $\{e_1,v_1,v_2\}$ where
\begin{align}
    v_1=\ & \sqrt{\frac{2}{3}}e_2+\frac{1}{\sqrt{3}}e_3,\\
    v_2=\ & \frac{1}{\sqrt{3}}e_2-\sqrt{\frac{2}{3}}e_3,
\end{align}
or
\begin{align}
    e_2=\ & \sqrt{\frac{2}{3}}v_1+\frac{1}{\sqrt{3}}v_2,\\
    e_3=\ & \frac{1}{\sqrt{3}}v_1-\sqrt{\frac{2}{3}}v_2.
\end{align}
The eigenvalues of $e_1$, $v_1$, and $v_2$ under $\rho_Z(b)$ are respectively $-1$, $1$, and $-1$.

Because the $\rho_Z(a)$ eigenvalues are distinct, it is useful to first parameterize each fusion in terms of the $\{u_1,u_2,e_3\}$ basis and then switch to the $\{e_1,v_1,v_2\}$ basis to determine further restrictions on the coefficients.  Thus we will need to go back and forth between the two bases,
\begin{align}
    e_1=\ & \frac{1}{\sqrt{2}}u_1+\frac{1}{\sqrt{2}}u_2,\\
    v_1=\ & \frac{i}{\sqrt{3}}u_1-\frac{i}{\sqrt{3}}u_2+\frac{1}{\sqrt{3}}e_3,\\
    v_2=\ & \frac{i}{\sqrt{6}}u_1-\frac{i}{\sqrt{6}}u_2-\sqrt{\frac{2}{3}}e_3,
\end{align}
or
\begin{align}
    u_1=\ & \frac{1}{\sqrt{2}}e_1-\frac{i}{\sqrt{3}}v_1-\frac{i}{\sqrt{6}}v_2,\\
    u_2=\ & \frac{1}{\sqrt{2}}e_1+\frac{i}{\sqrt{3}}v_1+\frac{i}{\sqrt{6}}v_2,\\
    e_3=\ & \frac{1}{\sqrt{3}}v_1-\sqrt{\frac{2}{3}}v_2.
\end{align}

For example, consider the fusion $XZ\rightarrow Z$.  By considering the action of $a$, we determine
\begin{align}
    \la_{X,Z}^Z(e_Xu_1)=\ & \al_1u_2,\\
    \la_{X,Z}^Z(e_Xu_2)=\ & \al_2e_3,\\
    \la_{X,Z}^Z(e_Xe_3)=\ & \al_3u_1.
\end{align}
Here the $\al_i$ are undetermined complex constants and we have matched eigenvalues under the action of $a$ (acting as $\rho_X(a)\otimes\rho_Z(a)$ on the left-hand side and $\rho_Z(a)$ on the right-hand side).  In terms of the other basis we then have
\begin{align}
    \la_{X,Z}^Z(e_Xe_1)=\ & \frac{1}{\sqrt{2}}\la_{X,Z}^Z(e_Xu_1)+\frac{1}{\sqrt{2}}\la_{X,Z}^Z(e_Xu_2)=\frac{1}{\sqrt{2}}\al_1u_2+\frac{1}{\sqrt{2}}\al_2e_3\non\\
    =\ & \hlf\al_1e_1+\lp\frac{i}{\sqrt{6}}\al_1+\frac{1}{\sqrt{6}}\al_2\rp v_1+\lp\frac{i}{2\sqrt{3}}\al_1-\frac{1}{\sqrt{3}}\al_2\rp v_2,\\
    \la_{X,Z}^Z(e_Xv_1)=\ & \frac{i}{\sqrt{3}}\al_1u_2-\frac{i}{\sqrt{3}}\al_2e_3+\frac{1}{\sqrt{3}}\al_3u_1\non\\
    =\ & \lp\frac{i}{\sqrt{6}}\al_1+\frac{1}{\sqrt{6}}\al_3\rp e_1+\lp -\frac{1}{3}\al_1-\frac{i}{3}\al_2-\frac{i}{3}\al_3\rp v_1+\lp -\frac{1}{3\sqrt{2}}\al_1+\frac{\sqrt{2}i}{3}\al_2-\frac{i}{3\sqrt{2}}\al_3\rp v_2,\\
    \la_{X,Z}^Z(e_Xv_2)=\ & \frac{i}{\sqrt{6}}\al_1u_2-\frac{i}{\sqrt{6}}\al_2e_3-\sqrt{\frac{2}{3}}\al_3u_1\non\\
    =\ & \lp\frac{i}{2\sqrt{3}}\al_1-\frac{1}{\sqrt{3}}\al_3\rp e_1+\lp -\frac{1}{3\sqrt{2}}\al_1-\frac{i}{3\sqrt{2}}\al_2+\frac{\sqrt{2}i}{3}\al_3\rp v_1+\lp -\frac{1}{6}\al_1+\frac{i}{3}\al_2+\frac{i}{3}\al_3\rp v_2.
\end{align}
Now $e_Xe_1$ and $e_Xv_2$ should both have eigenvalue $-1$ under the action of $b$, and hence should get mapped to a linear combination of $e_1$ and $v_2$, but the coefficient of $v_1$ should vanish, while $e_Xv_1$ has eigenvalue $+1$ and so should map to a multiple of $v_1$, with the coefficients of $e_1$ and $v_2$ vanishing.  These considerations give four equations,
\begin{align}
    \frac{i}{\sqrt{6}}\al_1+\frac{1}{\sqrt{6}}\al_2=\ & 0,\\
    -\frac{1}{3\sqrt{2}}\al_1-\frac{i}{3\sqrt{2}}\al_2+\frac{\sqrt{2}i}{3}\al_3=\ & 0,\\
    \frac{i}{\sqrt{6}}\al_1+\frac{1}{\sqrt{6}}\al_3=\ & 0,\\
    -\frac{1}{3\sqrt{2}}\al_1+\frac{\sqrt{2}i}{3}\al_2-\frac{i}{3\sqrt{2}}\al_3=\ & 0.
\end{align}
These equations are not all independent, and the general solution is simply
\be
\al_1=\beta_5,\qquad\al_2=-i\beta_5,\qquad\al_3=-i\beta_5.
\ee
So we conclude that the most general intertwiner $\la_{X,Z}^Z$ is
\begin{align}
    \la_{X,Z}^Z(e_Xe_1)=\ & \frac{1}{\sqrt{2}}\al_1u_2+\frac{1}{\sqrt{2}}\al_2e_3=\beta_5\lp\hlf e_1+\frac{i}{2}e_2-\frac{i}{\sqrt{2}}e_3\rp,\\
    \la_{X,Z}^Z(e_Xe_2)=\ & \frac{i}{\sqrt{2}}\al_1u_2-\frac{i}{\sqrt{2}}\al_2e_3=\beta_5\lp\frac{i}{2}e_1-\hlf e_2-\frac{1}{\sqrt{2}}e_3\rp,\\
    \la_{X,Z}^Z(e_Xe_3)=\ & \al_3u_1=\beta_5\lp -\frac{i}{\sqrt{2}}e_1-\frac{1}{\sqrt{2}}e_2\rp.
\end{align}

The calculation for $ZX\rightarrow Z$ is identical, leading to
\begin{align}
    \la_{Z,X}^Z(e_1e_X)=\ & \beta_6\lp\hlf e_1+\frac{i}{2}e_2-\frac{i}{\sqrt{2}}e_3\rp,\\
    \la_{Z,X}^Z(e_2e_X)=\ & \beta_6\lp\frac{i}{2}e_1-\hlf e_2-\frac{1}{\sqrt{2}}e_3\rp,\\
    \la_{Z,X}^Z(e_3e_X)=\ & \beta_6\lp -\frac{i}{\sqrt{2}}e_1-\frac{1}{\sqrt{2}}e_2\rp.
\end{align}

For computing the $YZ\rightarrow Z$ fusion we start with
\begin{align}
    \la_{Y,Z}^Z(e_Yu_1)=\ & \al_1e_3,\\
    \la_{Y,Z}^Z(e_Yu_2)=\ & \al_2u_1,\\
    \la_{Y,Z}^Z(e_Ye_3)=\ & \al_3u_2.
\end{align}
Then
\begin{align}
    \la_{Y,Z}^Z(e_Ye_1)=\ & \frac{1}{\sqrt{2}}\al_1e_3+\frac{1}{\sqrt{2}}\al_2u_1\non\\
    =\ & \hlf\al_2e_1+\lp\frac{1}{\sqrt{6}}\al_1-\frac{i}{\sqrt{6}}\al_2\rp v_1+\lp -\frac{1}{\sqrt{3}}\al_1-\frac{i}{2\sqrt{3}}\al_2\rp v_2,\\
    \la_{Y,Z}^Z(e_Yv_1)=\ & \frac{i}{\sqrt{3}}\al_1e_3-\frac{i}{\sqrt{3}}\al_2u_1+\frac{1}{\sqrt{3}}\al_3u_2\non\\
    =\ & \lp -\frac{i}{\sqrt{6}}\al_2+\frac{1}{\sqrt{6}}\al_3\rp e_1+\lp\frac{i}{3}\al_1-\frac{1}{3}\al_2+\frac{i}{3}\al_3\rp v_1+\lp -\frac{\sqrt{2}i}{3}\al_1-\frac{1}{3\sqrt{2}}\al_2+\frac{i}{3\sqrt{2}}\al_3\rp v_2,\\
    \la_{Y,Z}^Z(e_Yv_2)=\ & \frac{i}{\sqrt{6}}\al_1e_3-\frac{i}{\sqrt{6}}\al_2u_1-\sqrt{\frac{2}{3}}\al_3u_2\non\\
    =\ & \lp -\frac{i}{2\sqrt{3}}\al_2-\frac{1}{\sqrt{3}}\al_3\rp e_1+\lp\frac{i}{3\sqrt{2}}\al_1-\frac{1}{3\sqrt{2}}\al_2-\frac{\sqrt{2}i}{3}\al_3\rp v_1+\lp -\frac{i}{3}\al_1-\frac{1}{6}\al_2-\frac{i}{3}\al_3\rp v_2.
\end{align}
Demanding the correct behavior under the action of $b$ fixes
\be
\al_1=\beta_7,\qquad\al_2=-i\beta_7,\qquad\al_3=\beta_7,
\ee
and we have
\begin{align}
    \la_{Y,Z}^Z(e_Ye_1)=\ & \frac{1}{\sqrt{2}}\al_1e_3+\frac{1}{\sqrt{2}}\al_2u_1=\beta_7\lp -\frac{i}{2}e_1-\hlf e_2+\frac{1}{\sqrt{2}}e_3\rp,\\
    \la_{Y,Z}^Z(e_Ye_2)=\ & \frac{i}{\sqrt{2}}\al_1e_3-\frac{i}{\sqrt{2}}\al_2u_1=\beta_7\lp -\hlf e_1+\frac{i}{2}e_2+\frac{i}{\sqrt{2}}e_3\rp,\\
    \la_{Y,Z}^Z(e_Ye_3)=\ & \al_3u_2=\beta_7\lp\frac{1}{\sqrt{2}}e_1+\frac{i}{\sqrt{2}}e_2\rp.
\end{align}
Similarly,
\begin{align}
    \la_{Z,Y}^Z(e_1e_Y)=\ & \beta_8\lp -\frac{i}{2}e_1-\hlf e_2+\frac{1}{\sqrt{2}}e_3\rp,\\
    \la_{Z,Y}^Z(e_2e_Y)=\ & \beta_8\lp -\hlf e_1+\frac{i}{2}e_2+\frac{i}{\sqrt{2}}e_3\rp,\\
    \la_{Z,Y}^Z(e_3e_Y)=\ & \beta_8\lp\frac{1}{\sqrt{2}}e_1+\frac{i}{\sqrt{2}}e_2\rp.
\end{align}

Next we need to consider the various $ZZ$ fusions, starting with the case where the fusion product is a one-dimensional irrep.  In this case many of the fusions (in the $\{u_1,u_2,e_3\}$ basis) must vanish.  For instance, for $ZZ\rightarrow 1$, we have only
\begin{align}
    \la_{Z,Z}^1(u_1u_2)=\ & \al_1e,\\
    \la_{Z,Z}^1(u_2u_1)=\ & \al_2e,\\
    \la_{Z,Z}^1(e_3e_3)=\ & \al_3e,
\end{align}
with other products necessarily mapping to zero since they are not invariant under the action of $\rho_{ZZ}(a)$.  Switching to the $\{e_1,v_1,v_2\}$ basis then, we find that some combinations should vanish (since they are not $\rho_{ZZ}(b)$ invariants).  Computing only as much as we need,
\begin{align}
    0=\ & \la_{Z,Z}^1(e_1v_1)=\lp -\frac{i}{\sqrt{6}}\al_1+\frac{i}{\sqrt{6}}\al_2\rp e,\\
    0=\ & \la_{Z,Z}^1(v_1v_2)=\lp\frac{1}{3\sqrt{2}}\al_1+\frac{1}{3\sqrt{2}}\al_2-\frac{\sqrt{2}}{3}\al_3\rp e.
\end{align}
This is already enough to conclude that we must have
\be
\al_1=\al_2=\al_3=\beta_9,
\ee
This determines
\begin{align}
    \la_{Z,Z}^1(e_1e_1)=\ & \lp\hlf\al_1+\hlf\al_2\rp e=\beta_9e,\\
    \la_{Z,Z}^1(e_1e_2)=\ & \lp -\frac{i}{2}\al_1+\frac{i}{2}\al_2\rp e=0,\\
    \la_{Z,Z}^1(e_1e_3)=\ & 0,\\
    \la_{Z,Z}^1(e_2e_1)=\ & \lp -\frac{i}{2}\al_1+\frac{i}{2}\al_2\rp e=0,\\
    \la_{Z,Z}^1(e_2e_2)=\ & \lp\hlf\al_1+\hlf\al_2\rp e=\beta_9e,\\
    \la_{Z,Z}^1(e_2e_3)=\ & 0,\\
    \la_{Z,Z}^1(e_3e_1)=\ & 0,\\
    \la_{Z,Z}^1(e_3e_2)=\ & 0,\\
    \la_{Z,Z}^1(e_3e_3)=\ & \al_3e=\beta_9e,
\end{align}
which we probably could have simply guessed.

For $ZZ\rightarrow X$ we start with
\begin{align}
    \la_{Z,Z}^X(u_1e_3)=\ & \al_1e_X,\\
    \la_{Z,Z}^X(u_2u_2)=\ & \al_2e_X,\\
    \la_{Z,Z}^X(e_3u_1)=\ & \al_3e_X.
\end{align}
And then
\begin{align}
    0=\ & \la_{Z,Z}^X(e_1v_1)=\lp\frac{1}{\sqrt{6}}\al_1-\frac{i}{\sqrt{6}}\al_2\rp e_X,\\
    0=\ & \la_{Z,Z}^X(v_1e_1)=\lp -\frac{i}{\sqrt{6}}\al_2+\frac{1}{\sqrt{6}}\al_3\rp e_X,
\end{align}
which is already enough to determine
\be
\al_1=\beta_{10},\qquad\al_2=-i\beta_{10},\qquad\al_3=\beta_{10},
\ee
and hence
\begin{align}
    \la_{Z,Z}^X(e_1e_1)=\ & \hlf\al_2e_X=-\frac{i}{2}\beta_{10}e_X,\\
    \la_{Z,Z}^X(e_1e_2)=\ & -\frac{i}{2}\al_2e_X=-\hlf\beta_{10}e_X,\\
    \la_{Z,Z}^X(e_1e_3)=\ & \frac{1}{\sqrt{2}}\al_1e_X=\frac{1}{\sqrt{2}}\beta_{10}e_X,\\
    \la_{Z,Z}^X(e_2e_1)=\ & -\frac{i}{2}\al_2e_X=-\hlf\beta_{10}e_X,\\
    \la_{Z,Z}^X(e_2e_2)=\ & -\hlf\al_2e_X=\frac{i}{2}\beta_{10}e_X,\\
    \la_{Z,Z}^X(e_2e_3)=\ & \frac{i}{\sqrt{2}}\al_1e_X=\frac{i}{\sqrt{2}}\beta_{10}e_X,\\
    \la_{Z,Z}^X(e_3e_1)=\ & \frac{1}{\sqrt{2}}\al_3e_X=\frac{1}{\sqrt{2}}\beta_{10}e_X,\\
    \la_{Z,Z}^X(e_3e_2)=\ & \frac{i}{\sqrt{2}}\al_3e_X=\frac{i}{\sqrt{2}}\beta_{10}e_X,\\
    \la_{Z,Z}^X(e_3e_3)=\ & 0.
\end{align}

Following the same procedure for $ZZ\rightarrow Y$,
\begin{align}
    \la_{Z,Z}^Y(u_1u_1)=\ & \al_1e_Y,\\
    \la_{Z,Z}^Y(u_2e_3)=\ & \al_2e_Y,\\
    \la_{Z,Z}^Y(e_3u_2)=\ & \al_3e_Y.
\end{align}
\begin{align}
    0=\ & \la_{Z,Z}^Y(e_1v_1)=\lp\frac{i}{\sqrt{6}}\al_1+\frac{1}{\sqrt{6}}\al_2\rp e_Y,\\
    0=\ & \la_{Z,Z}^Y(v_1e_1)=\lp\frac{i}{\sqrt{6}}\al_1+\frac{1}{\sqrt{6}}\al_3\rp e_Y,
\end{align}
\be
\Rightarrow\qquad\al_1=\beta_{11},\qquad\al_2=-i\beta_{11},\qquad\al_3=-i\beta_{11}.
\ee
So
\begin{align}
    \la_{Z,Z}^Y(e_1e_1)=\ & \hlf\al_1e_Y=\hlf\beta_{11}e_Y,\\
    \la_{Z,Z}^Y(e_1e_2)=\ & \frac{i}{2}\al_1e_Y=\frac{i}{2}\beta_{11}e_Y,\\
    \la_{Z,Z}^Y(e_1e_3)=\ & \frac{1}{\sqrt{2}}\al_2e_Y=-\frac{i}{\sqrt{2}}\beta_{11}e_Y,\\
    \la_{Z,Z}^Y(e_2e_1)=\ & \frac{i}{2}\al_1e_Y=\frac{i}{2}\beta_{11}e_Y,\\
    \la_{Z,Z}^Y(e_2e_2)=\ & -\hlf\al_1e_Y=-\hlf\beta_{11}e_Y,\\
    \la_{Z,Z}^Y(e_2e_3)=\ & -\frac{i}{\sqrt{2}}\al_2e_Y=-\frac{1}{\sqrt{2}}\beta_{11}e_Y,\\
    \la_{Z,Z}^Y(e_3e_1)=\ & \frac{1}{\sqrt{2}}\al_3e_Y=-\frac{i}{\sqrt{2}}\beta_{11}e_Y,\\
    \la_{Z,Z}^Y(e_3e_2)=\ & -\frac{i}{\sqrt{2}}\al_3e_Y=-\frac{1}{\sqrt{2}}\beta_{11}e_Y,\\
    \la_{Z,Z}^Y(e_3e_3)=\ & 0.
\end{align}

Finally we need to determine the two-dimensional space of intertwiners corresponding to the $ZZ\rightarrow Z$ fusion.  We start with
\begin{align}
    \la_{Z,Z}^Z(u_1u_1)=\ & \al_1u_2,\\
    \la_{Z,Z}^Z(u_1u_2)=\ & \al_2e_3,\\
    \la_{Z,Z}^Z(u_1e_3)=\ & \al_3u_1,\\
    \la_{Z,Z}^Z(u_2u_1)=\ & \al_4e_3,\\
    \la_{Z,Z}^Z(u_2u_2)=\ & \al_5u_1,\\
    \la_{Z,Z}^Z(u_2e_3)=\ & \al_6u_2,\\
    \la_{Z,Z}^Z(e_3u_1)=\ & \al_7u_1,\\
    \la_{Z,Z}^Z(e_3u_2)=\ & \al_8u_2,\\
    \la_{Z,Z}^Z(e_3e_3)=\ & \al_9e_3.
\end{align}
Switching bases,
\begin{align}
    \la_{Z,Z}^Z(e_1e_1)=\ & \hlf\al_5u_1+\hlf\al_1u_2+\frac{\al_2+\al_4}{2}e_3\non\\
    =\ & \frac{\al_1+\al_5}{2\sqrt{2}}e_1+\frac{i\al_1+\al_2+\al_4-i\al_5}{2\sqrt{3}}v_1+\frac{i\al_1-2\al_2-2\al_4-i\al_5}{2\sqrt{6}}v_2,\\
    \la_{Z,Z}^Z(e_1v_1)=\ & \frac{\al_3-i\al_5}{\sqrt{6}}u_1+\frac{i\al_1+\al_6}{\sqrt{6}}u_2+\frac{-i\al_2+i\al_4}{\sqrt{6}}e_3=\frac{i\al_1+\al_3-i\al_5+\al_6}{2\sqrt{3}}e_1\non\\
    & \quad +\frac{-\al_1-i\al_2-i\al_3+i\al_4-\al_5+i\al_6}{3\sqrt{2}}v_1+\frac{-\al_1+2i\al_2-i\al_3-2i\al_4-\al_5+i\al_6}{6}v_2,\\
    \la_{Z,Z}^Z(e_1v_2)=\ & \frac{-2\al_3-i\al_5}{2\sqrt{3}}u_1+\frac{i\al_1-2\al_6}{2\sqrt{3}}u_2+\frac{-i\al_2+i\al_4}{2\sqrt{3}}e_3=\frac{i\al_1-2\al_3-i\al_5-2\al_6}{2\sqrt{6}}e_1\\
    & \quad +\frac{-\al_1-i\al_2+2i\al_3+i\al_4-\al_5-2i\al_6}{6}v_1+\frac{-\al_1+2i\al_2+2i\al_3-2i\al_4-\al_5-2i\al_6}{6\sqrt{2}}v_2,\non\\
    \la_{Z,Z}^Z(v_1e_1)=\ & \frac{-i\al_5+\al_7}{\sqrt{6}}u_1+\frac{i\al_1+\al_8}{\sqrt{6}}u_2+\frac{i\al_2-i\al_4}{\sqrt{6}}e_3=\frac{i\al_1-i\al_5+\al_7+\al_8}{2\sqrt{3}}e_1\\
    & \quad +\frac{-\al_1+i\al_2-i\al_4-\al_5-i\al_7+i\al_8}{3\sqrt{2}}v_1+\frac{-\al_1-2i\al_2+2i\al_4-\al_5-i\al_7+i\al_8}{6}v_2,\non\\
    \la_{Z,Z}^Z(v_1v_1)=\ & \frac{i\al_3-\al_5+i\al_7}{3}u_1+\frac{-\al_1-i\al_6-i\al_8}{3}u_2+\frac{\al_2+\al_4+\al_9}{3}e_3\non\\
    =\ & \frac{-\al_1+i\al_3-\al_5-i\al_6+i\al_7-i\al_8}{3\sqrt{2}}e_1+\frac{-i\al_1+\al_2+\al_3+\al_4+i\al_5+\al_6+\al_7+\al_8+\al_9}{3\sqrt{3}}v_1\non\\
    & \quad +\frac{-i\al_1-2\al_2+\al_3-2\al_4+i\al_5+\al_6+\al_7+\al_8-2\al_9}{3\sqrt{6}}v_2,\\
    \la_{Z,Z}^Z(v_1v_2)=\ & \frac{-2i\al_3-\al_5+i\al_7}{3\sqrt{2}}u_1+\frac{-\al_1+2i\al_6-i\al_8}{3\sqrt{2}}u_2+\frac{\al_2+\al_4-2\al_9}{3\sqrt{2}}e_3\non\\
    =\ & \frac{-\al_1-2i\al_3-\al_5+2i\al_6+i\al_7-i\al_8}{6}e_1+\frac{-i\al_1+\al_2-2\al_3+\al_4+i\al_5-2\al_6+\al_7+\al_8-2\al_9}{3\sqrt{6}}v_1\non\\
    & \quad +\frac{-i\al_1-2\al_2-2\al_3-2\al_4+i\al_5-2\al_6+\al_7+\al_8+4\al_9}{6\sqrt{3}}v_2,\\
    \la_{Z,Z}^Z(v_2e_1)=\ & \frac{-i\al_5-2\al_7}{2\sqrt{3}}u_1+\frac{i\al_1-2\al_8}{2\sqrt{3}}u_2+\frac{i\al_2-i\al_4}{2\sqrt{3}}e_3=\frac{i\al_1-i\al_5-2\al_7-2\al_8}{2\sqrt{6}}e_1\\
    & \quad +\frac{-\al_1+i\al_2-i\al_4-\al_5+2i\al_7-2i\al_8}{6}v_1+\frac{-\al_1-2i\al_2+2i\al_4-\al_5+2i\al_7-2i\al_8}{6\sqrt{2}}v_2,\non\\
    \la_{Z,Z}^Z(v_2v_1)=\ & \frac{i\al_3-\al_5-2i\al_7}{3\sqrt{2}}u_1+\frac{-\al_1-i\al_6+2i\al_8}{3\sqrt{2}}u_2+\frac{\al_2+\al_4-2\al_9}{3\sqrt{2}}e_3\non\\
    =\ & \frac{-\al_1+i\al_3-\al_5-i\al_6-2i\al_7+2i\al_8}{6}e_1+\frac{-i\al_1+\al_2+\al_3+\al_4+i\al_5+\al_6-2\al_7-2\al_8-2\al_9}{3\sqrt{6}}v_1\non\\
    & \quad +\frac{-i\al_1-2\al_2+\al_3-2\al_4+i\al_5+\al_6-2\al_7-2\al_8+4\al_9}{6\sqrt{3}}v_2,\\
    \la_{Z,Z}^Z(v_2v_2)=\ & \frac{-2i\al_3-\al_5-2i\al_7}{6}u_1+\frac{-\al_1+2i\al_6+2i\al_8}{6}u_2+\frac{\al_2+\al_4+4\al_9}{6}e_3\non\\
    =\ & \frac{-\al_1-2i\al_3-\al_5+2i\al_6-2i\al_7+2i\al_8}{6\sqrt{2}}e_1+\frac{-i\al_1+\al_2-2\al_3+\al_4+i\al_5-2\al_6-2\al_7-2\al_8+4\al_9}{6\sqrt{3}}v_1\non\\
    & \quad +\frac{-i\al_1-2\al_2-2\al_3-2\al_4+i\al_5-2\al_6-2\al_7-2\al_8-8\al_9}{6\sqrt{3}}v_2.
\end{align}
Demanding appropriate behavior under the action of $\rho_Z(b)$ reduces us to a two-parameter set of solutions, as expected,
\be
\al_3=\al_4=\al_8=i\al_1-\al_2,\qquad \al_5=-\al_1,\qquad \al_6=\al_7=\al_2,\qquad \al_9=-i\al_1.
\ee
Putting $\al_1=\beta_{12}$, $\al_2=\beta_{13}$ defines the first of our two fusion maps,
\begin{align}
    \label{eq:A4lambdaZZZ1e1e1}
    \lp\la_{Z,Z}^Z\rp_1(e_1e_1)=\ & \frac{\al_5}{2}u_1+\frac{\al_1}{2}u_2+\frac{\al_2+\al_4}{2}e_3=\beta_{12}\lp\frac{i}{\sqrt{2}}e_2+\frac{i}{2}e_3\rp,\\
    \lp\la_{Z,Z}^Z\rp_1(e_1e_2)=\ & \frac{-i\al_5}{2}u_1+\frac{i\al_1}{2}u_2+\frac{-i\al_2+i\al_4}{2}e_3=\beta_{12}\lp\frac{i}{\sqrt{2}}e_1-\hlf e_3\rp-i\beta_{13}e_3,\\
    \lp\la_{Z,Z}^Z\rp_1(e_1e_3)=\ & \frac{\al_3}{\sqrt{2}}u_1+\frac{\al_6}{\sqrt{2}}u_2=\beta_{12}\lp\frac{i}{2}e_1+\hlf e_2\rp+i\beta_{13}e_2,\\
    \lp\la_{Z,Z}^Z\rp_1(e_2e_1)=\ & \frac{-i\al_5}{2}u_1+\frac{i\al_1}{2}u_2+\frac{i\al_2-i\al_4}{2}e_3=\beta_{12}\lp\frac{i}{\sqrt{2}}e_1+\hlf e_3\rp+i\beta_{13}e_3,\\
    \lp\la_{Z,Z}^Z\rp_1(e_2e_2)=\ & \frac{-\al_5}{2}u_1+\frac{-\al_1}{2}u_1+\frac{\al_2+\al_4}{2}e_3=\beta_{12}\lp -\frac{i}{\sqrt{2}}e_2+\frac{i}{2}e_3\rp,\\
    \lp\la_{Z,Z}^Z\rp_1(e_2e_3)=\ & \frac{i\al_3}{\sqrt{2}}u_1-\frac{i\al_6}{\sqrt{2}}u_2=\beta_{12}\lp -\hlf e_1+\frac{i}{2}e_2\rp-i\beta_{13}e_1,\\
    \lp\la_{Z,Z}^Z\rp_1(e_3e_1)=\ & \frac{\al_7}{\sqrt{2}}u_1+\frac{\al_8}{\sqrt{2}}u_2=\beta_{12}\lp\frac{i}{2}e_1-\hlf e_2\rp-i\beta_{13}e_2,\\
    \lp\la_{Z,Z}^Z\rp_1(e_3e_2)=\ & \frac{i\al_7}{\sqrt{2}}u_1+\frac{-i\al_8}{\sqrt{2}}u_2=\beta_{12}\lp\hlf e_1+\frac{i}{2}e_2\rp+i\beta_{13}e_1,\\
    \label{eq:A4lambdaZZZ1e3e3}
    \lp\la_{Z,Z}^Z\rp_1(e_3e_3)=\ & \al_9e_3=-i\beta_{12}e_3.
\end{align}
For the second fusion map, $\lp\la_{Z,Z}^Z\rp_2$, we will have the same form but with $\beta_{12}$ replaced by $\beta_{14}$ and $\beta_{13}$ replaced by $\beta_{15}$.

We have now parameterized a basis for the fusion intertwiners.  $\beta_1$ through $\beta_{11}$ are any elements of $\C^\times\cong\GL(1,\C)$, while the matrix $\lp\begin{matrix} \beta_{12} & \beta_{13} \\ \beta_{14} & \beta_{15} \end{matrix}\rp$ should be an element of $\GL(2,\C)$.

\subsection{Evaluation, co-evaluation, and co-fusion maps}
\label{appsub:A4CoFusionMaps}

Following the conventions in~\cite{Perez-Lona:2023djo}, we can define a set of evaluation maps using the $\la_{A,\ov{A}}^1$ basis,
\begin{equation}
    \e_1(ee)=\ov{\e}_1(ee)=1,\quad\e_X(e_Ye_X)=\ov{\e}_Y(e_Ye_X)=\beta_3,\quad\e_Y(e_Xe_Y)=\ov{\e}_X(e_Xe_Y)=\beta_2,
\end{equation}
and for $Z$, we have
\begin{equation}
    \e_Z(e_ie_j)=\ov{\e}_Z(e_ie_j)=\beta_9\d_{ij},
\end{equation}
where $i$ and $j$ run over $1,2,3$.  This then also determines a set of co-evaluation maps,
\begin{equation}
    \g_1(1)=\ov{\g}_1(1)=ee,\quad\g_X(1)=\ov{\g}_Y(1)=\beta_3^{-1}e_Xe_Y,\quad\g_Y(1)=\ov{\g}_X(1)=\beta_2^{-1}e_Ye_X,
\end{equation}
and
\begin{equation}
    \g_Z(1)=\ov{\g}_Z(1)=\beta_9^{-1}(e_1e_1+e_2e_2+e_3e_3).
\end{equation}

Using the fusion basis and the evaluation and co-evaluation maps, one can define a corresponding co-fusion basis (the definition is somewhat conventional) by
\begin{equation}
    \lp\d_{R_1}^{R_2,R_3}\rp_i(v_1)=\left[\left(\lp\la_{R_1,R_3^\ast}^{R_2}\rp_i\otimes 1_{R_3}\right)\circ\al^{-1}_{R_1,R_3^\ast,R_3}\right](v_1\otimes\ov{\g}_{R_3}(1)),\quad v_1\in R_1,
\end{equation}
If either superscript is $1$, then the intertwiner is trivial,
\begin{equation}
    \d_R^{1,R}(v)=ev,\qquad\d_R^{R,1}(v)=ve,
\end{equation}
and for the remaining cases we find
\begin{align}
    \d_1^{X,Y}(e)=\ & \beta_3^{-1}e_Xe_Y,\\
    \d_1^{Y,X}(e)=\ & \beta_2^{-1}e_Ye_X,\\
    \d_1^{Z,Z}(e)=\ & \beta_9^{-1}\lp e_1e_1+e_2e_2+e_3e_3\rp,\\
    \d_X^{Y,Y}(e_X)=\ & \beta_1\beta_3^{-1}e_Ye_Y,\\
    \d_X^{Z,Z}(e_X)=\ & \beta_5\beta_9^{-1}\lp\hlf e_1e_1+\frac{i}{2}e_1e_2-\frac{i}{\sqrt{2}}e_1e_3+\frac{i}{2}e_2e_1-\hlf e_2e_2-\frac{1}{\sqrt{2}}e_2e_3-\frac{i}{\sqrt{2}}e_3e_1-\frac{1}{\sqrt{2}}e_3e_2\rp,\\
    \d_Y^{X,X}(e_Y)=\ & \beta_4\beta_2^{-1}e_Xe_X,\\
    \d_Y^{Z,Z}(e_Y)=\ & \beta_7\beta_9^{-1}\lp -\frac{i}{2}e_1e_1-\hlf e_1e_2+\frac{1}{\sqrt{2}}e_1e_3-\hlf e_2e_1+\frac{i}{2}e_2e_2+\frac{i}{\sqrt{2}}e_2e_3+\frac{1}{\sqrt{2}}e_3e_1+\frac{i}{\sqrt{2}}e_3e_2\rp,\\
    \d_Z^{X,Z}(e_1)=\ & \beta_{10}\beta_9^{-1}\lp -\frac{i}{2}e_Xe_1-\hlf e_Xe_2+\frac{1}{\sqrt{2}}e_Xe_3\rp,\\
    \d_Z^{X,Z}(e_2)=\ & \beta_{10}\beta_9^{-1}\lp -\hlf e_Xe_1+\frac{i}{2}e_Xe_2+\frac{i}{\sqrt{2}}e_Xe_3\rp,\\
    \d_Z^{X,Z}(e_3)=\ & \beta_{10}\beta_9^{-1}\lp \frac{1}{\sqrt{2}}e_Xe_1+\frac{i}{\sqrt{2}}e_Xe_2\rp,\\
    \d_Z^{Y,Z}(e_1)=\ & \beta_{11}\beta_9^{-1}\lp \hlf e_Ye_1+\frac{i}{2}e_Ye_2-\frac{i}{\sqrt{2}}e_Ye_3\rp,\\
    \d_Z^{Y,Z}(e_2)=\ & \beta_{11}\beta_9^{-1}\lp\frac{i}{2}e_Ye_1-\hlf e_Ye_2-\frac{1}{\sqrt{2}}e_Ye_3\rp,\\
    \d_Z^{Y,Z}(e_3)=\ & \beta_{11}\beta_9^{-1}\lp -\frac{i}{\sqrt{2}}e_Ye_1-\frac{1}{\sqrt{2}}e_Ye_2\rp,\\
    \d_Z^{Z,X}(e_1)=\ & \beta_8\beta_2^{-1}\lp-\frac{i}{2}e_1e_X-\hlf e_2e_X+\frac{1}{\sqrt{2}}e_3e_X\rp,\\
    \d_Z^{Z,X}(e_2)=\ & \beta_8\beta_2^{-1}\lp -\hlf e_1e_X+\frac{i}{2}e_2e_X+\frac{i}{\sqrt{2}}e_3e_X\rp,\\
    \d_Z^{Z,X}(e_3)=\ & \beta_8\beta_2^{-1}\lp\frac{1}{\sqrt{2}}e_1e_X+\frac{i}{\sqrt{2}}e_2e_X\rp,\\
    \d_Z^{Z,Y}(e_1)=\ & \beta_6\beta_3^{-1}\lp\hlf e_1e_Y+\frac{i}{2}e_2e_Y-\frac{i}{\sqrt{2}}e_3e_Y\rp,\\
    \d_Z^{Z,Y}(e_2)=\ & \beta_6\beta_3^{-1}\lp\frac{i}{2}e_1e_Y-\hlf e_2e_Y-\frac{1}{\sqrt{2}}e_3e_Y\rp,\\
    \d_Z^{Z,Y}(e_3)=\ & \beta_6\beta_3^{-1}\lp -\frac{i}{\sqrt{2}}e_1e_Y-\frac{1}{\sqrt{2}}e_2e_Y\rp,
\end{align}
and
\begin{align}
    \lp\d_Z^{Z,Z}\rp_1(e_1)=\ & \beta_{12}\beta_9^{-1}\lp\frac{i}{\sqrt{2}}e_1e_2+\frac{i}{2}e_1e_3+\frac{i}{\sqrt{2}}e_2e_1+\hlf e_2e_3+\frac{i}{2}e_3e_1-\hlf e_3e_2\rp+\beta_{13}\beta_9^{-1}\lp ie_2e_3-ie_3e_2\rp,\\
    \lp\d_Z^{Z,Z}\rp_1(e_2)=\ & \beta_{12}\beta_9^{-1}\lp\frac{i}{\sqrt{2}}e_1e_1-\hlf e_1e_3-\frac{i}{\sqrt{2}}e_2e_2+\frac{i}{2}e_2e_3+\hlf e_3e_1+\frac{i}{2}e_3e_2\rp+\beta_{13}\beta_9^{-1}\lp -ie_1e_3+ie_3e_1\rp,\\
    \lp\d_Z^{Z,Z}\rp_1(e_3)=\ & \beta_{12}\beta_9^{-1}\lp\frac{i}{2}e_1e_1+\hlf e_1e_2-\hlf e_2e_1+\frac{i}{2}e_2e_2-ie_3e_3\rp+\beta_{13}\beta_9^{-1}\lp ie_1e_2-ie_2e_1\rp,\\
    \lp\d_Z^{Z,Z}\rp_2(e_1)=\ & \beta_{14}\beta_9^{-1}\lp\frac{i}{\sqrt{2}}e_1e_2+\frac{i}{2}e_1e_3+\frac{i}{\sqrt{2}}e_2e_1+\hlf e_2e_3+\frac{i}{2}e_3e_1-\hlf e_3e_2\rp+\beta_{15}\beta_9^{-1}\lp ie_2e_3-ie_3e_2\rp,\\
    \lp\d_Z^{Z,Z}\rp_2(e_2)=\ & \beta_{14}\beta_9^{-1}\lp\frac{i}{\sqrt{2}}e_1e_1-\hlf e_1e_3-\frac{i}{\sqrt{2}}e_2e_2+\frac{i}{2}e_2e_3+\hlf e_3e_1+\frac{i}{2}e_3e_2\rp+\beta_{15}\beta_9^{-1}\lp -ie_1e_3+ie_3e_1\rp,\\
    \lp\d_Z^{Z,Z}\rp_2(e_3)=\ & \beta_{14}\beta_9^{-1}\lp\frac{i}{2}e_1e_1+\hlf e_1e_2-\hlf e_2e_1+\frac{i}{2}e_2e_2-ie_3e_3\rp+\beta_{15}\beta_9^{-1}\lp ie_1e_2-ie_2e_1\rp.
\end{align}

\subsection{Computing the associator}
\label{appsub:A4Associators}

If any of $A$, $B$, and $C$ are the trivial representation $1$, then the result is trivial.  For instance, if $A=1$ we have (applying the map to $(ev_B)v_C$) the result $\lp\la_{B,C}^D\rp_j(v_Bv_C)$ on the left and $\sum_\ell\mathsf{F}^{(1\,B\,C)\,D}_{Dj,B\ell}\lp\la_{B,C}^D\rp_\ell(v_Bv_C)$ on the right (note that we omit indices like $i$ if they can only take one value, as happens in a fusion involving the identity).  This determines
\be
\mathsf{F}^{(1\,B\,C)\,D}_{Dj,B\ell}=\d_{j\ell}.
\ee
Similarly if $B=1$ we fix
\be
\mathsf{F}^{(A\,1\,C)\,D}_{iC,A\ell}=\d_{i\ell},
\ee
and if $C=1$ then
\be
\mathsf{F}^{(A\,B\,1)\,D}_{iB,kD}=\d_{ik}.
\ee

Next suppose we have $A=B=C=X$.  The fusion rules then fix $D=1$ and also forces $E=F=Y$, with no multiplicities present.  Then our defining equation is
\be
\la_{X,Y}^1\circ\lp 1_X\otimes\la_{X,X}^Y\rp\circ\al_{X,X,X}=\mathsf{F}^{(X\,X\,X)\,1}_{Y,Y}\la_{Y,X}^1\circ\lp\la_{X,X}^Y\otimes 1_X\rp.
\ee
Applying it to the basis vector $(e_Xe_X)e_X$, the left-hand side is
\be
\ls\la_{X,Y}^1\circ\lp 1_X\otimes\la_{X,X}^Y\rp\circ\al_{X,X,X}\rs((e_Xe_X)e_X)=\ls\la_{X,Y}^1\circ\lp 1_X\otimes\la_{X,X}^Y\rp\rs(e_X(e_Xe_X))=\la_{X,Y}^1(e_X(\beta_1e_Y))=\beta_1\beta_2e,
\ee
while the right-hand side (excluding the coefficient) is
\be
\ls\la_{Y,X}^1\circ\lp\la_{X,X}^Y\otimes 1_X\rp\rs((e_Xe_X)e_X)=\la_{Y,X}^1((\beta_1e_Y)e_X)=\beta_1\beta_3e.
\ee
Comparing the two we conclude
\be
\mathsf{F}^{(X\,X\,X)\,1}_{Y,Y}=\frac{\beta_2}{\beta_3}.
\ee

Similar calculations give us all the other components where $A,B,C$ are either $X$ or $Y$,
\begin{align}
    \mathsf{F}^{(X\,X\,Y)\,X}_{1,Y}=\ & \frac{\beta_2}{\beta_1\beta_4},\\
    \mathsf{F}^{(X\,Y\,X)\,X}_{1,1}=\ & \frac{\beta_3}{\beta_2},\\
    \mathsf{F}^{(X\,Y\,Y)\,Y}_{X,1}=\ & \frac{\beta_1\beta_4}{\beta_2},\\
    \mathsf{F}^{(Y\,X\,X)\,X}_{Y,1}=\ & \frac{\beta_1\beta_4}{\beta_3},\\
    \mathsf{F}^{(Y\,X\,Y)\,Y}_{1,1}=\ & \frac{\beta_2}{\beta_3},\\
    \mathsf{F}^{(Y\,Y\,X)\,Y}_{1,X}=\ & \frac{\beta_3}{\beta_1\beta_4},\\
    \mathsf{F}^{(Y\,Y\,Y)\,1}_{X,X}=\ & \frac{\beta_3}{\beta_2}.
\end{align}

Next let's consider $A=B=X$ and $C=Z$.  Then we must have $D=E=Z$ and $F=Y$, and we don't have any multiplicities to worry about.  Now there are three basis vectors we could act on, but we only need one in order to fix the value of the associator component.  For instance, acting on $(e_Xe_X)e_1$ the left-hand side gives
\be
\ls\la_{X,Z}^Z\circ\lp 1_X\otimes\la_{X,Z}^Z\rp\circ\al_{X,X,Z}\rs((e_Xe_X)e_1)=\beta_5^2\lp -\hlf e_1+\frac{i}{2}e_2-\frac{i}{\sqrt{2}}e_3\rp,
\ee
and the right-hand side gives
\be
\mathsf{F}^{(X\,X\,Z)\,Z}_{Z,Y}\ls\la_{Y,Z}^Z\circ\lp\la_{X,X}^Y\otimes 1_Z\rp\rs((e_Xe_X)e_1)=\mathsf{F}^{(X\,X\,Z)\,Z}_{Z,Y}\beta_1\beta_7\lp -\frac{i}{2}e_1-\hlf e_2+\frac{1}{\sqrt{2}}e_3\rp.
\ee
Comparing the two fixes
\be
\mathsf{F}^{(X\,X\,Z)\,Z}_{Z,Y}=-\frac{i\beta_5^2}{\beta_1\beta_7}.
\ee

Similar calculations involving just one $Z$ among $A$, $B$, and $C$ give
\begin{align}
    \mathsf{F}^{(X\,Y\,Z)\,Z}_{Z,1}=\ & -\frac{i\beta_5\beta_7}{\beta_2},\\
    \mathsf{F}^{(Y\,X\,Z)\,Z}_{Z,1}=\ & -\frac{i\beta_5\beta_7}{\beta_3},\\
    \mathsf{F}^{(Y\,Y\,Z)\,Z}_{Z,X}=\ & \frac{\beta_7^2}{\beta_4\beta_5},\\
    \mathsf{F}^{(X\,Z\,X)\,Z}_{Z,Z}=\ & 1,\\
    \mathsf{F}^{(X\,Z\,Y)\,Z}_{Z,Z}=\ & 1,\\
    \mathsf{F}^{(Y\,Z\,X)\,Z}_{Z,Z}=\ & 1,\\
    \mathsf{F}^{(Y\,Z\,Y)\,Z}_{Z,Z}=\ & 1,\\
    \mathsf{F}^{(Z\,X\,X)\,Z}_{Y,Z}=\ & \frac{i\beta_1\beta_8}{\beta_6^2},\\
    \mathsf{F}^{(Z\,X\,Y)\,Z}_{1,Z}=\ & \frac{i\beta_2}{\beta_6\beta_8},\\
    \mathsf{F}^{(Z\,Y\,X)\,Z}_{1,Z}=\ & \frac{i\beta_3}{\beta_6\beta_8},\\
    \mathsf{F}^{(Z\,Y\,Y)\,Z}_{X,Z}=\ & \frac{\beta_4\beta_6}{\beta_8^2}.
\end{align}

Next, suppose $A=X$ but $B=C=Z$.  Then $D$ can be anything (since the fusion $XZ^2=1+X+Y+2Z$ contains all possible irreps).  If $D=1$, then $E=Y$ and $F=Z$ and we are still in the multiplicity free case.  We have to compare
\be
\ls\la_{X,Y}^1\circ\lp 1_X\otimes\la_{Z,Z}^Y\rp\circ\al_{X,Z,Z}\rs((e_Xe_1)e_1)=\hlf\beta_2\beta_{11}e,
\ee
to
\be
\ls\la_{Z,Z}^1\circ\lp\la_{X,Z}^Z\otimes 1_Z\rp\rs((e_Xe_1)e_1)=\hlf\beta_5\beta_9e,
\ee
which determines
\be
\mathsf{F}^{(X\,Z\,Z)\,1}_{Y,Z}=\frac{\beta_2\beta_{11}}{\beta_5\beta_9}.
\ee
Similarly,
\begin{align}
    \mathsf{F}^{(X\,Z\,Z)\,X}_{1,Z}=\ & \frac{i\beta_9}{\beta_5\beta_{10}},\\
    \mathsf{F}^{(X\,Z\,Z)\,Y}_{X,Z}=\ & \frac{i\beta_1\beta_{10}}{\beta_5\beta_{11}},\\
    \mathsf{F}^{(Y\,Z\,Z)\,1}_{X,Z}=\ & \frac{\beta_3\beta_{10}}{\beta_7\beta_9},\\
    \mathsf{F}^{(Y\,Z\,Z)\,X}_{Y,Z}=\ & \frac{\beta_4\beta_{11}}{\beta_7\beta_{10}},\\
    \mathsf{F}^{(Y\,Z\,Z)\,Y}_{1,Z}=\ & \frac{i\beta_9}{\beta_7\beta_{11}},\\
    \mathsf{F}^{(Z\,X\,Z)\,1}_{Z,Z}=\ & \frac{\beta_5}{\beta_6},\\
    \mathsf{F}^{(Z\,X\,Z)\,X}_{Z,Z}=\ & \frac{\beta_5}{\beta_6},\\
    \mathsf{F}^{(Z\,X\,Z)\,Y}_{Z,Z}=\ & \frac{\beta_5}{\beta_6},\\
    \mathsf{F}^{(Z\,Y\,Z)\,1}_{Z,Z}=\ & \frac{\beta_7}{\beta_8},\\
    \mathsf{F}^{(Z\,Y\,Z)\,X}_{Z,Z}=\ & \frac{\beta_7}{\beta_8},\\
    \mathsf{F}^{(Z\,Y\,Z)\,Y}_{Z,Z}=\ & \frac{\beta_7}{\beta_8},\\
    \mathsf{F}^{(Z\,Z\,X)\,1}_{Z,Y}=\ & \frac{\beta_6\beta_9}{\beta_3\beta_{11}},\\
    \mathsf{F}^{(Z\,Z\,X)\,X}_{Z,1}=\ & -\frac{i\beta_6\beta_{10}}{\beta_9},\\
    \mathsf{F}^{(Z\,Z\,X)\,Y}_{Z,X}=\ & -\frac{i\beta_6\beta_{11}}{\beta_1\beta_{10}},\\
    \mathsf{F}^{(Z\,Z\,Y)\,1}_{Z,X}=\ & \frac{\beta_8\beta_9}{\beta_2\beta_{10}},\\
    \mathsf{F}^{(Z\,Z\,Y)\,X}_{Z,Y}=\ & \frac{\beta_8\beta_{10}}{\beta_4\beta_{11}},\\
    \mathsf{F}^{(Z\,Z\,Y)\,Y}_{Z,1}=\ & -\frac{i\beta_8\beta_{11}}{\beta_9}.
\end{align}

More difficult is the case where $A=X$ and $B=C=D=Z$.  This case will be the first to exhibit the intricacies of higher multiplicity.  We will necessarily have $E=F=Z$ and then the indices $j$ and $\ell$ in (\ref{eq:FundamentalAssociatorRelation}) will run over two values ($i$ and $k$ will only take one value each and will be omitted).

With a bit of hindsight, let's apply (\ref{eq:FundamentalAssociatorRelation}) to the vector $((e_Xe_1)e_2)$.  For $j=1$ the left-hand side gives
\begin{align}
    \ls\la_{X,Z}^Z\circ\lp 1_X\otimes\lp\la_{Z,Z}^Z\rp_1\rp\circ\al_{X,Z,Z}\rs((e_Xe_1)e_2)=\ & \la_{X,Z}^Z(e_X(\beta_{12}(\frac{i}{\sqrt{2}}e_1-\hlf e_3)-i\beta_{13}e_3))\non\\
    =\ & \beta_5\beta_{12}\lp\frac{i}{\sqrt{2}}e_1+\hlf e_3\rp+\beta_5\beta_{13}\lp -\frac{1}{\sqrt{2}}e_1+\frac{i}{\sqrt{2}}e_2\rp,
\end{align}
and $j=2$ would give
\begin{align}
    \ls\la_{X,Z}^Z\circ\lp 1_X\otimes\lp\la_{Z,Z}^Z\rp_1\rp\circ\al_{X,Z,Z}\rs((e_Xe_1)e_2)=\ & \la_{X,Z}^Z(e_X(\beta_{14}(\frac{i}{\sqrt{2}}e_1-\hlf e_3)-i\beta_{15}e_3))\non\\
    =\ & \beta_5\beta_{14}\lp\frac{i}{\sqrt{2}}e_1+\hlf e_3\rp+\beta_5\beta_{15}\lp -\frac{1}{\sqrt{2}}e_1+\frac{i}{\sqrt{2}}e_2\rp.
\end{align}
The right-hand side is a linear combination of two pieces ($\ell=1$ and $\ell=2$),
\begin{align}
    \ls\lp\la_{Z,Z}^Z\rp_1\circ\lp\la_{X,Z}^Z\otimes 1_Z\rp\rs((e_Xe_1)e_2)=\ & \lp\la_{Z,Z}^Z\rp_1(\beta_5(\hlf e_1+\frac{i}{2}e_2-\frac{i}{\sqrt{2}}e_3)e_2)\non\\
    =\ & \beta_5\beta_{12}\lp\frac{1}{\sqrt{2}}e_2-\hlf e_3\rp+\beta_5\beta_{13}\lp\frac{1}{\sqrt{2}}e_1-\frac{i}{2}e_3\rp,\\
    \ls\lp\la_{Z,Z}^Z\rp_2\circ\lp\la_{X,Z}^Z\otimes 1_Z\rp\rs((e_Xe_1)e_2)=\ & \lp\la_{Z,Z}^Z\rp_2(\beta_5(\hlf e_1+\frac{i}{2}e_2-\frac{i}{\sqrt{2}}e_3)e_2)\non\\
    =\ & \beta_5\beta_{14}\lp\frac{1}{\sqrt{2}}e_2-\hlf e_3\rp+\beta_5\beta_{15}\lp\frac{1}{\sqrt{2}}e_1-\frac{i}{2}e_3\rp.
\end{align}
If we focus on just the coefficients of $e_1$, we would have (with the vector components corresponding to $j$)
\be
\lp\begin{matrix} \frac{\beta_5}{\sqrt{2}}\lp i\beta_{12}-\beta_{13}\rp \\ \frac{\beta_5}{\sqrt{2}}\lp i\beta_{14}-\beta_{15}\rp \end{matrix}\rp=\lp\mathsf{F}^{(X\,Z\,Z)\,Z}_{Z,Z}\rp\cdot\lp\begin{matrix} \frac{\beta_5\beta_{13}}{\sqrt{2}} \\ \frac{\beta_5\beta_{15}}{\sqrt{2}} \end{matrix}\rp,
\ee
where we have defined the two-by-two matrix
\be
\lp\mathsf{F}^{(X\,Z\,Z)\,Z}_{Z,Z}\rp:=\lp\begin{matrix} \mathsf{F}^{(X\,Z\,Z)\,Z}_{Z1,Z1} & \mathsf{F}^{(X\,Z\,Z)\,Z}_{Z1,Z2} \\ \mathsf{F}^{(X\,Z\,Z)\,Z}_{Z2,Z1} & \mathsf{F}^{(X\,Z\,Z)\,Z}_{Z2,Z2} \end{matrix}\rp.
\ee
If we focus instead on the coefficients of $e_2$ we get
\be
\lp\begin{matrix} \frac{i\beta_5\beta_{13}}{\sqrt{2}} \\ \frac{i\beta_5\beta_{15}}{\sqrt{2}} \end{matrix}\rp=\lp\mathsf{F}^{(X\,Z\,Z)\,Z}_{Z,Z}\rp\cdot\lp\begin{matrix} \frac{\beta_5\beta_{12}}{\sqrt{2}} \\ \frac{\beta_5\beta_{14}}{\sqrt{2}} \end{matrix}\rp.
\ee
Combining these, we can solve for $(\mathsf{F}^{(X\,Z\,Z)\,Z}_{Z,Z})$,
\begin{align}
\label{eq:FXZZZ}
    \lp\mathsf{F}^{(X\,Z\,Z)\,Z}_{Z,Z}\rp=\ & \lp\begin{matrix} \frac{\beta_5}{\sqrt{2}}\lp i\beta_{12}-\beta_{13}\rp & \frac{i\beta_5\beta_{13}}{\sqrt{2}} \\ \frac{\beta_5}{\sqrt{2}}\lp i\beta_{14}-\beta_{15}\rp & \frac{i\beta_5\beta_{15}}{\sqrt{2}} \end{matrix}\rp\lp\begin{matrix} \frac{\beta_5\beta_{13}}{\sqrt{2}} & \frac{\beta_5\beta_{12}}{\sqrt{2}} \\ \frac{\beta_5\beta_{15}}{\sqrt{2}} & \frac{\beta_5\beta_{14}}{\sqrt{2}} \end{matrix}\rp^{-1}\non\\
    =\ & \frac{1}{\beta_{12}\beta_{15}-\beta_{13}\beta_{14}}\lp\begin{matrix} \beta_{13}\beta_{14}-i\beta_{12}\beta_{14}+i\beta_{13}\beta_{15} & -\beta_{12}\beta_{13}+i\beta_{12}^2-i\beta_{13}^2 \\ \beta_{14}\beta_{15}-i\beta_{14}^2+i\beta_{15}^2 & -\beta_{12}\beta_{15}+i\beta_{12}\beta_{14}-i\beta_{13}\beta_{15} \end{matrix}\rp.
\end{align}
As a check, if we had replaced either column by coefficients of $e_3$ instead, we obtain the same matrix.  Moreover, if we had applied our equation to the vector $(e_Xe_1)e_1$ instead of $(e_Xe_1)e_2$ we would have only found one independent relation (the reason is that $(e_Xe_1)e_1$ is invariant under the action of $b$ and hence the result must simply be proportional to $v_1$).  But of course things still work with our result, i.e.~
\begin{align}
    \ls\la_{X,Z}^Z\circ\lp 1_X\otimes\lp\la_{Z,Z}^Z\rp_i\rp\circ\al_{X,Z,Z}\rs((e_Xe_1)e_1)=\ & \lp\begin{matrix} \la_{X,Z}^Z(e_X(\beta_{12}(\frac{i}{\sqrt{2}}e_2+\frac{i}{2}e_3))) \\ \la_{X,Z}^Z(e_X(\beta_{14}(\frac{i}{\sqrt{2}}e_2+\frac{i}{2}e_3))) \end{matrix}\rp\non\\
    =\ & \lp\begin{matrix} \beta_5\beta_{12}\lp -\frac{i}{\sqrt{2}}e_2-\frac{i}{2}e_3\rp \\ \beta_5\beta_{14}\lp -\frac{i}{\sqrt{2}}e_2-\frac{i}{2}e_3\rp \end{matrix}\rp,
\end{align}
\begin{align}
    \ls\lp\la_{Z,Z}^Z\rp_i\circ\lp\la_{X,Z}^Z\otimes 1_Z\rp\rs((e_Xe_1)e_1)=\ & \lp\begin{matrix} \lp\la_{Z,Z}^Z\rp_1(\beta_5(\hlf e_1+\frac{i}{2}e_2-\frac{i}{\sqrt{2}}e_3)e_1) \\ \lp\la_{Z,Z}^Z\rp_2(\beta_5(\hlf e_1+\frac{i}{2}e_2-\frac{i}{\sqrt{2}}e_3)e_1) \end{matrix}\rp\non\\
    =\ & \lp\begin{matrix} \beta_5\beta_{12}\lp\frac{i}{\sqrt{2}}e_2+\frac{i}{2}e_3\rp+\beta_5\beta_{13}\lp -\frac{1}{\sqrt{2}}e_2-\hlf e_3\rp \\ \beta_5\beta_{14}\lp\frac{i}{\sqrt{2}}e_2+\frac{i}{2}e_3\rp+\beta_5\beta_{15}\lp -\frac{1}{\sqrt{2}}e_2-\hlf e_3\rp \end{matrix}\rp.
\end{align}
Factoring out the vector $\beta_5(\frac{1}{\sqrt{2}}e_2+\hlf e_3)$, we get the equation
\be
\lp\begin{matrix} -i\beta_{12} \\ -i\beta_{14} \end{matrix}\rp=\lp\mathsf{F}^{(X\,Z\,Z)\,Z}_{Z,Z}\rp\cdot\lp\begin{matrix} i\beta_{12}-\beta_{13} \\ i\beta_{14}-\beta_{15} \end{matrix}\rp,
\ee
which is consistent with the solution (\ref{eq:FXZZZ}).

Similar steps lead to (we define the determinant $\Omega=\beta_{12}\beta_{15}-\beta_{13}\beta_{14}\ne 0$)
\begin{align}
    \lp\mathsf{F}^{(Y\,Z\,Z)\,Z}_{Zj,Z\ell}\rp=\ & \frac{1}{\Omega}\lp\begin{matrix} -\beta_{12}\beta_{15}+i\beta_{12}\beta_{14}-i\beta_{13}\beta_{15} & \beta_{12}\beta_{13}-i\beta_{12}^2+i\beta_{13}^2 \\ -\beta_{14}\beta_{15}+i\beta_{14}^2-i\beta_{15}^2 & \beta_{13}\beta_{14}-i\beta_{12}\beta_{14}+i\beta_{13}\beta_{15} \end{matrix}\rp,\\
    \lp\mathsf{F}^{(Z\,X\,Z)\,Z}_{iZ,Z\ell}\rp=\ & \frac{\beta_5}{\beta_6\Omega}\lp\begin{matrix} -\beta_{12}\beta_{15}+i\beta_{12}\beta_{14}-i\beta_{13}\beta_{15} & \beta_{12}\beta_{13}-i\beta_{12}^2+i\beta_{13}^2 \\ -\beta_{14}\beta_{15}+i\beta_{14}^2-i\beta_{15}^2 & \beta_{13}\beta_{14}-i\beta_{12}\beta_{14}+i\beta_{13}\beta_{15} \end{matrix}\rp,\\
    \lp\mathsf{F}^{(Z\,Y\,Z)\,Z}_{iZ,Z\ell}\rp=\ & \frac{\beta_7}{\beta_8\Omega}\lp\begin{matrix} \beta_{13}\beta_{14}-i\beta_{12}\beta_{14}+i\beta_{13}\beta_{15} & -\beta_{12}\beta_{13}+i\beta_{12}^2-i\beta_{13}^2 \\ \beta_{14}\beta_{15}-i\beta_{14}^2+i\beta_{15}^2 & -\beta_{12}\beta_{15}+i\beta_{12}\beta_{14}-i\beta_{13}\beta_{15} \end{matrix}\rp,\\
    \lp\mathsf{F}^{(Z\,Z\,X)\,Z}_{iZ,kZ}\rp=\ & \frac{1}{\Omega}\lp\begin{matrix} \beta_{13}\beta_{14}-i\beta_{12}\beta_{14}+i\beta_{13}\beta_{15} & -\beta_{12}\beta_{13}+i\beta_{12}^2-i\beta_{13}^2 \\ \beta_{14}\beta_{15}-i\beta_{14}^2+i\beta_{15}^2 & -\beta_{12}\beta_{15}+i\beta_{12}\beta_{14}-i\beta_{13}\beta_{15} \end{matrix}\rp,\\
    \lp\mathsf{F}^{(Z\,Z\,Y)\,Z}_{iZ,kZ}\rp=\ & \frac{1}{\Omega}\lp\begin{matrix} -\beta_{12}\beta_{15}+i\beta_{12}\beta_{14}-i\beta_{13}\beta_{15} & \beta_{12}\beta_{13}-i\beta_{12}^2+i\beta_{13}^2 \\ -\beta_{14}\beta_{15}+i\beta_{14}^2-i\beta_{15}^2 & \beta_{13}\beta_{14}-i\beta_{12}\beta_{14}+i\beta_{13}\beta_{15} \end{matrix}\rp.
\end{align}

The case $A=B=C=Z$ and $D$ is a one-dimensional irrep is a similar level of effort.  For instance, taking $D=1$, we must have $E=F=Z$ and $j$ and $k$ can run over two values each.  Computing (note that we need to go up to $(e_1e_1)e_2$ and $(e_1e_2)e_3$ to get two linearly independent results), we find
\begin{align}
    \ls\la_{Z,Z}^1\circ\lp 1_Z\otimes\lp\la_{Z,Z}^Z\rp_1\rp\circ\al_{Z,Z,Z}\rs((e_1e_1)e_2)=\ & \frac{i}{\sqrt{2}}\beta_9\beta_{12}e,\\
    \ls\la_{Z,Z}^1\circ\lp 1_Z\otimes\lp\la_{Z,Z}^Z\rp_1\rp\circ\al_{Z,Z,Z}\rs((e_1e_2)e_3)=\ & \lp -\hlf\beta_9\beta_{12}-i\beta_9\beta_{13}\rp e,\\
    \ls\la_{Z,Z}^1\circ\lp\lp\la_{Z,Z}^Z\rp_1\otimes 1_Z\rp\rs((e_1e_1)e_2)=\ & \frac{i}{\sqrt{2}}\beta_9\beta_{12}e,\\
    \ls\la_{Z,Z}^1\circ\lp\lp\la_{Z,Z}^Z\rp_1\otimes 1_Z\rp\rs((e_1e_2)e_3)=\ & \lp -\hlf\beta_9\beta_{12}-i\beta_9\beta_{13}\rp e,
\end{align}
and similarly for $j=2$ or $k=2$ if we replace $(\beta_{12},\beta_{13})$ by $(\beta_{14},\beta_{15})$.  Then
\be
\lp\mathsf{F}^{(Z\,Z\,Z)\,1}_{Zj,kZ}\rp=\lp\begin{matrix} \frac{i}{\sqrt{2}}\beta_9\beta_{12} & -\hlf\beta_9\beta_{12}-i\beta_9\beta_{13} \\ \frac{i}{\sqrt{2}}\beta_9\beta_{14} & -\hlf\beta_9\beta_{14}-i\beta_9\beta_{15} \end{matrix}\rp\lp\begin{matrix} \frac{i}{\sqrt{2}}\beta_9\beta_{12} & -\hlf\beta_9\beta_{12}-i\beta_9\beta_{13} \\ \frac{i}{\sqrt{2}}\beta_9\beta_{14} & -\hlf\beta_9\beta_{14}-i\beta_9\beta_{15} \end{matrix}\rp^{-1}=\lp\begin{matrix} 1 & 0 \\ 0 & 1 \end{matrix}\rp.
\ee

Similar calculations yield
\begin{align}
    \lp\mathsf{F}^{(Z\,Z\,Z)\,X}_{Zj,kZ}\rp=\ & \frac{1}{\Om}\lp\begin{matrix} \beta_{13}\beta_{14}-i\beta_{12}\beta_{14}+i\beta_{13}\beta_{15} & -\beta_{12}\beta_{13}+i\beta_{12}^2-i\beta_{13}^2 \\ \beta_{14}\beta_{15}-i\beta_{14}^2+i\beta_{15}^2 & -\beta_{12}\beta_{15}+i\beta_{12}\beta_{14}-i\beta_{13}\beta_{15} \end{matrix}\rp,\\
    \lp\mathsf{F}^{(Z\,Z\,Z)\,Y}_{Zj,kZ}\rp=\ & \frac{1}{\Om}\lp\begin{matrix} -\beta_{12}\beta_{15}+i\beta_{12}\beta_{14}-i\beta_{13}\beta_{15} & \beta_{12}\beta_{13}-i\beta_{12}^2+i\beta_{13}^2 \\ -\beta_{14}\beta_{15}+i\beta_{14}^2-i\beta_{15}^2 & \beta_{13}\beta_{14}-i\beta_{12}\beta_{14}+i\beta_{13}\beta_{15} \end{matrix}\rp.
\end{align}

Finally we turn to the case $A=B=C=D=Z$.  If $E$ is $1$, $X$, or $Y$ then $i$ and $j$ only take one label, while if $E=Z$ then $i$ and $j$ can take two values each, and similarly for $F$, $k$ and $\ell$.  All together, this means that $(\mathsf{F}^{(Z\,Z\,Z)\,Z})$ will be a seven-by-seven matrix.  To compute it, we first need to compute $(\la_{Z,E}^Z)_i\circ(1_Z\otimes(\la_{Z,Z}^E)_j)\circ\al_{Z,Z,Z}$ and $(\la_{F,Z}^Z)_\ell\circ((\la_{Z,Z}^F)_k\otimes 1_Z)$ on seven different vectors.  

To start, let's evaluate each of these on the vector $(u_1u_1)u_1$.  Since this is an eigenvector of $\rho(a)$ with eigenvalue $1$, each case must give a multiple of $e_3$.  Computing, we find
\begin{align}
    \la_{Z,1}^Z\circ\lp 1_Z\otimes\la_{Z,Z}^1\rp\circ\al_{Z,Z,Z}((u_1u_1)u_1)=\ & 0,\\
    \la_{Z,X}^Z\circ\lp 1_Z\otimes\la_{Z,Z}^X\rp\circ\al_{Z,Z,Z}((u_1u_1)u_1)=\ & 0,\\
    \la_{Z,Y}^Z\circ\lp 1_Z\otimes\la_{Z,Z}^Y\rp\circ\al_{Z,Z,Z}((u_1u_1)u_1)=\ & \beta_8\beta_{11}e_3,\\
    \lp\la_{Z,Z}^Z\rp_1\circ\lp 1_Z\otimes\lp\la_{Z,Z}^Z\rp_1\rp\circ\al_{Z,Z,Z}((u_1u_1)u_1)=\ & \beta_{12}\beta_{13}e_3,\\
    \lp\la_{Z,Z}^Z\rp_1\circ\lp 1_Z\otimes\lp\la_{Z,Z}^Z\rp_2\rp\circ\al_{Z,Z,Z}((u_1u_1)u_1)=\ & \beta_{13}\beta_{14}e_3,\\
    \lp\la_{Z,Z}^Z\rp_2\circ\lp 1_Z\otimes\lp\la_{Z,Z}^Z\rp_1\rp\circ\al_{Z,Z,Z}((u_1u_1)u_1)=\ & \beta_{12}\beta_{15}e_3,\\
    \lp\la_{Z,Z}^Z\rp_2\circ\lp 1_Z\otimes\lp\la_{Z,Z}^Z\rp_2\rp\circ\al_{Z,Z,Z}((u_1u_1)u_1)=\ & \beta_{14}\beta_{15}e_3,
\end{align}
and
\begin{align}
    \la_{1,Z}^Z\circ\lp\la_{Z,Z}^1\otimes 1_Z\rp((u_1u_1)u_1)=\ & 0,\\
    \la_{X,Z}^Z\circ\lp\la_{Z,Z}^X\otimes 1_Z\rp((u_1u_1)u_1)=\ & 0,\\
    \la_{Y,Z}^Z\circ\lp\la_{Z,Z}^Y\otimes 1_Z\rp((u_1u_1)u_1)=\ & \beta_7\beta_{11}e_3,\\
    \lp\la_{Z,Z}^Z\rp_1\circ\lp\lp\la_{Z,Z}^Z\rp_1\otimes 1_Z\rp((u_1u_1)u_1)=\ & \beta_{12}\lp i\beta_{12}-\beta_{13}\rp e_3,\\
    \lp\la_{Z,Z}^Z\rp_2\circ\lp\lp\la_{Z,Z}^Z\rp_1\otimes 1_Z\rp((u_1u_1)u_1)=\ & \beta_{12}\lp i\beta_{14}-\beta_{15}\rp e_3,\\
    \lp\la_{Z,Z}^Z\rp_1\circ\lp\lp\la_{Z,Z}^Z\rp_2\otimes 1_Z\rp((u_1u_1)u_1)=\ & \beta_{14}\lp i\beta_{12}-\beta_{13}\rp e_3,\\
    \lp\la_{Z,Z}^Z\rp_2\circ\lp\lp\la_{Z,Z}^Z\rp_2\otimes 1_Z\rp((u_1u_1)u_1)=\ & \beta_{14}\lp i\beta_{14}-\beta_{15}\rp e_3.
\end{align}
This translates to an equation
\begin{equation}
    \lp\begin{matrix} 0 \\ 0 \\ \beta_8\beta_{11} \\ \beta_{12}\beta_{13} \\ \beta_{13}\beta_{14} \\ \beta_{12}\beta_{15} \\ \beta_{14}\beta_{15} \end{matrix}\rp=\lp\mathsf{F}^{(Z\,Z\,Z)\,Z}\rp\cdot\lp\begin{matrix} 0 \\ 0 \\ \beta_7\beta_{11} \\ \beta_{12}(i\beta_{12}-\beta_{13}) \\ \beta_{12}(i\beta_{14}-\beta_{15}) \\ \beta_{14}(i\beta_{12}-\beta_{13}) \\\beta_{14}(i\beta_{14}-\beta_{15}) \end{matrix}\rp.
\end{equation}

Repeating this exercise with six more vectors ($(u_1u_1)u_2$, $(u_1u_1)e_3$, $(u_1u_2)u_1$, $(u_1u_2)u_2$, $(u_1u_2)e_3$, and $(u_1e_3)u_1$) yields a matrix equation which can be solved for $(\mathsf{F}^{(Z\,Z\,Z)\,Z})$,
\begin{multline}
    \lp\mathsf{F}^{(Z\,Z\,Z)\,Z}\rp=\\
    \lp\begin{matrix} 0 & \beta_9 & 0 & \beta_9 & 0 & 0 & 0 \\ 0 & 0 & \beta_6\beta_{10} & 0 & -i\beta_6\beta_{10} & 0 & \beta_6\beta_{10} \\ \beta_8\beta_{11} & 0 & 0 & 0 & 0 & -i\beta_8\beta_{11} & 0 \\ \beta_{12}\beta_{13} & \beta_{13}(i\beta_{12}-\beta_{13}) & \beta_{12}(i\beta_{12}-\beta_{13}) & (i\beta_{12}-\beta_{13})^2 & -\beta_{12}^2 & \beta_{13}^2 & \beta_{12}\beta_{13} \\ \beta_{13}\beta_{14} & \beta_{15}(i\beta_{12}-\beta_{13}) & \beta_{12}(i\beta_{14}-\beta_{15}) & (i\beta_{12}-\beta_{13})(i\beta_{14}-\beta_{15}) & -\beta_{12}\beta_{14} & \beta_{13}\beta_{15} & \beta_{12}\beta_{15} \\ \beta_{12}\beta_{15} & \beta_{12}(i\beta_{14}-\beta_{15}) & \beta_{14}(i\beta_{12}-\beta_{13}) & (i\beta_{12}-\beta_{13})(i\beta_{14}-\beta_{15}) & -\beta_{12}\beta_{14} & \beta_{13}\beta_{15} & \beta_{13}\beta_{14} \\ \beta_{14}\beta_{15} & \beta_{15}(i\beta_{14}-\beta_{15}) & \beta_{14}(i\beta_{14}-\beta_{15}) & (i\beta_{14}-\beta_{15})^2 & -\beta_{14}^2 & \beta_{15}^2 & \beta_{14}\beta_{15} \end{matrix}\rp\\
    \times\lp\begin{matrix} 0 & 0 & 0 & \beta_9 & \beta_9 & \beta_9 & 0 \\ 0 & 0 & 0 & 0 & 0 & 0 & \beta_5\beta_{10} \\ \beta_7\beta_{11} & -i\beta_7\beta_{11} & \beta_7\beta_{11} & 0 & 0 & 0 & 0 \\ \beta_{12}(i\beta_{12}-\beta_{13}) & -\beta_{12}^2 & \beta_{12}\beta_{13} & \beta_{13}^2 & \beta_{13}(i\beta_{12}-\beta_{13}) & -i\beta_{12}\beta_{13} & \beta_{12}(i\beta_{12}-\beta_{13}) \\ \beta_{12}(i\beta_{14}-\beta_{15}) & -\beta_{12}\beta_{14} & \beta_{12}\beta_{15} & \beta_{13}\beta_{15} & \beta_{13}(i\beta_{14}-\beta_{15}) & -i\beta_{13}\beta_{14} & \beta_{14}(i\beta_{12}-\beta_{13}) \\ \beta_{14}(i\beta_{12}-\beta_{13}) & -\beta_{12}\beta_{14} & \beta_{13}\beta_{14} & \beta_{13}\beta_{15} & \beta_{15}(i\beta_{12}-\beta_{13}) & -i\beta_{12}\beta_{15} & \beta_{12}(i\beta_{14}-\beta_{15}) \\ \beta_{14}(i\beta_{14}-\beta_{15}) & -\beta_{14}^2 & \beta_{14}\beta_{15} & \beta_{15}^2 & \beta_{15}(i\beta_{14}-\beta_{15}) & -i\beta_{14}\beta_{15} & \beta_{14}(i\beta_{14}-\beta_{15} \end{matrix}\rp^{-1}\\
    =\lp\begin{array}{c:c} A & B \\ \hdashline C & D \end{array}\rp,
\end{multline}
where in the last line we have written the result as a block matrix.  Here
\begin{equation}
    A=\frac{1}{3}\lp\begin{matrix} 1 & i\beta_9\beta_5^{-1}\beta_{10}^{-1} & i\beta_9\beta_7^{-1}\beta_{11}^{-1} \\ -i\beta_6\beta_{10}\beta_9^{-1} & \beta_6\beta_5^{-1} & \beta_6\beta_{10}\beta_7^{-1}\beta_{11}^{-1} \\ -i\beta_8\beta_{11}\beta_9^{-1} & \beta_8\beta_{11}\beta_5^{-1}\beta_{10}^{-1} & \beta_8\beta_7^{-1} \end{matrix}\rp,
\end{equation}
\begin{equation}
    B=\frac{1}{3\Om^2}\lp\begin{matrix} 2\beta_9b_{22} & -\beta_9(b_{12}+b_{21}) & -\beta_9(b_{12}+b_{21}) & 2\beta_9b_{11} \\ i\beta_6\beta_{10}b_{22} & i\beta_6\beta_{10}(-2b_{12}+b_{21}) & i\beta_6\beta_{10}(b_{12}-2b_{21}) & i\beta_6\beta_{10}b_{11} \\ i\beta_8\beta_{11}b_{22} & i\beta_8\beta_{11}(b_{12}-2b_{21}) & i\beta_8\beta_{11}(-2b_{12}+b_{21}) & i\beta_8\beta_{11}b_{11} \end{matrix}\rp,
\end{equation}
\begin{equation}
    C=\frac{1}{3}\lp\begin{matrix} -2\beta_9^{-1}b_{11} & i\beta_5^{-1}\beta_{10}^{-1}b_{11} & i\beta_7^{-1}\beta_{11}^{-1}b_{11} \\ -\beta_9^{-1}(b_{12}+b_{21}) & i\beta_5^{-1}\beta_{10}^{-1}(-b_{12}+2b_{21}) & i\beta_7^{-1}\beta_{11}^{-1}(2b_{12}-b_{21}) \\ -\beta_9^{-1}(b_{12}+b_{21}) & i\beta_5^{-1}\beta_{10}^{-1}(2b_{12}-b_{21}) & i\beta_7^{-1}\beta_{11}^{-1}(-b_{12}+2b_{21}) \\ -2\beta_9^{-1}b_{22} & i\beta_5^{-1}\beta_{10}^{-1}b_{22} & i\beta_7^{-1}\beta_{11}^{-1}b_{22} \end{matrix}\rp,
\end{equation}
and
\begin{equation}
    D=\frac{1}{3\Om^2}\lp\begin{matrix} -c_3 & c_2 & c_2 & -c_1 \\ -c_4 & c_3 & c_3 & -c_2 \\ -c_4 & c_3 & c_3 & -c_2 \\ -c_5 & c_4 & c_4 & -c_3 \end{matrix}\rp,
\end{equation}
where we have also defined
\begin{align}
    b_{11}=\ & \beta_{12}^2+i\beta_{12}\beta_{13}-\beta_{13}^2,\\
    b_{12}=\ & \beta_{12}\beta_{14}+i\beta_{12}\beta_{15}-\beta_{13}\beta_{15},\\
    b_{21}=\ & \beta_{12}\beta_{14}+i\beta_{13}\beta_{14}-\beta_{13}\beta_{15},\\
    b_{22}=\ & \beta_{14}^2+i\beta_{14}\beta_{15}-\beta_{15}^2,
\end{align}
\begin{align}
    c_1=\ & \beta_{12}^4+8i\beta_{12}^3\beta_{13}-6\beta_{12}^2\beta_{13}^2+4i\beta_{12}\beta_{13}^3-2\beta_{13}^4,\\
    c_2=\ & \beta_{12}^3\beta_{14}+2i\beta_{12}^3\beta_{15}+6i\beta_{12}^2\beta_{13}\beta_{14}-3\beta_{12}^2\beta_{13}\beta_{15}-3\beta_{12}\beta_{13}^2\beta_{14}+3i\beta_{12}\beta_{13}^2\beta_{15}+i\beta_{13}^3\beta_{14}-2\beta_{13}^3\beta_{15},\\
    c_3=\ & \beta_{12}^2\beta_{14}^2+4i\beta_{12}^2\beta_{14}\beta_{15}-\beta_{12}^2\beta_{15}^2+4i\beta_{12}\beta_{13}\beta_{14}^2-4\beta_{12}\beta_{13}\beta_{14}\beta_{15}\non\\
    & \quad +2i\beta_{12}\beta_{13}\beta_{15}^2-\beta_{13}^2\beta_{14}^2+2i\beta_{13}^2\beta_{14}\beta_{15}-2\beta_{13}^2\beta_{15}^2,\\
    c_4=\ & \beta_{12}\beta_{14}^3+6i\beta_{12}\beta_{14}^2\beta_{15}-3\beta_{12}\beta_{14}\beta_{15}^2+i\beta_{12}\beta_{15}^3+2i\beta_{13}\beta_{14}^3-3\beta_{13}\beta_{14}^2\beta_{15}+3i\beta_{13}\beta_{14}\beta_{15}^2-2\beta_{13}\beta_{15}^3,\\
    c_5=\ & \beta_{14}^4+8i\beta_{14}^3\beta_{15}-6\beta_{14}^2\beta_{15}^2+4i\beta_{14}\beta_{15}^3-2\beta_{15}^4.
\end{align}

For later reference, we also note that the inverse matrix is given by
\begin{equation}
    \lp\mathsf{F}^{(Z\,Z\,Z)\,Z}\rp^{-1}=\lp\begin{array}{c:c} \widetilde{A} & \widetilde{B} \\ \hdashline \widetilde{C} & \widetilde{D} \end{array}\rp,
\end{equation}
with
\begin{equation}
    \widetilde{A}=\frac{1}{3}\lp\begin{matrix} 1 & i\beta_9\beta_6^{-1}\beta_{10}^{-1} & i\beta_9\beta_8^{-1}\beta_{11}^{-1} \\ -i\beta_5\beta_{10}\beta_9^{-1} & \beta_5\beta_6^{-1} & \beta_5\beta_{10}\beta_8^{-1}\beta_{11}^{-1} \\ -i\beta_7\beta_{11}\beta_9^{-1} & \beta_7\beta_{11}\beta_6^{-1}\beta_{10}^{-1} & \beta_7\beta_8^{-1} \end{matrix}\rp,
\end{equation}
\begin{equation}
    \widetilde{B}=\frac{1}{3\Om^2}\lp\begin{matrix} 2\beta_9b_{22} & -\beta_9\lp b_{12}+b_{21}\rp & -\beta_9\lp b_{12}+b_{21}\rp & 2\beta_9b_{11} \\ i\beta_5\beta_{10}b_{22} & i\beta_5\beta_{10}\lp -2b_{12}+b_{21}\rp & i\beta_5\beta_{10}\lp b_{12}-2b_{21}\rp & i\beta_5\beta_{10}b_{11} \\ i\beta_7\beta_{11}b_{22} & i\beta_7\beta_{11}\lp b_{12}-2b_{21}\rp & i\beta_7\beta_{11}\lp -2b_{12}+b_{21}\rp & i\beta_7\beta_{11}b_{11} \end{matrix}\rp,
\end{equation}
\begin{equation}
    \widetilde{C}=\frac{1}{3}\lp\begin{matrix} -2\beta_9^{-1}b_{11} & i\beta_6^{-1}\beta_{10}^{-1}b_{11} & i\beta_8^{-1}\beta_{11}^{-1}b_{11} \\ -\beta_9^{-1}\lp b_{12}+b_{21}\rp & i\beta_6^{-1}\beta_{10}^{-1}\lp -b_{12}+2b_{21}\rp & i\beta_8^{-1}\beta_{11}^{-1}\lp 2b_{12}-b_{21}\rp \\ -\beta_9^{-1}\lp b_{12}+b_{21}\rp & i\beta_6^{-1}\beta_{10}^{-1}\lp 2b_{12}-b_{21}\rp & i\beta_8^{-1}\beta_{11}^{-1}\lp -b_{12}+2b_{21}\rp \\ -2\beta_9^{-1}b_{22} & i\beta_6^{-1}\beta_{10}^{-1}b_{22} & i\beta_8^{-1}\beta_{11}^{-1}b_{22} \end{matrix}\rp,
\end{equation}
and $\widetilde{D}=D$.

\subsection{A simplifying gauge choice}
\label{appsub:A4GaugeChoice}

Though we will leave the arbitrary constants $\beta_i$ explicit in all of our $\Rep(A_4)$ calculations in this paper, because we find they add a nice additional consistency check, it can be useful to make simplifying choices for these numbers.  In particular, some nice properties that we might want to hold (see, e.g.~the discussion in~\cite{Diatlyk:2023fwf}) would be that the matrices $(\mathsf{F}^{(A\,B\,C)\,D})$ are all unitary, and that the pairing between $(\la_{A,B}^C)_i$ and $(\d_A^{B,C})_j$ obeys the nice relation
\begin{equation}
\label{eq:GaugeOrthonormality}
    \lp\la_{A,B}^C\rp_i\circ\lp\d_C^{A,B}\rp_j=\sqrt{\left|\frac{c_Ac_B}{c_C}\right|}\d_{ij}\operatorname{id}_C,
\end{equation}
where $c_A$ is the quantum dimension of the simple line $A$.  In $A_4$ we have $c_1=c_X=c_Y=1$ and $c_Z=3$.  These conditions do not fix a gauge - in fact they don't even completely fix the components of $\mathsf{F}$ - but reduce the choices significantly.  Here we present one specific gauge choice which satisfies the conditions above.

First, imposing (\ref{eq:GaugeOrthonormality}), we require
\begin{equation}
    \beta_2=\beta_3=\beta_1\beta_4=-i\beta_6\beta_8,\qquad\beta_5\beta_{10}=\beta_7\beta_{11}=i\beta_9,
\end{equation}
and
\begin{align}
    2\beta_{12}\beta_{14}+i\beta_{12}\beta_{15}+i\beta_{13}\beta_{14}-2\beta_{13}\beta_{15}=\ & 0,\\
    \beta_{12}^2+i\beta_{12}\beta_{13}-\beta_{13}^2=\beta_{14}^2+i\beta_{14}\beta_{15}-\beta_{15}^2=\ & -\frac{\sqrt{3}}{2}\beta_9.
\end{align}
Before looking at conditions imposed by unitarity of $\mathsf{F}$, we'll make a couple of other choices.  Demanding that an alternative construction of the co-fusion basis (putting the co-evaluation map on the other side of the incoming line) agrees with the given one requires that we have
\begin{equation}
    \beta_5=\beta_6,\qquad\beta_7=\beta_8.
\end{equation}
And we would always have the freedom of rotating between the two basis vectors in the $\Hom(Z\otimes Z,Z)$ space.  We can use that freedom to set, e.g., $\beta_{14}=0$.  Then the conditions above fix $\beta_{13}=\frac{i}{2}\beta_{12}$ and $\beta_{15}^2=-\frac{3}{4}\beta_{12}^2=\frac{\sqrt{3}}{2}\beta_9$.  Once all of these choices are made, the only additional requirement from unitarity is that $|\beta_1\beta_2|=|\beta_5|^3$.

Defining $\om=-\beta_1\beta_2\beta_5^{-3}$, which should now be pure phase, the non-trivial (i.e.~not equal to $1$) components of $\mathsf{F}$ are
\begin{equation}
    \mathsf{F}^{(X\,X\,Z)\,Z}_{Z,Y}=\mathsf{F}^{(Z\,Y\,Y)\,Z}_{X,Z}=\mathsf{F}^{(Y\,Z\,Z)\,X}_{Y,Z}=\mathsf{F}^{(Z\,Z\,X)\,Y}_{Z,X}=\om^{-1},
\end{equation}
\begin{equation}
    \mathsf{F}^{(Y\,Y\,Z)\,Z}_{Z,X}=\mathsf{F}^{(Z\,X\,X)\,Z}_{Y,Z}=\mathsf{F}^{(X\,Z\,Z)\,Y}_{X,Z}=\mathsf{F}^{(Z\,Z\,Y)\,X}_{Z,Y}=\om,
\end{equation}
\begin{equation}
    \lp\mathsf{F}^{(X\,Z\,Z)\,Z}\rp=\lp\mathsf{F}^{(Z\,Y\,Z)\,Z}\rp=\lp\mathsf{F}^{(Z\,Z\,X)\,Z}\rp=\lp\mathsf{F}^{(Z\,Z\,Z)\,X}\rp=\lp\begin{matrix} -\hlf & \frac{\sqrt{3}}{2}\eta \\ -\frac{\sqrt{3}}{2}\eta & -\hlf \end{matrix}\rp,
\end{equation}
\begin{equation}
    \lp\mathsf{F}^{(Y\,Z\,Z)\,Z}\rp=\lp\mathsf{F}^{(Z\,X\,Z)\,Z}\rp=\lp\mathsf{F}^{(Z\,Z\,Y)\,Z}\rp=\lp\mathsf{F}^{(Z\,Z\,Z)\,Y}\rp=\lp\begin{matrix} -\hlf & -\frac{\sqrt{3}}{2}\eta \\ \frac{\sqrt{3}}{2}\eta & -\hlf \end{matrix}\rp,
\end{equation}
\begin{equation}
    \lp\mathsf{F}^{(Z\,Z\,Z)\,Z}\rp=\lp\begin{matrix} \frac{1}{3} & \frac{1}{3} & \frac{1}{3} & \frac{1}{\sqrt{3}} & 0 & 0 & \frac{1}{\sqrt{3}} \\ \frac{1}{3} & \frac{1}{3} & \frac{1}{3} & -\frac{1}{2\sqrt{3}} & -\hlf\eta & \hlf\eta & -\frac{1}{2\sqrt{3}} \\ \frac{1}{3} & \frac{1}{3} & \frac{1}{3} & -\frac{1}{2\sqrt{3}} & \hlf\eta & -\hlf\eta & -\frac{1}{2\sqrt{3}} \\ \frac{1}{\sqrt{3}} & -\frac{1}{2\sqrt{3}} & -\frac{1}{2\sqrt{3}} & \hlf & 0 & 0 & -\hlf \\ 0 & -\hlf\eta & \hlf\eta & 0 & -\hlf & -\hlf & 0 \\ 0 & \hlf\eta & -\hlf\eta & 0 & -\hlf & -\hlf & 0 \\ \frac{1}{\sqrt{3}} & -\frac{1}{2\sqrt{3}} & -\frac{1}{2\sqrt{3}} & -\hlf & 0 & 0 & \hlf \end{matrix}\rp,
\end{equation}
where $\eta=\pm 1$ appears in $\beta_{15}=\frac{\sqrt{3}i}{2}\eta\beta_{12}$.  Note that the two-by-two matrices all cube to the identity matrix, while the seven-by-seven matrix squares to the identity.

A very concrete set of choices (corresponding to $\om=\eta=1$) would be
\begin{equation}
    \beta_1=\beta_2=\beta_3=\beta_4=1,\quad\beta_5=\beta_6=-1,\quad\beta_7=\beta_8=-i,\quad\beta_9=-\frac{\sqrt{3}}{2},\quad\beta_{10}=\frac{\sqrt{3}i}{2}\quad\beta_{11}=\frac{\sqrt{3}}{2},
\end{equation}
\begin{equation}
    \beta_{12}=1,\quad\beta_{13}=\frac{i}{2},\quad\beta_{14}=0,\quad\beta_{15}=\frac{\sqrt{3}i}{2}.
\end{equation}

\subsection{Modular transformations of $A_4$ partial traces}
\label{appsub:A4ModTrans}

As explained in section~\ref{sect:genl}, we can use the associators to compute the modular transformations of our partial traces, using
\begin{align}
    \lp Z_{A,B}^C\rp_{ij}(\tau+1,\bar{\tau}+1)=\ & \sum_{D,k,\ell}\ls \mathsf{F}^{(A\,B\,\ov{A})\,B}\rs^{-1}_{iCj,kD\ell}\,\lp Z_{A,D}^B\rp_{k\ell}(\tau,\bar{\tau}),\\
    \lp Z_{A,B}^C\rp_{ij}(-1/\tau,-1/\bar{\tau})=\ & \sum_{D,k,\ell,m}\ls\mathsf{F}^{(A\,B\,\ov{A})\,B}\rs^{-1}_{iCj,kD\ell}\,\ls\mathsf{F}^{(A\,D\,\ov{B})\,1}\rs^{-1}_{kB,\ov{A}m}\,\lp Z_{B,\ov{A}}^D\rp_{\ell m}(\tau,\bar{\tau}).
\end{align}

Plugging in the results for $\Rep(A_4)$ gives us
\begin{align}
    Z_{1,1}^1(\tau+1,\btau+1)=\ & Z_{1,1}^1,\\
    Z_{1,X}^X(\tau+1,\btau+1)=\ & Z_{1,X}^X,\\
    Z_{1,Y}^Y(\tau+1,\btau+1)=\ & Z_{1,Y}^Y,\\
    Z_{1,Z}^Z(\tau+1,\btau+1)=\ & Z_{1,Z}^Z,\\
    Z_{X,1}^X(\tau+1,\btau+1)=\ & Z_{X,Y}^1,\\
    Z_{X,X}^Y(\tau+1,\btau+1)=\ & \frac{\beta_1\beta_4}{\beta_2}Z_{X,1}^X,\\
    Z_{X,Y}^1(\tau+1,\btau+1)=\ & \frac{\beta_2}{\beta_1\beta_4}Z_{X,X}^Y,\\
    Z_{X,Z}^Z(\tau+1,\btau+1)=\ & Z_{X,Z}^Z,\\
    Z_{Y,1}^Y(\tau+1,\btau+1)=\ & Z_{Y,X}^1,\\
    Z_{Y,X}^1(\tau+1,\btau+1)=\ & \frac{\beta_3}{\beta_1\beta_4}Z_{Y,Y}^X,\\
    Z_{Y,Y}^X(\tau+1,\btau+1)=\ & \frac{\beta_1\beta_4}{\beta_3}Z_{Y,1}^Y,\\
    Z_{Y,Z}^Z(\tau+1,\btau+1)=\ & Z_{Y,Z}^Z,\\
    Z_{Z,1}^Z(\tau+1,\btau+1)=\ & Z_{Z,Z}^1,\\
    Z_{Z,X}^Z(\tau+1,\btau+1)=\ & \frac{\beta_6}{\beta_5}Z_{Z,Z}^X,\\
    Z_{Z,Y}^Z(\tau+1,\btau+1)=\ & \frac{\beta_8}{\beta_7}Z_{Z,Z}^Y,
\end{align}
\begin{align}
    Z_{Z,Z}^1(\tau+1,\btau+1)=\ & \frac{1}{3}\ls Z_{Z,1}^Z+\frac{i\beta_9}{\beta_6\beta_{10}}Z_{Z,X}^Z+\frac{i\beta_9}{\beta_8\beta_{11}}Z_{Z,Y}^Z\rs\\
    & \quad +\frac{\beta_9}{3\Om^2}\ls 2b_{22}\lp Z_{Z,Z}^Z\rp_{11}-\lp b_{12}+b_{21}\rp\lp\lp Z_{Z,Z}^Z\rp_{12}+\lp Z_{Z,Z}^Z\rp_{21}\rp+2b_{11}\lp Z_{Z,Z}^Z\rp_{22}\rs,\non\\
    Z_{Z,Z}^X(\tau+1,\btau+1)=\ & \frac{\beta_5}{3}\ls -\frac{i\beta_{10}}{\beta_9}Z_{Z,1}^Z+\frac{1}{\beta_6}Z_{Z,X}^Z+\frac{\beta_{10}}{\beta_8\beta_{11}}Z_{Z,Y}^Z\rs+\frac{i\beta_5\beta_{10}}{3\Om^2}\ls b_{22}\lp Z_{Z,Z}^Z\rp_{11}\right.\non\\
    & \qquad\left. +\lp -2b_{12}+b_{21}\rp\lp Z_{Z,Z}^Z\rp_{12}+\lp b_{12}-2b_{21}\rp\lp Z_{Z,Z}^Z\rp_{21}+b_{11}\lp Z_{Z,Z}^Z\rp_{22}\rs,\\
    Z_{Z,Z}^Y(\tau+1,\btau+1)=\ & \frac{\beta_7}{3}\ls -\frac{i\beta_{11}}{\beta_9}Z_{Z,1}^Z+\frac{\beta_{11}}{\beta_6\beta_{10}}Z_{Z,X}^Z+\frac{1}{\beta_8}Z_{Z,Y}^Z\rs+\frac{i\beta_7\beta_{11}}{3\Om^2}\ls b_{22}\lp Z_{Z,Z}^Z\rp_{11}\right.\non\\
    & \qquad\left. +\lp b_{12}-2b_{21}\rp\lp Z_{Z,Z}^Z\rp_{12}+\lp -2b_{12}+b_{21}\rp\lp Z_{Z,Z}^Z\rp_{21}+b_{11}\lp Z_{Z,Z}^Z\rp_{22}\rs,\\
    \lp Z_{Z,Z}^Z\rp_{11}(\tau+1,\btau+1)=\ & \frac{b_{11}}{3}\ls -\frac{2}{\beta_9}Z_{Z,1}^Z+\frac{i}{\beta_6\beta_{10}}Z_{Z,X}^Z+\frac{i}{\beta_8\beta_{11}}Z_{Z,Y}^Z\rs\non\\
    & \quad +\frac{1}{3\Om^2}\ls -c_3\lp Z_{Z,Z}^Z\rp_{11}+c_2\lp Z_{Z,Z}^Z\rp_{12}+c_2\lp Z_{Z,Z}^Z\rp_{21}-c_1\lp Z_{Z,Z}^Z\rp_{22}\rs,\\
    \lp Z_{Z,Z}^Z\rp_{12}(\tau+1,\btau+1)=\ & \frac{1}{3}\ls -\frac{b_{12}+b_{21}}{\beta_9}Z_{Z,1}^Z+\frac{i\lp -b_{12}+2b_{21}\rp}{\beta_6\beta_{10}}Z_{Z,X}^Z+\frac{i\lp 2b_{12}-b_{21}\rp}{\beta_8\beta_{11}}Z_{Z,Y}^Z\rs\non\\
    & \quad +\frac{1}{3\Om^2}\ls -c_4\lp Z_{Z,Z}^Z\rp_{11}+c_3\lp Z_{Z,Z}^Z\rp_{12}+c_3\lp Z_{Z,Z}^Z\rp_{21}-c_2\lp Z_{Z,Z}^Z\rp_{22}\rs,\\
    \lp Z_{Z,Z}^Z\rp_{21}(\tau+1,\btau+1)=\ & \frac{1}{3}\ls -\frac{b_{12}+b_{21}}{\beta_9}Z_{Z,1}^Z+\frac{i\lp 2b_{12}-b_{21}\rp}{\beta_6\beta_{10}}Z_{Z,X}^Z+\frac{i\lp -b_{12}+2b_{21}\rp}{\beta_8\beta_{11}}Z_{Z,Y}^Z\rs\non\\
    & \quad +\frac{1}{3\Om^2}\ls -c_4\lp Z_{Z,Z}^Z\rp_{11}+c_3\lp Z_{Z,Z}^Z\rp_{12}+c_3\lp Z_{Z,Z}^Z\rp_{21}-c_2\lp Z_{Z,Z}^Z\rp_{22}\rs,\\
    \lp Z_{Z,Z}^Z\rp_{22}(\tau+1,\btau+1)=\ & \frac{b_{22}}{3}\ls -\frac{2}{\beta_9}Z_{Z,1}^Z+\frac{i}{\beta_6\beta_{10}}Z_{Z,X}^Z+\frac{i}{\beta_8\beta_{11}}Z_{Z,Y}^Z\rs\non\\
    & \quad +\frac{1}{3\Om^2}\ls -c_5\lp Z_{Z,Z}^Z\rp_{11}+c_4\lp Z_{Z,Z}^Z\rp_{12}+c_4\lp Z_{Z,Z}^Z\rp_{21}-c_3\lp Z_{Z,Z}^Z\rp_{22}\rs.
\end{align}
and
\begin{align}
    Z_{1,1}^1(-1/\tau,-1/\btau)=\ & Z_{1,1}^1,\\
    Z_{1,X}^X(-1/\tau,-1/\btau)=\ & Z_{X,1}^X,\\
    Z_{1,Y}^Y(-1/\tau,-1/\btau)=\ & Z_{Y,1}^Y,\\
    Z_{1,Z}^Z(-1/\tau,-1/\btau)=\ & Z_{Z,1}^Z,\\
    Z_{X,1}^X(-1/\tau,-1/\btau)=\ & Z_{1,Y}^Y,\\
    Z_{X,X}^Y(-1/\tau,-1/\btau)=\ & \frac{\beta_1\beta_4}{\beta_2}Z_{X,Y}^1,\\
    Z_{X,Y}^1(-1/\tau,-1/\btau)=\ & \frac{\beta_3}{\beta_1\beta_4}Z_{Y,Y}^X,\\
    Z_{X,Z}^Z(-1/\tau,-1/\btau)=\ & \frac{\beta_5\beta_9}{\beta_2\beta_{11}}Z_{Z,Y}^Z,\\
    Z_{Y,1}^Y(-1/\tau,-1/\btau)=\ & Z_{1,X}^X,\\
    Z_{Y,X}^1(-1/\tau,-1/\btau)=\ & \frac{\beta_2}{\beta_1\beta_4}Z_{X,X}^Y,\\
    Z_{Y,Y}^X(-1/\tau,-1/\btau)=\ & \frac{\beta_1\beta_4}{\beta_3}Z_{Y,X}^1,\\
    Z_{Y,Z}^Z(-1/\tau,-1/\btau)=\ & \frac{\beta_7\beta_9}{\beta_3\beta_{10}}Z_{Z,X}^Z,\\
    Z_{Z,1}^Z(-1/\tau,-1/\btau)=\ & Z_{1,Z}^Z,\\
    Z_{Z,X}^Z(-1/\tau,-1/\btau)=\ & \frac{\beta_2\beta_6\beta_{10}}{\beta_5\beta_8\beta_9}Z_{X,Z}^Z,\\
    Z_{Z,Y}^Z(-1/\tau,-1/\btau)=\ & \frac{\beta_3\beta_8\beta_{11}}{\beta_6\beta_7\beta_9}Z_{Y,Z}^Z,
\end{align}
\begin{align}
    Z_{Z,Z}^1(-1/\tau,-1/\btau)=\ & \frac{1}{3}\ls Z_{Z,Z}^1+\frac{i\beta_9}{\beta_5\beta_{10}}Z_{Z,Z}^X+\frac{i\beta_9}{\beta_7\beta_{11}}Z_{Z,Z}^Y\rs\\
    & \quad +\frac{\beta_9}{3\Om^2}\ls 2b_{22}\lp Z_{Z,Z}^Z\rp_{11}-\lp b_{12}+b_{21}\rp\lp\lp Z_{Z,Z}^Z\rp_{12}+\lp Z_{Z,Z}^Z\rp_{21}\rp+2b_{11}\lp Z_{Z,Z}^Z\rp_{22}\rs,\non\\
    Z_{Z,Z}^X(-1/\tau,-1/\btau)=\ & \frac{1}{3}\ls -\frac{i\beta_5\beta_{10}}{\beta_9}Z_{Z,Z}^1+Z_{Z,Z}^X+\frac{\beta_5\beta_{10}}{\beta_7\beta_{11}}Z_{Z,Z}^Y\rs+\frac{i\beta_5\beta_{10}}{3\Om^2}\ls b_{22}\lp Z_{Z,Z}^Z\rp_{11}\right.\non\\
    & \qquad\left. +\lp -2b_{12}+b_{21}\rp\lp Z_{Z,Z}^Z\rp_{12}+\lp b_{12}-2b_{21}\rp\lp Z_{Z,Z}^Z\rp_{21}+b_{11}\lp Z_{Z,Z}^Z\rp_{22}\rs,\\
    Z_{Z,Z}^Y(-1/\tau,-1/\btau)=\ & \frac{1}{3}\ls -\frac{i\beta_7\beta_{11}}{\beta_9}Z_{Z,Z}^1+\frac{\beta_7\beta_{11}}{\beta_5\beta_{10}}Z_{Z,Z}^X+Z_{Z,Z}^Y\rs+\frac{i\beta_7\beta_{11}}{3\Om^2}\ls b_{22}\lp Z_{Z,Z}^Z\rp_{11}\right.\non\\
    & \qquad\left. +\lp b_{12}-2b_{21}\rp\lp Z_{Z,Z}^Z\rp_{12}+\lp -2b_{12}+b_{21}\rp\lp Z_{Z,Z}^Z\rp_{21}+b_{11}\lp Z_{Z,Z}^Z\rp_{22}\rs,\\
    \lp Z_{Z,Z}^Z\rp_{11}(-1/\tau,-1/\btau)=\ & \frac{b_{11}}{3}\ls -\frac{2}{\beta_9}Z_{Z,Z}^1+\frac{i}{\beta_5\beta_{10}}Z_{Z,Z}^X+\frac{i}{\beta_7\beta_{11}}Z_{Z,Z}^Y\rs\non\\
    & \quad +\frac{1}{3\Om^2}\ls -c_3\lp Z_{Z,Z}^Z\rp_{11}+c_2\lp Z_{Z,Z}^Z\rp_{12}+c_2\lp Z_{Z,Z}^Z\rp_{21}-c_1\lp Z_{Z,Z}^Z\rp_{22}\rs,\\
    \lp Z_{Z,Z}^Z\rp_{12}(-1/\tau,-1/\btau)=\ & \frac{1}{3}\ls -\frac{b_{12}+b_{21}}{\beta_9}Z_{Z,Z}^1+\frac{i\lp -b_{12}+2b_{21}\rp}{\beta_5\beta_{10}}Z_{Z,Z}^X+\frac{i\lp 2b_{12}-b_{21}\rp}{\beta_7\beta_{11}}Z_{Z,Z}^Y\rs\non\\
    & \quad +\frac{1}{3\Om^2}\ls -c_4\lp Z_{Z,Z}^Z\rp_{11}+c_3\lp Z_{Z,Z}^Z\rp_{12}+c_3\lp Z_{Z,Z}^Z\rp_{21}-c_2\lp Z_{Z,Z}^Z\rp_{22}\rs,\\
    \lp Z_{Z,Z}^Z\rp_{21}(-1/\tau,-1/\btau)=\ & \frac{1}{3}\ls -\frac{b_{12}+b_{21}}{\beta_9}Z_{Z,Z}^1+\frac{i\lp 2b_{12}-b_{21}\rp}{\beta_5\beta_{10}}Z_{Z,Z}^X+\frac{i\lp -b_{12}+2b_{21}\rp}{\beta_7\beta_{11}}Z_{Z,Z}^Y\rs\non\\
    & \quad +\frac{1}{3\Om^2}\ls -c_4\lp Z_{Z,Z}^Z\rp_{11}+c_3\lp Z_{Z,Z}^Z\rp_{12}+c_3\lp Z_{Z,Z}^Z\rp_{21}-c_2\lp Z_{Z,Z}^Z\rp_{22}\rs,\\
    \lp Z_{Z,Z}^Z\rp_{22}(-1/\tau,-1/\btau)=\ & \frac{b_{22}}{3}\ls -\frac{2}{\beta_9}Z_{Z,Z}^1+\frac{i}{\beta_5\beta_{10}}Z_{Z,Z}^X+\frac{i}{\beta_7\beta_{11}}Z_{Z,Z}^Y\rs\non\\
    & \quad +\frac{1}{3\Om^2}\ls -c_5\lp Z_{Z,Z}^Z\rp_{11}+c_4\lp Z_{Z,Z}^Z\rp_{12}+c_4\lp Z_{Z,Z}^Z\rp_{21}-c_3\lp Z_{Z,Z}^Z\rp_{22}\rs.
\end{align}

One can then check that there are five independent modular invariant combinations of these partial traces,
\begin{equation}
    Z_{1,1}^1,
\end{equation}
\begin{equation}
    Z_{1,X}^X+Z_{1,Y}^Y+Z_{X,1}^X+\frac{\beta_2}{\beta_1\beta_4}Z_{X,X}^Y+Z_{X,Y}^1+Z_{Y,1}^Y+Z_{Y,X}^1+\frac{\beta_3}{\beta_1\beta_4}Z_{Y,Y}^X,
\end{equation}
\begin{equation}
    Z_{1,Z}^Z+\frac{i\beta_2}{\beta_5\beta_8}Z_{X,Z}^Z+\frac{i\beta_3}{\beta_6\beta_7}Z_{Y,Z}^Z+Z_{Z,1}^Z+\frac{i\beta_9}{\beta_6\beta_{10}}Z_{Z,X}^Z+\frac{i\beta_9}{\beta_8\beta_{11}}Z_{Z,Y}^Z+Z_{Z,Z}^1+\frac{i\beta_9}{\beta_5\beta_{10}}Z_{Z,Z}^X+\frac{i\beta_9}{\beta_7\beta_{11}}Z_{Z,Z}^Y,
\end{equation}
\begin{multline}
    Z_{1,Z}^Z+Z_{Z,1}^Z+Z_{Z,Z}^1+\frac{\beta_9}{3\Om^2}\ls\lp\beta_{14}+2i\beta_{15}\rp^2\lp Z_{Z,Z}^Z\rp_{11}-\lp\beta_{12}+2i\beta_{13}\rp\lp\beta_{14}+2i\beta_{15}\rp\lp\lp Z_{Z,Z}^Z\rp_{12}+\lp Z_{Z,Z}^Z\rp_{21}\rp\right.\\
    \left. +\lp\beta_{12}+2i\beta_{13}\rp^2\lp Z_{Z,Z}^Z\rp_{22}\rs,
\end{multline}
\begin{multline}
    \frac{\beta_9}{\Om^2}\ls\lp\beta_{14}^2-2i\beta_{14}\beta_{15}+2\beta_{15}^2\rp\lp Z_{Z,Z}^Z\rp_{11}-\lp\beta_{12}\beta_{14}-i\beta_{12}\beta_{15}-i\beta_{13}\beta_{14}+2\beta_{13}\beta_{15}\rp\lp\lp Z_{Z,Z}^Z\rp_{12}+\lp Z_{Z,Z}^Z\rp_{21}\rp\right.\\
    \left. +\lp\beta_{12}^2-2i\beta_{12}\beta_{13}+2\beta_{13}^2\rp\lp Z_{Z,Z}^Z\rp_{22}\rs.
\end{multline}
If we make the simplifying gauge choice of the previous section, these become
\begin{equation}
    Z_{1,1}^1,
\end{equation}
\begin{equation}
    Z_{1,X}^X+Z_{1,Y}^Y+Z_{X,1}^X+Z_{X,X}^Y+Z_{X,Y}^1+Z_{Y,1}^Y+Z_{Y,X}^1+Z_{Y,Y}^X,
\end{equation}
\begin{equation}
    Z_{1,Z}^Z+Z_{X,Z}^Z+Z_{Y,Z}^Z+Z_{Z,1}^Z+Z_{Z,X}^Z+Z_{Z,Y}^Z+Z_{Z,Z}^1+Z_{Z,Z}^X+Z_{Z,Z}^Y,
\end{equation}
\begin{equation}
    Z_{1,Z}^Z+Z_{Z,1}^Z+Z_{Z,Z}^1+\frac{2}{\sqrt{3}}\lp Z_{Z,Z}^Z\rp_{11},
\end{equation}
\begin{equation}
    -\sqrt{3}\lp Z_{Z,Z}^Z\rp_{11}+\sqrt{3}\lp Z_{Z,Z}^Z\rp_{22}.
\end{equation}

\section{Brauer-Picard group of Rep$(A_4)$}\label{app: BrPic}

In this appendix, we briefly review some background on the Brauer-Picard group, which we utilize in the main text to summarize the Rep$(A_4)$ gaugings. We refer the reader to \cite{eno} for more details.

We start with the notion of the \emph{Brauer-Picard groupoid}. The Brauer-Picard groupoid of a fusion category $\mathcal{C}$ is a 3-groupoid\footnote{Groupoids can be understood as special cases of categories, where every morphism is invertible.}. It contains the following information
\begin{itemize}
    \item Objects (i.e., 0-morphisms) are fusion categories that are Morita equivalent to $\mathcal{C}$.
    \item 1-morphisms are invertible bimodule categories between such fusion categories.
    \item  2-morphisms are equivalences of such bimodule categories.
    \item 3-morphisms are isomorphisms of such equivalences.
\end{itemize}
  From the physics perspective, we are interested in the 1-truncation of this 3-groupoid. Namely, objects correspond to categorical symmetries connected by gaugings, while 1-morphisms correspond to these gauging manipulations, or, equivalently, correspond to topological interfaces via half-space gaugings. 

Among all gauging manipulations, there are cases when the fusion category $\mathcal{C}$ is self-dual. Mathematically, these self-dual gaugings are associated with $\mathcal{C}-\mathcal{C}$ bimodule categories.  For a given object $\mathcal{C}$ in the Brauer-Picard groupoid, the equivalence classes of these bimodule categories build the \emph{Brauer-Picard} group, denoted as $\mathfrak{BrPic}(\mathcal{C})$\footnote{Recall that a groupoid with only one object is equivalent to an ordinary group.}. An alternative way to understand this group is by regarding it as the symmetry group of the 3D topological field theory \cite{Turaev:1992}, also known as the symmetry theory. Mathematically, this is translated in 
\begin{equation}
	\mathfrak{BrPic}(\mathcal{C})\cong \text{Aut}^{\text{br}}(\mathcal{Z}(\mathcal{C})).
\end{equation}
$\mathcal{Z}(\mathcal{C})$ is the Drinfeld center \cite{Drinfeld:1986, Muger:2003, Etingof:2002} for the fusion category $\mathcal{C}$, which contains the information of the 3D symmetry topological theory. 

When $\mathcal{C}=\text{Rep}(G)$ with G is a finite group, one has the following condition 
\begin{equation}
	\mathfrak{BrPic}(\text{Vec}_G)=\mathfrak{BrPic}(\text{Rep}(G))
\end{equation}
where $\text{Vec}_G$ can be physically understood as the group $G$ symmetry. 
Let $\mathbb{L}(G)$ denote the categorical Lagrangian Grassmannian of $\mathcal{Z}(\text{Vec}_G)$, and 
\begin{equation}
	\mathbb{L}_0(G)=\{ \mathcal{L}\in \mathbb{L}(G) | \mathcal{L}\cong \text{Rep}(G)~\text{as a braided fusion category} \}.
\end{equation}
The action of $\text{Aut}^{\text{br}}(\mathcal{Z}(\mathcal{C}))$ on $\mathbb{L}_0(G)$ is transitive \cite{nr2013}. Therefore, the image of $\text{Aut}^{\text{br}}(\mathcal{Z}(\mathcal{C}))$ is a transitive subgroup of $\text{Sym}(\mathbb{L}_0(G))$. Denote the stabilizer of the canonical Lagrangian subcategory Rep$(G)\subset \mathcal{Z}(\text{Vec}_G)$ in $\text{Aut}^{\text{br}}(\mathcal{Z}(\mathcal{C}))$ as $\mathfrak{Stab}(\text{Rep}(G))$, which can be computed by\footnote{$k^{\times}$ indicates the multiplicative group of nonzero elements. Here the second group cohomology can be intuitively understood as with the coefficient $U(1)$, i.e. the discrete torsion of the group $G$.}
\begin{equation}
	\mathfrak{Stab}(\text{Rep}(G))\cong H^2(G, k^{\times})\rtimes \text{Out}(G).
\end{equation}
This allows one to find the order of the Brauer-Picard group \cite{nr2013}
\begin{equation}
	|\mathfrak{BrPic}(\text{Rep}(G))|=|H^2(G, k^{\times})|\times |\text{Out}(G)| \times |\mathbb{L}_0(G)|.
\end{equation}

For our case of interest, $G=A_4$, for which 
\begin{equation}
	H^2(A_4,k^{\times})\cong \mathbb{Z}_2, ~\text{Out}(A_4)=\mathbb{Z}_2
\end{equation}
 This gives rise to the stabilizer for Rep$(A_4)$ 
\begin{equation}
	\mathfrak{Stab}(\text{Rep}(G))=\mathbb{Z}_2\times \mathbb{Z}_2.
\end{equation}
The set $\mathbb{L}_0(A_4)$ consists of three Lagrangian subcategories \cite{Naidu:2008}
\begin{equation}
	\mathcal{L}_{(1,1)},~\mathcal{L}_{(\mathbb{Z}_2\times\mathbb{Z}_2, 1)},~\mathcal{L}_{(\mathbb{Z}_2\times \mathbb{Z}_2, \mu)}.
\end{equation}
In total, this $\mathbb{L}_0(A_4)$ together with $\mathfrak{Stab}(\text{Rep}(G))$ leads to a order 12 group. In \cite{nr2013}, it is shown that this group is $D_6$.

\end{document}